\documentclass{article}
    \PassOptionsToPackage{numbers, compress}{natbib}
\usepackage{wrapfig, blindtext} 
\usepackage[final]{neurips_2024}
\usepackage[utf8]{inputenc} 
\usepackage[T1]{fontenc}    
\usepackage{url}            
\usepackage{booktabs}       
\usepackage{amsfonts}       
\usepackage{nicefrac}       
\usepackage{microtype}      
\usepackage{xcolor}         

\usepackage[pagebackref=true,breaklinks=true,colorlinks,bookmarks=false]{hyperref}
\usepackage{amsmath}
\usepackage{graphicx}
\usepackage{adjustbox}
\usepackage{orcidlink}
\usepackage{bbding}
\usepackage{bm} 

\usepackage{algpseudocode}
\usepackage[linesnumbered,ruled]{algorithm2e} 
\usepackage{arydshln} 
\usepackage{subcaption}
\newif\ifdraft
\draftfalse

\ifdraft

\newcommand{\smc}[1]{{\color{olive}[\textbf{SM:} \textit{#1}]}}

\newcommand{\bl}[1]{{\color{purple}#1}}
\newcommand{\jw}[1]{{\color{blue}#1}}

\newcommand{\sm}[1]{{\color{olive}#1}}

\else
\newcommand{\zwc}[1]{}
\newcommand{\yzc}[1]{}
\newcommand{\smc}[1]{}
\newcommand{\bl}[1]{{\color{black}#1}}

\newcommand{\jw}[1]{{\color{black}#1}}

\newcommand{\sm}[1]{{\color{black}#1}}
\fi

\newcommand{\fakescore}[1]{\textcolor{gray}{#1}}

\title{Sagiri: Low Dynamic Range Image Enhancement with \\Generative Diffusion Prior}
\author{
    Baiang Li\textsuperscript{1,5},
    Sizhuo Ma\textsuperscript{3},
    Yanhong Zeng\textsuperscript{1},
    Xiaogang Xu\textsuperscript{2,4}, \\
    \textbf{
    Youqing Fang\textsuperscript{1},
    Zhao Zhang\textsuperscript{5},
    Jian Wang\textsuperscript{3}\thanks{Co-corresponding authors. J. Wang initialized the project.},  \,
    Kai Chen\textsuperscript{1}\footnotemark[1]}
    \\
    \textsuperscript{1}Shanghai AI Lab, 
    \textsuperscript{2}The Chinese University of Hong Kong, \\
    \textsuperscript{3}Snap Inc., 
    \textsuperscript{4}Zhejiang University,
    \textsuperscript{5}Hefei University of Technology
     }

\begin{document}
\maketitle

\begin{figure*}[h]
    \centering
    \includegraphics[width=1\textwidth]{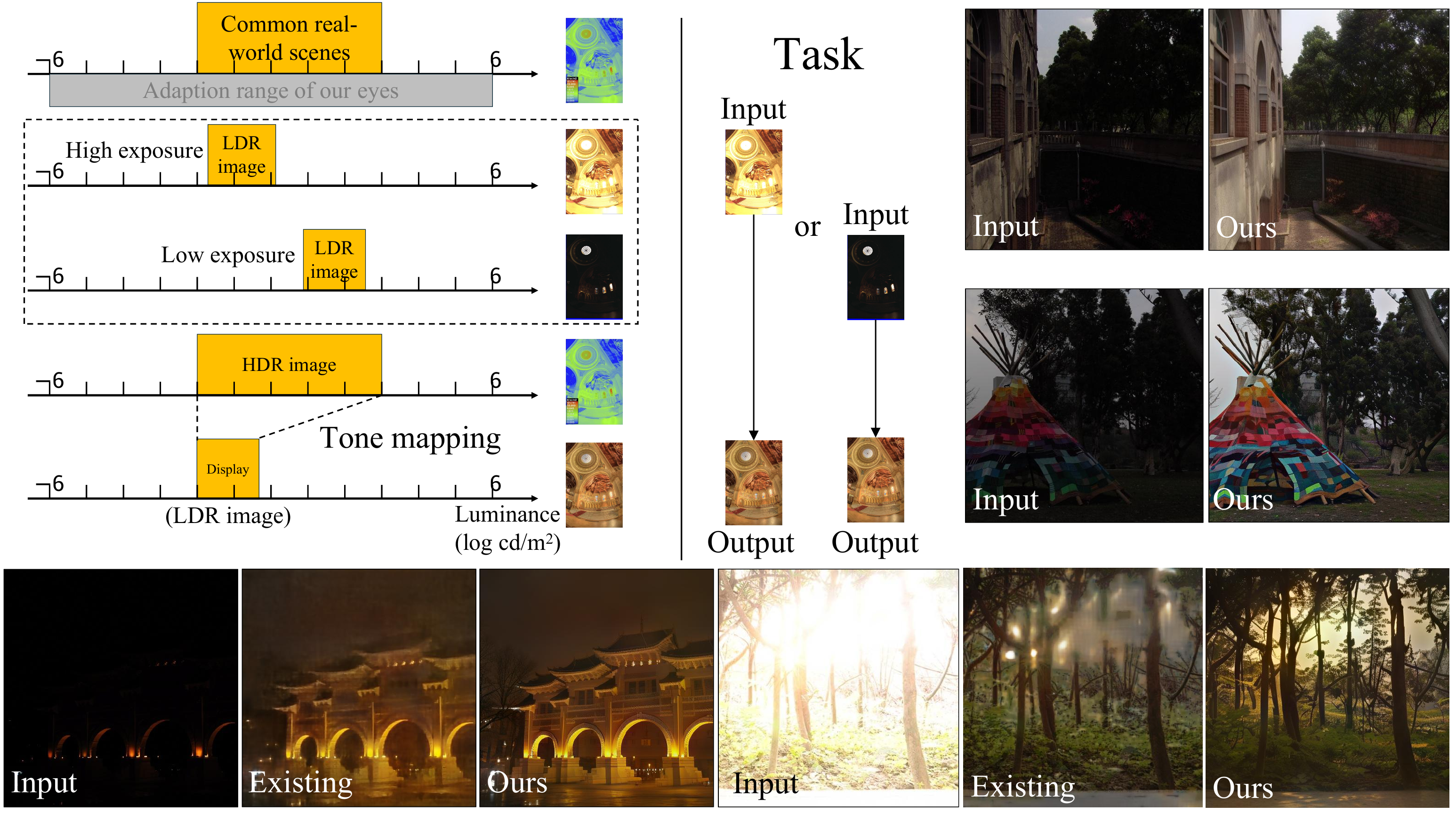}
    \caption{
    \sm{Common real-world scenes have broad dynamic ranges. 
    A typical 8-bit camera captures a limited dynamic range, where the exposure value determines which part of the scene's dynamic range is captured, often resulting in either oversaturated bright regions or quantized dark areas overlwhelmed by noise.
    Traditionally, multiple exposures are merged into an HDR image (32-bit or 64-bit) to accurately represent the scene, which is subsequently tone-mapped to an 8-bit image for LDR displays.
    In our method, we directly learn to generate the final output from a single LDR image with generative diffusion prior, which includes (1) color mapping, (2) generating reasonable content for saturated/black regions, (3) enhancing details in low bit-depth regions, (4) dark region denoising.}
    }
    \label{fig:problem}
\end{figure*}
\begin{abstract}

\sm{Capturing High Dynamic Range (HDR) scenery using 8-bit cameras often suffers from over-/underexposure, loss of fine details due to low bit-depth compression, skewed color distributions, and strong noise in dark areas. Traditional LDR image enhancement methods primarily focus on color mapping, which enhances the visual representation by expanding the image's color range and adjusting the brightness. However, these approaches fail to effectively restore content in \emph{dynamic range extremes}, which are regions with pixel values close to 0 or 255.
To address the full scope of challenges in HDR imaging and surpass the limitations of current models, we propose a novel two-stage approach. The first stage maps the color and brightness to an appropriate range while keeping the existing details, and the second stage utilizes a diffusion prior to generate content in dynamic range extremes lost during capture. This generative refinement module can also be used as a \emph{plug-and-play} module to enhance and complement existing LDR enhancement models.
The proposed method markedly improves the quality and details of LDR images, demonstrating superior performance through rigorous experimental validation. 
The project page is at \url{https://sagiri0208.github.io}.}
\end{abstract}

\section{Introduction}
\label{sec:intro}
\vspace{-1mm}

\sm{Scenes with broad dynamic ranges are common in the real world. However, most common cameras such as those on phones usually make use of 8-bit image sensors, which have limited dynamic ranges. Thus, they cannot capture image details of both the bright sun and leaves in the shadow. Exposure bracketing~\cite{debevec1997recovering} has been proposed to create high dynamic range (HDR) images from multiple low dynamic range (LDR) ones. However, it has to be enabled at capture time, requires long capture duration, and can be computationally heavy for motion compensation.}
\sm{Deep learning models have made progress in recoering details in overexposed and underexposed regions from a \emph{single} LDR image, }
while they do not always provide a fully satisfactory visual experience \cite{zou2023rawhdr, wang2023glowgan}. This limitation is particularly noticeable in regions with brightness levels \sm{close to 0 or 255, which we term \emph{dynamic range extremes} in this paper}, as highlighted in \sm{Figure}~\ref{fig:problem}.

\sm{In this work, we aim to push the boundaries of \emph{single LDR image enhancement}, which we define}
\bl{as the process of enhancing details that are lost or obscured due to the camera's limited dynamic range. This task includes improving tone mapping, reducing noise, enhancing details in regions affected by low bit-depth, and generating content in oversaturated and dark areas to achieve a closer representation of the original scene's dynamic range, as shown in Figure~\ref{fig:problem}.}
\sm{While Convolutional Neural Networks (CNNs) and transformers have been shown to be capable of tone mapping and denoising, they face challenges with recovering content in dynamic range extremes where the information is almost completely lost at capture. Fortunately,}
trained on vast datasets with abundant texts and images~\cite{ho2020denoising, song2020denoising}, Stable Diffusion~\cite{rombach2022high} features exceptional generative abilities and provides a novel and promising approach to these challenges.

We introduce a two-stage model tailored to LDR image enhancement.
Initially, an LDR image is processed by Latent-SwinIR$_{c}$ (LS), a transformer-based model~\cite{liang2021swinir}, which is designed to harmonize the uneven color distribution of LDR images. \sm{In this way, extremely bright or dark regions are mapped reasonable brightness ranges for human viewers\footnote{This process is similar to tone mapping except that the input is also an LDR image.}.}
This is achieved through a specially formulated color mapping loss \sm{computed over color histograms}.
Following this initial enhancement, the image is further refined by our Sagiri model, 
which leverages the powerful generative capabilities of ControlNet~\cite{zhang2023adding}. Sagiri utilizs the previously restored image as a reference in a parallel encoder configuration \sm{to effectively enhance content that was inadequately recovered in the initial restoration stage, and generate new image details completely lost due to over/underexposure, offering a robust solution for fine-grained image enhancement}.
\sm{Additionally, we propose an adaptive regional processing approach during the sampling process, enabling users to direct content generation through customized prompts such as text or pixel masks.} 
Our contributions in this research can be summarized as follows:
\begin{itemize}
    \item \jw{We introduce the LS-Sagiri framework, a novel two-stage model specifically tailored for single LDR image enhancement, where Stage 1  adjusts the overall color and brightness, and Stage 2 enhances/generates the content details.}
    \item \jw{Our Stage 2 model Sagiri employs a generative diffusion prior to create plausible content in saturated and black areas, and to enhance details particularly in regions suffering from low bit-depth. A two-step strategy is proposed to train the model such that it can function as a plug-and-play component for enhancing existing methods.}
    \item \jw{Comprehensive experiments show our method's superior performance in both quantitative and visual results, as well as  Sagiri's versatility in enhancing existing methods.}
\end{itemize}
\section{Related Work}
\subsection{HDR Image Reconstruction and LDR Image Enhancement}
Various restoration-based models have been proposed~\cite{wang2021deep} for HDR image reconstruction.
\sm{While multi-image methods~\cite{liu2022ghost} may achieve higher fidelity to the actual scene, we focus on single-image methods which reconstruct HDR or tone mapped LDR images from one image and are thus more flexible. }
SingleHDR~\cite{liu2020single} incorporates domain knowledge of the LDR image formation pipeline into their model, tackling the reconstruction problem by reversing the image formation process. 
\sm{However, error accumulation can happen at each stage of the pipeline.}
\sm{Multi-exposure generation~\cite{le2023single} synthesizes multiple images at different exposure values and then fuse them using conventional HDR methods.}
HDRUNet~\cite{chen2021hdrunet} learns an end-to-end mapping strategy for single-image HDR reconstruction with denoising and dequantization.
Wang et al.~\cite{wang2022local} observe that local color distributions of an image suffer from both over- and under-exposure and propose a method to enhance the two types of regions. RawHDR~\cite{zou2023rawhdr} focuses on raw images, learning exposure masks to separate challenging regions in high dynamic scenes.
Due to their limited generative ability, the methods above struggle with dynamic range extremes.
Generative models offer alternative solutions for HDR image reconstruction. A recent method~\cite{fei2023generative} leverages a diffusion prior
for unified unsupervised image restoration and enhancement, employing hierarchical guidance and patch-based methods to improve the quality of natural image outputs. However, it requires an extremely long inference time and needs multiple LDR images as inputs. GlowGAN~\cite{wang2023glowgan} trains a generative adversarial network to generate HDR images from in-the-wild LDR image collections in an unsupervised manner. Despite its novelty, GlowGAN still faces challenges in generating satisfactory results for \emph{large} over-exposed areas, 
\sm{which our second-stage Sagiri model excels by utilizing a diffusion prior.}
\subsection{Conditional Generation Based on Stable Diffusion}
Stable diffusion models~\cite{rombach2022high} have made significant strides in conditional generation. RePaint~\cite{lugmayr2022repaint} introduces a DDPM-based inpainting approach, utilizing a pretrained unconditional DDPM as the generative prior. \bl{Zhang et al.~\cite{zhang2023adding}  introduces a new architecture to add spatial conditioning controls to large stable diffusion models.}
Chu et al.~\cite{chu2023rethinking} proposes an Unbiased Fast Fourier Convolution module for efficient frequency information capture and artifact-free reconstruction. Uni-paint~\cite{yang2023uni} presents a multimodal inpainting method, which is based on stable diffusion v1.4, and offers various modes of guidance without requiring task-specific finetuning. The Pixel Spread Model~\cite{li2023image} iteratively employs a decoupled probabilistic model to selectively spread informative pixels throughout the image in a few iterations. MagicRemover~\cite{yang2023magicremover} proposes a tuning-free method leveraging powerful diffusion models for text-guided image inpainting, further introducing a classifier optimization algorithm to enhance denoising stability within fewer sampling steps. 
Although these stable diffusion-based models possess content generation capabilities, their abilities are primarily confined to completing missing areas. Besides, they lack the capacity to utilize the existing texture and color information within dynamic range extreme region. 
Furthermore, they are unable to generate content with high relevance based on the texture and color information of the dynamic range extreme region, nor can they perform 
\sm{fine adjustment} on the entire image to enhance overall details.

\section{Our Method}
\vspace{-2mm}
Given an LDR image input, our approach initially employs a restoration model in the first stage to adjust the overall brightness and corresponding color, aiming to achieve an output with a color distribution closely resembling the ground truth (GT). However, 
the limited generative capability of the restoration model falls short in restoring or generating the lost details in \sm{dynamic range extremes, necessitating our second-stage model which makes use of a diffusion prior.}
To effectively guide 
\sm{the learning at each stage} and leverage the advantages of each model in a more targeted manner,
we design specialized loss functions for each stage. In the first stage, the color reconstruction loss focuses on color restoration and brightness adjustment \sm{by aligning the color histograms of the predicted and target images.}
In the second stage, the content enhancement loss is employed to generate finer texture details and align the content distribution of the generated image closer to that of the detail-rich images, which is instrumental in facilitating the generation of missing details. \textbf{Details of the designed losses are included in the supplementary material.}
\vspace{-1mm}
\subsection{Color Restoration and Brightness Adjustment}
\label{sec:blind}
In our approach, we utilize SwinIR~\cite{liang2021swinir} as the color restoration and brightness adjustment module, 
with modifications to the pre-processing and post-processing functions. Specifically, we employ a pixel 
\begin{figure*}[h]
	\centering
	\includegraphics[width=1.0\textwidth]{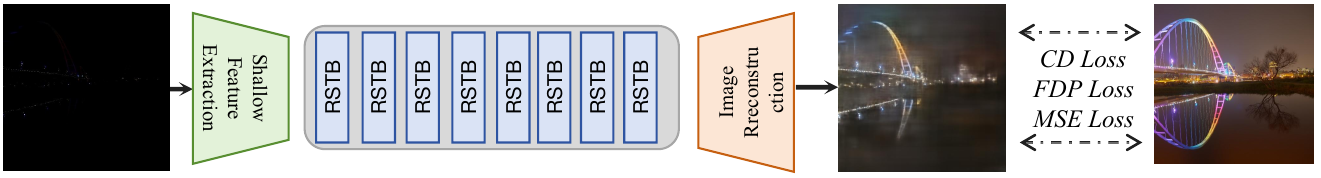}
 	\caption{Overview of Latent-SwinIR$_{c}$ (LS) and color reconstruction loss. 
\bl{Through our unique design, it is able to capture color distribution with higher fidelity.}}
	\label{swinir}
 \vspace{-2mm}
\end{figure*}
unshuffle operation to downsample the original low-quality input by a scale factor of 8. For pre-processing, we incorporate a $3 \times 3$ convolution layer for shallow feature extraction and color space feature extraction. This is followed by Residual Swin Transformer Blocks (RSTB) for feature processing. The features are then upsampled back to the original image space using nearest interpolation and a $3 \times 3$ convolutional layer, repeated three times to take the features back to its original image space. We call this model \emph{Latent-SwinIR$_{c}$} (Figure.~\ref{swinir}),
which focuses on adjusting the overall color distribution while possessing some degree of image content recovery capabilities.
\subsection{Content Generation}
\begin{figure*}[t]
	\centering
	\includegraphics[width=1.0\textwidth]{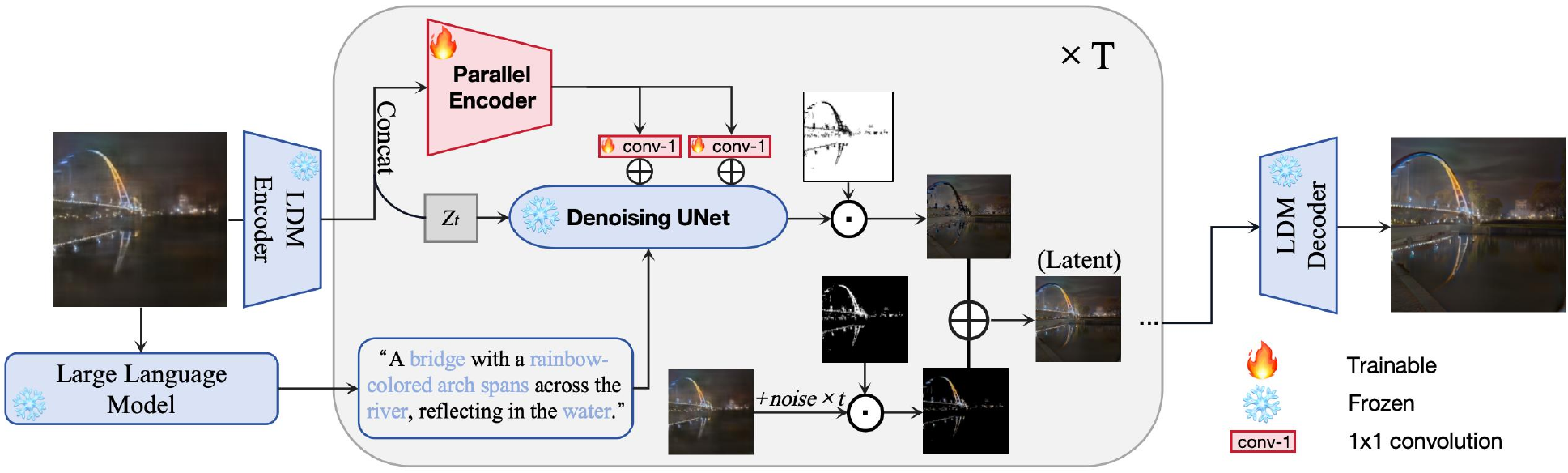}
 	\caption{\sm{Overview of Sagiri. Our model takes the output of the previous stage as input, with an optional text prompt input generated using a large language model. It uses a pretrained VAE encoder to map  previous result into the latent space. The obtained latent feature is concatenated with time-step noise to serve as condition. 
 An unknown region mask (pixels with values of 0 or 255) is used to combine the input latent feature with the denoised feature map.}}
 \label{sagiri}
 \vspace{-4mm}
\end{figure*}

After obtaining a restored image with a balanced color distribution in the first stage, we still need to address 
missing details in areas with poor visual qualities, especially in dynamic range extremes. Previous methods have attempted to handle these extreme cases, but often fail to generate \cite{wang2022local,zou2023rawhdr}
\begin{wrapfigure}{r}{0.4\linewidth}
    \centering
    \vspace{-4mm}
	\includegraphics[width=0.4\textwidth]{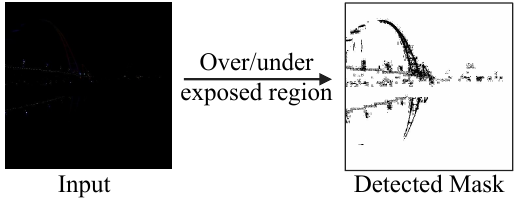}
 \caption{\sm{Unknown region mask. Pixels with values of 0 or 255 are detected as unknown regions. 
 The mask is downsampled and broadcasted to match the shape of the latent feature maps.
 }}
 \vspace{-4mm}
 \label{mask_generation}
\end{wrapfigure}
high-quality details when content is missing. Leveraging the diffusion model's ability to generate high-quality images, we propose a generative approach,
which is shown in Figure \ref{sagiri}. 
In this stage, the restored result from the previous stage is first processed through a Variational Autoencoder (VAE)~\cite{kingma2019introduction} to obtain its latent representation.
We employ a parallel module containing the same encoder and middle block as the U-Net denoiser. The latent feature, concatenated with noise, is fed into this parallel encoder. The outputs of different encoder blocks serve as latent controls, concatenated with the denoising U-Net's decoder part as conditions. Newly added parameters are initialized to zero, while other weights are inherited from the pre-trained denoising U-Net. A $1 \times 1$ convolution is added before each concatenation, with the new parameters and module being trainable and others kept frozen.
For the denoising process, we differentiate between known regions (where content generation is
not desired) and unknown regions (where content generation is needed). \sm{This is determined by a binary mask where pixels with values of 0 or 255 are marked as unknown regions, as shown in Figure~\ref{mask_generation}.} Inspired by RePaint \cite{lugmayr2022repaint}, \sm{we preserve the known regions by directly predicting the $t$-th step from the initial latent feature map utilizing the properties of a Markov chain of added Gaussian noise. }
For unknown regions, we use the predicted denoised results at step $t$. The denoised latent feature at step $t$ can be expressed as:
\begin{align}
x_{t-1}^{known} &\sim \mathcal{N}(\sqrt{\overline\alpha_{t}}x_0, (1-\overline\alpha_{t})I), \\
x_{t-1}^{unknown} &\sim \mathcal{N}(\mu_{\theta}(x_{t},t), {\textstyle \sum_{}^{}} _{\theta}(x_t, t)), \\
x_{t-1} = m_{latent}\odot& x_{t-1}^{known} + (1-m_{latent})\odot x_{t-1}^{unknown}, 
\end{align}
where $x_{t-1}^{known}$ is sampled using the known pixels in the given image $m_{latent} \odot x_{0}$, and $x_{t-1}^{unknown}$ is sampled from the model. The combined new sample is $x_{t-1}$, with $x_{t}$ being the previous sampling iteration. 
Different training strategies are employed at various stages of the training process, which will be discussed in subsequent subsections. After $t$ steps' denoising, the predicted feature is sent to the LDM Decoder to obtain the final result.\\
\vspace{-5mm}
\subsection{Training Strategy}
\label{strateg}
\textbf{Pipeline.} 
Latent-SwinIR$_{c}$ is directly trained on HDR-Real \cite{liu2020single} dataset 
to learn the adjustment of color
\begin{wrapfigure}{r}{0.5\linewidth}
    \centering
    \begin{minipage}[t]{0.32\linewidth}
        \centering
        \captionsetup{font={tiny}}
        \includegraphics[height=2.3cm,width=2.3cm]{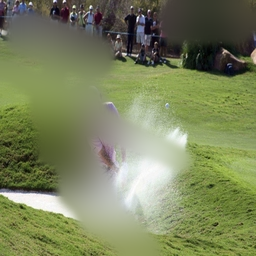}
      \vspace{-5mm}
    \end{minipage}
    \begin{minipage}[t]{0.32\linewidth}
        \centering
        \captionsetup{font={tiny}}
        \includegraphics[height=2.3cm,width=2.3cm]{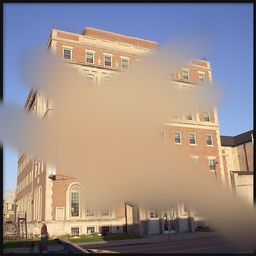}
            \vspace{-5mm}
    \end{minipage}
    \begin{minipage}[t]{0.32\linewidth}
        \centering
        \includegraphics[height=2.3cm,width=2.3cm]{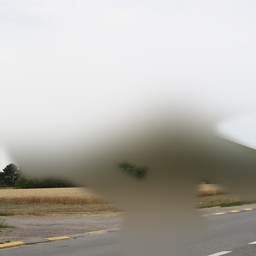}
        \captionsetup{font={tiny}}
           \vspace{-5mm}
    \end{minipage}\\
     \begin{minipage}[t]{0.32\linewidth}
        \centering
        \captionsetup{font={tiny}}
        \includegraphics[height=2.3cm,width=2.3cm]{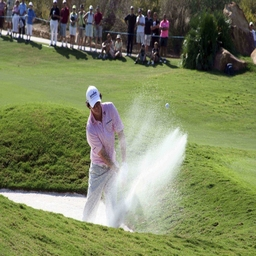}
           \vspace{-7mm}
    \end{minipage}
    \begin{minipage}[t]{0.32\linewidth}
        \centering
        \captionsetup{font={tiny}}
        \includegraphics[height=2.3cm,width=2.3cm]{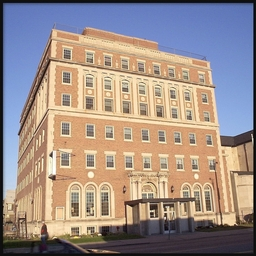}
           \vspace{-7mm}
    \end{minipage}
    \begin{minipage}[t]{0.32\linewidth}
        \centering
        \includegraphics[height=2.3cm,width=2.3cm]{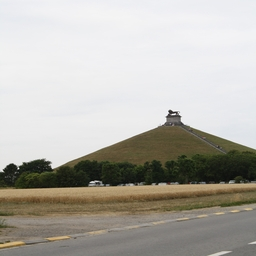}
        \captionsetup{font={tiny}}
           \vspace{-7mm}
    \end{minipage}
    \caption{The first row is the result obtained using our degradation strategy, while the second row is the reference images. \bl{We aim to simulate the degradation caused by other models in dynamic range extremes during LDR enhancement and train Sagiri to handle these situations effectively.}
 }
    \label{inpain}
 \vspace{-5mm}
\end{wrapfigure}
distribution. For Sagiri, we first use the large-scale Places365 dataset~\cite{zhou2017places} for pretraining to enhance its ability in generating different scenes, \sm{and then finetune it on HDR-Real}. 

\noindent\textbf{Degradation generation in pre-training.} 
\sm{During the pre-training of Sagiri on Places365, it is necessary to simulate the results from Latent-SwinIR$_c$ to minimize the domain gap.}
We devise a degradation generation method which involves applying random degradation to high-quality images, introducing \jw{blur-like artifacts}  that mimic over-exposed and under-exposed areas. Specifically, we  create a degradation mask by drawing random lines of varying thickness and positions, which then undergoes dilation and Gaussian blurring to produce smooth, realistic degradation patterns. This mask is used to blend the original image with a heavily blurred version of itself, 
resulting in an image that combines clear and degraded areas in a manner that reflects the challenges encountered in current image restoration models \sm{including but not limited to SwinIR which tend to generate oversmoothed images. As a result, the trained Sagiri can be applied not only to Latent-SwinIR$_c$ but also as a \emph{plug-and-play} module for other LDR enhancement methods.}
Examples of generated images are shown in Figure~\ref{inpain}.

\noindent\textbf{Unknown region mask.}
During pre-training, we do not apply the unknown region mask because we want the model to perceive and judge high-quality and low-quality areas on its own, and avoid introducing unrealistic details as much as possible by learning more real-world scenes. During fine-tuning, we apply the unknown region mask to guide the model's attention to the unique challenges of inpainting over/under-exposed regions.

\vspace{-3mm}
\section{Experiments}

\subsection{Training and Inference Settings}
\label{sett}
\vspace{-1mm}
\textbf{Training.} 
We train Latent-SwinIR  on the HDR-Real \jw{training set}~\cite{liu2020single}  with a batch size of 
16 for 150,000 iterations.
We utilize pretrained stable diffusion v2.1 as the base checkpoint for our Sagiri model.
We first pre-train Sagiri on 250,000 randomly selected images from Places365~\cite{zhou2017places} for 70,000 steps, and then fine-tune it on HDR-Real \jw{training set} for another 20,000 steps.
We use the Adam \cite{kingma2014adam} optimizer with a learning rate of 1e-4 for all training stages, conducted on 4 NVIDIA A100 GPUs.

\begin{table*}[t]
\caption{Quantitative results on HDR-Real \cite{liu2020single}, NTIRE \cite{gu2022ntire}, HDR-Eye\cite{nemoto2015eye}, Eye-over and Eye-under datasets. The latter two datasets are made by uniformly adjusting the exposure value of HDR-Eye dataset to synthesize datasets with large areas at dynamic range extremes.
\bl{In addition to comparing the performance of our pipeline with existing methods, we plugged Sagiri into each model to see performance improvements. The results show that (1) Sagiri enhances the performance of each method, and (2) LS-Sagiri achieves the best overall results.}}
\centering
\footnotesize
\begin{tabular}{c}
\resizebox{\linewidth}{!}{
\hspace{-4mm}\begin{tabular}{ccccccccc}
\toprule
Datasets & \multicolumn{4}{c}{HDR-Real} & \multicolumn{4}{c}{NTIRE}  \\
\midrule
Metrics & BRISQUE$\downarrow$ & NIQE$\downarrow$ & MANIQA$\uparrow$ & CLIP-IQA$\uparrow$ & BRISQUE$\downarrow$ & NIQE$\downarrow$ & MANIQA$\uparrow$ & CLIP-IQA$\uparrow$\\
\midrule
SingleHDR \cite{liu2020single} &23.597  &20.839  &0.367  &0.387  &22.730  & \textbf{21.399} &0.250 &0.411   \\
SingleHDR+Sagiri  &\textbf{19.855}  &\textbf{20.326}  &\textbf{0.556}  &\textbf{0.649} &\textbf{10.211} &21.622 & \textbf{0.385}& \textbf{0.676} \\
\midrule
LCDPNet \cite{wang2022local} &30.704  &20.660  & 0.344 &0.383  & 19.237 & \textbf{20.978} & 0.267& 0.415\\
LCDPNet+Sagiri &\textbf{24.464}&\textbf{20.318}  & \textbf{0.542}&\textbf{0.641} & \textbf{9.951} & 21.622  &\textbf{0.385} & \textbf{0.674} \\
\midrule
HDRUNet \cite{chen2021hdrunet}& 41.521 & 21.388 & 0.341 & 0.361 & 52.898 & 22.752 &0.229 & 0.377 \\
HDRUNet+Sagiri  & \textbf{24.935} & \textbf{20.704} & \textbf{0.503} & \textbf{0.609} & \textbf{21.353} & \textbf{21.749} &\textbf{0.397} & \textbf{0.650} \\
\midrule
GlowGAN \cite{wang2023glowgan}&36.727  &21.774  & \textbf{0.470} &0.503  & 21.769 & \textbf{24.053} & \textbf{0.403}&  0.478 \\
GlowGAN+Sagiri & \textbf{22.840} &\textbf{21.602}  & 0.443 &\textbf{0.554} & \textbf{15.549}& 24.078 & 0.354  & \textbf{0.511}\\
\midrule
\fakescore{Latent-SwinIR$_{c}$} & \fakescore{35.407}  & \fakescore{21.457}  & \fakescore{0.291} & \fakescore{0.303}  & \fakescore{31.298}  & \fakescore{22.000} & \fakescore{0.224} & \fakescore{0.392} \\
LS-Sagiri &\textbf{19.725} &\textbf{20.309}  & \textbf{0.569} &\textbf{0.670}  & \textbf{9.724} &\textbf{21.652} & \textbf{0.395}& \textbf{0.671} \\
\bottomrule
\end{tabular}
}
\\ \\
\resizebox{\linewidth}{!}{
\hspace{-4mm}\begin{tabular}{ccccccccccccc}
\toprule
\multicolumn{1}{c}{Datasets} & \multicolumn{3}{c}{HDR-Eye} & \multicolumn{3}{c}{Eye-over} & \multicolumn{3}{c}{Eye-under}  \\
\midrule
\multicolumn{1}{c}{Metrics} & BRISQUE$\downarrow$ & MANIQA$\uparrow$  & CLIP-IQA$\uparrow$ & BRISQUE$\downarrow$ & MANIQA$\uparrow$  & CLIP-IQA$\uparrow$ & BRISQUE$\downarrow$ & MANIQA$\uparrow$  & CLIP-IQA$\uparrow$ \\
\midrule
SingleHDR \cite{liu2020single}& 18.338 & 0.452 & 0.466 & 20.573 & 0.447&0.428  &33.675 & 0.244 &0.244\\
SingleHDR+Sagiri  & \textbf{15.092} & \textbf{0.570} & \textbf{0.697} & \textbf{14.969}&  \textbf{0.557}&\textbf{0.676}&\textbf{13.477} & \textbf{0.339} &\textbf{0.523}\\
\midrule
LCDPNet \cite{wang2022local}& 20.672 & 0.453 & 0.475 & 26.374 & 0.398 & 0.365 &54.493 & 0.311 &0.335\\
LCDPNet+Sagiri & \textbf{14.137} & \textbf{0.543} & \textbf{0.665} & \textbf{14.973} & \textbf{0.478} & \textbf{0.638} & \textbf{37.825} & \textbf{0.382} & \textbf{0.552} \\
\midrule
HDRUNet \cite{chen2021hdrunet}&27.672  &0.418 & 0.390 & 24.545 & 0.454 & 0.410 &72.920 & 0.364 &0.403\\
HDRUNet+Sagiri  &\textbf{14.846}  &\textbf{0.555} &\textbf{0.662} & \textbf{15.905} & \textbf{0.560} &\textbf{0.668}  &\textbf{40.954} &\textbf{0.460}  &\textbf{0.610} \\
\midrule
GlowGAN \cite{wang2023glowgan}& \textbf{16.042} & \textbf{0.506} & \textbf{0.536} & \textbf{16.930} &  \textbf{0.503}&\textbf{0.561}  &46.667 & \textbf{0.356} & \textbf{0.483} \\
GlowGAN+Sagiri & 19.775 & 0.430 & 0.473 & 20.040 &  0.401 & 0.466 & \textbf{37.745} & 0.286 &0.432\\
\midrule
\fakescore{Latent-SwinIR$_{c}$} & \fakescore{25.870} & \fakescore{0.329} & \fakescore{0.286} & \fakescore{25.345} & \fakescore{0.321} & \fakescore{0.286} & \fakescore{45.168} & \fakescore{0.256} & \fakescore{0.252} \\
LS-Sagiri & \textbf{14.777} & \textbf{0.538} & \textbf{0.675} & \textbf{14.667} & \textbf{0.535} & \textbf{0.669 }& \textbf{12.066} & \textbf{0.462} & \textbf{0.660} \\
\bottomrule
\end{tabular}
}
\end{tabular}
\label{tab1}
\end{table*}

\noindent\textbf{Inference.} 
During inference, the model takes an LDR image with an unknown region masks obtained by detecting pixel values of 0 and 255. Our model operates efficiently, requiring only 30 steps of DDPM sampling~\cite{nichol2021improved}.

 \begin{figure}[h]
	\centering
 	\begin{minipage}[t]{0.136\linewidth}
		\centering
		\captionsetup{font={tiny}}
		\includegraphics[height=2.1cm,width=1.95cm]{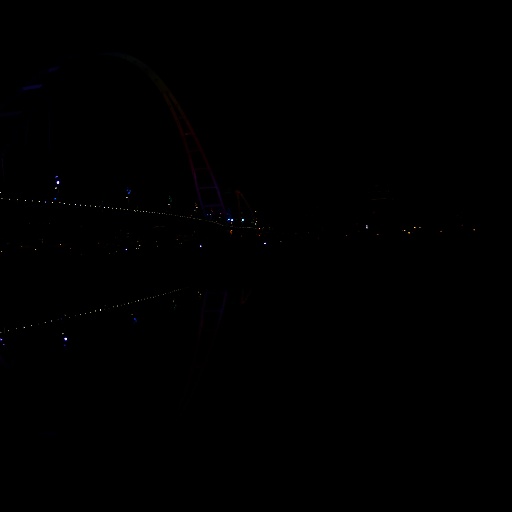}
  \vspace{-6mm}
	\end{minipage}
  	\begin{minipage}[t]{0.136\linewidth}
		\centering
		\captionsetup{font={tiny}}
		\includegraphics[height=2.1cm,width=1.95cm]{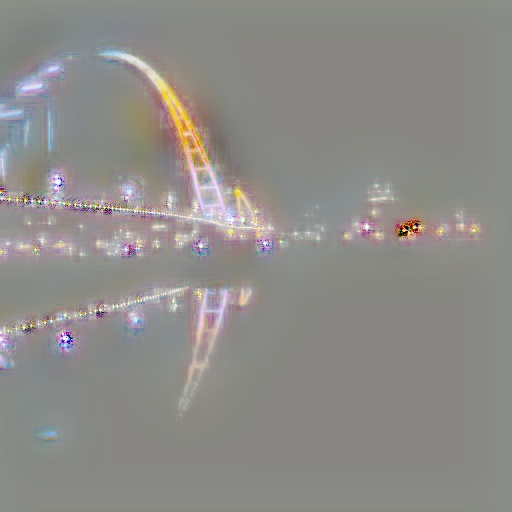}
         \vspace{-6mm}
	\end{minipage}
	\begin{minipage}[t]{0.136\linewidth}
		\centering
		\includegraphics[height=2.1cm,width=1.95cm]{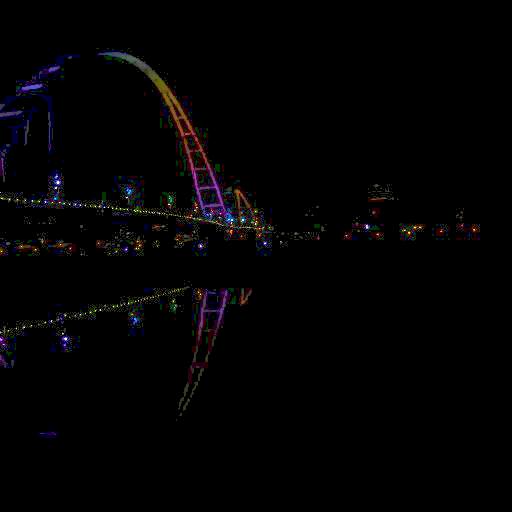}
		\captionsetup{font={tiny}}
         \vspace{-6mm}
	\end{minipage}
  	\begin{minipage}[t]{0.136\linewidth}
		\centering
		\captionsetup{font={tiny}}
		\includegraphics[height=2.1cm,width=1.95cm]{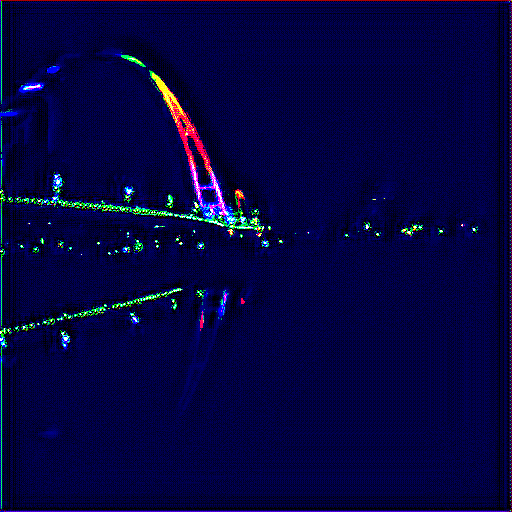}
         \vspace{-6mm}
	\end{minipage}
 	\begin{minipage}[t]{0.136\linewidth}
		\centering
		\captionsetup{font={tiny}}
		\includegraphics[height=2.1cm,width=1.95cm]{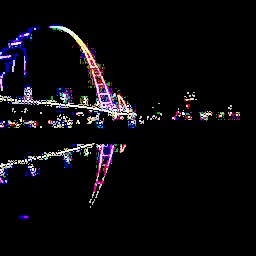}
          \vspace{-6mm}
	\end{minipage}
	\begin{minipage}[t]{0.136\linewidth}
		\centering
		\captionsetup{font={tiny}}
		\includegraphics[height=2.1cm,width=1.95cm]{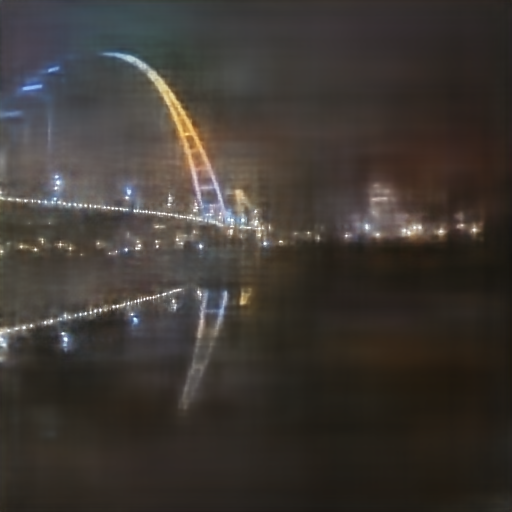}
         \vspace{-6mm}
	\end{minipage}
 	\begin{minipage}[t]{0.136\linewidth}
		\centering
		\captionsetup{font={tiny}}
		\includegraphics[height=2.1cm,width=1.95cm]{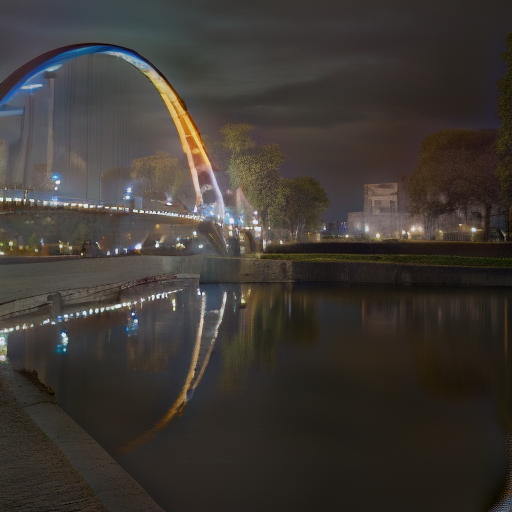}
         \vspace{-6mm}
	\end{minipage}
 	\begin{minipage}[t]{0.136\linewidth}
		\centering
		\captionsetup{font={tiny}}
		\includegraphics[height=2.1cm,width=1.95cm]{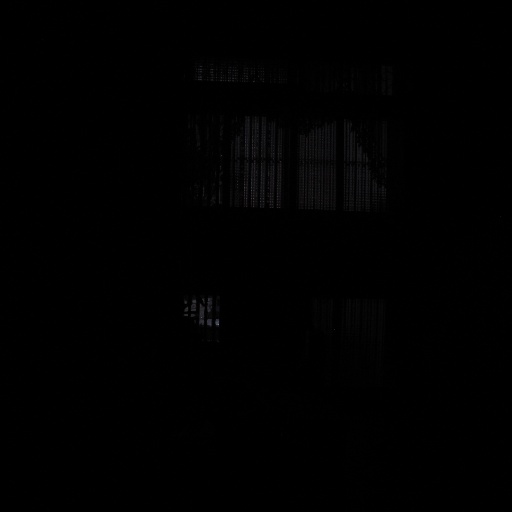}
  \vspace{-6mm}
	\end{minipage}
  	\begin{minipage}[t]{0.136\linewidth}
		\centering
		\captionsetup{font={tiny}}
		\includegraphics[height=2.1cm,width=1.95cm]{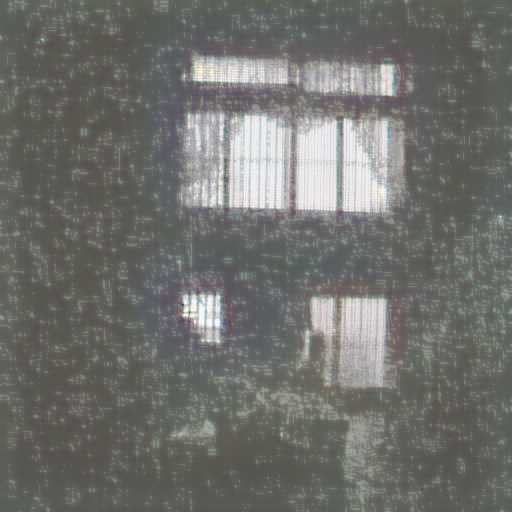}
         \vspace{-6mm}
	\end{minipage}
	\begin{minipage}[t]{0.136\linewidth}
		\centering
		\includegraphics[height=2.1cm,width=1.95cm]{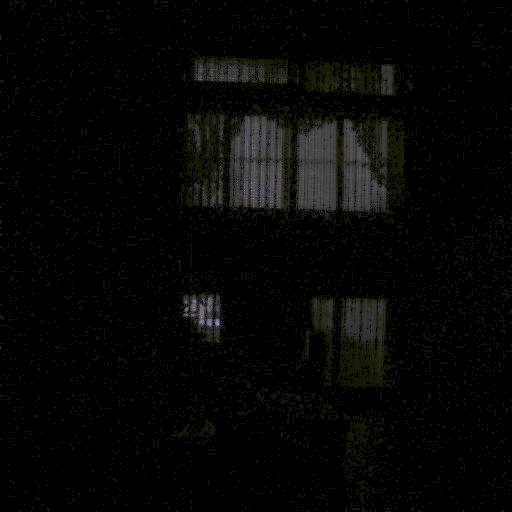}
		\captionsetup{font={tiny}}
         \vspace{-6mm}
	\end{minipage}
  	\begin{minipage}[t]{0.136\linewidth}
		\centering
		\captionsetup{font={tiny}}
		\includegraphics[height=2.1cm,width=1.95cm]{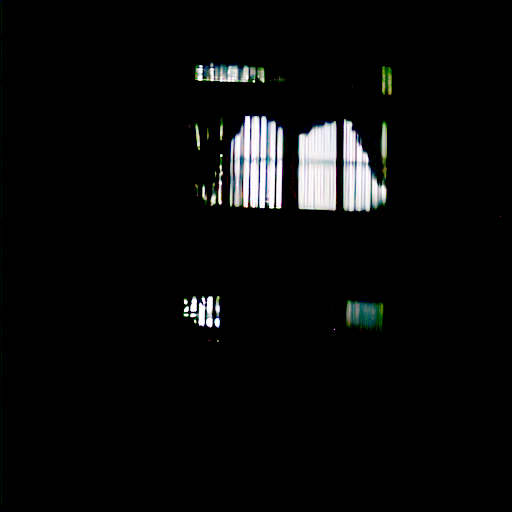}
         \vspace{-6mm}
	\end{minipage}
 	\begin{minipage}[t]{0.136\linewidth}
		\centering
		\captionsetup{font={tiny}}
		\includegraphics[height=2.1cm,width=1.95cm]{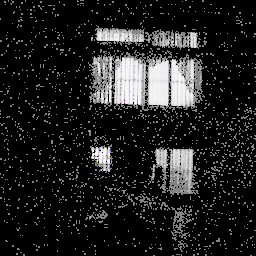}
          \vspace{-6mm}
	\end{minipage}
	\begin{minipage}[t]{0.136\linewidth}
		\centering
		\captionsetup{font={tiny}}
		\includegraphics[height=2.1cm,width=1.95cm]{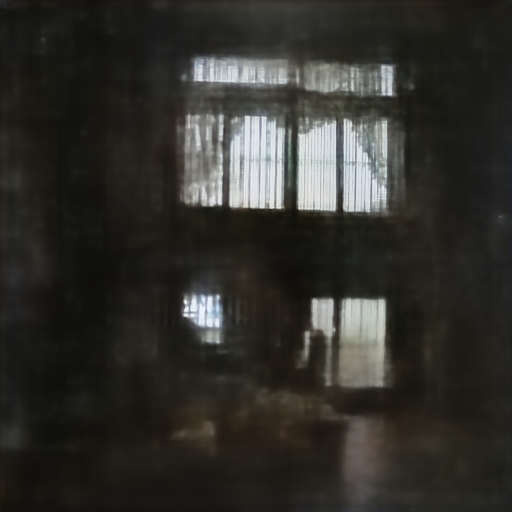}
         \vspace{-6mm}
	\end{minipage}
 	\begin{minipage}[t]{0.136\linewidth}
		\centering
		\captionsetup{font={tiny}}
		\includegraphics[height=2.1cm,width=1.95cm]{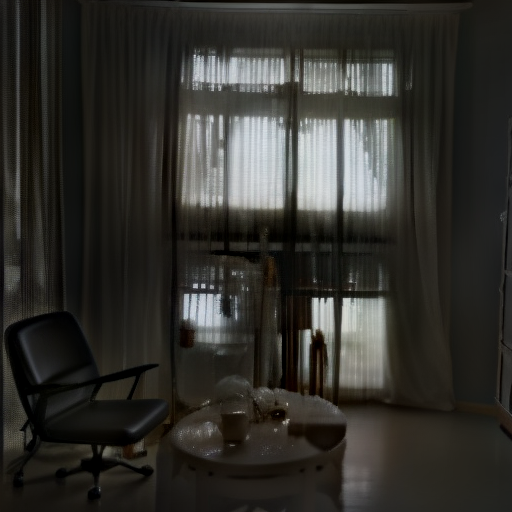}
         \vspace{-6mm}
	\end{minipage}
	\begin{minipage}[t]{0.136\linewidth}
		\centering
		\captionsetup{font={tiny}}
		\includegraphics[height=2.1cm,width=1.95cm]{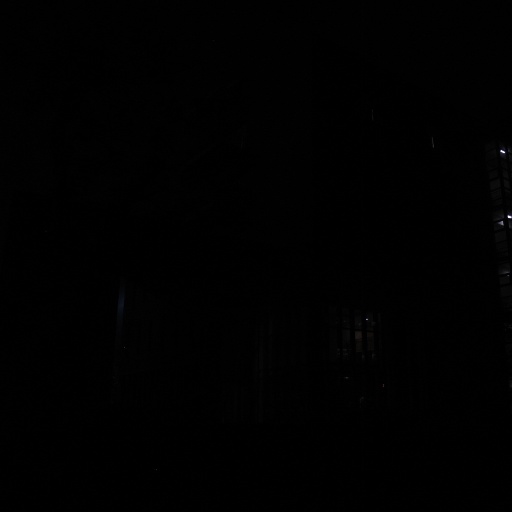}
  \vspace{-6mm}
        \caption*{(a) LQ}
	\end{minipage}
  	\begin{minipage}[t]{0.136\linewidth}
		\centering
		\captionsetup{font={tiny}}
		\includegraphics[height=2.1cm,width=1.95cm]{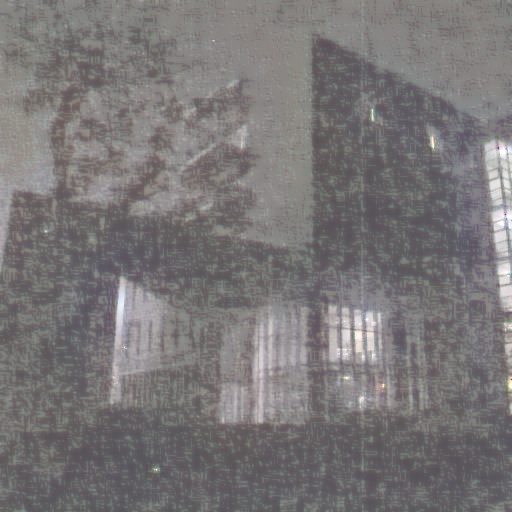}
         \vspace{-6mm}
 \caption*{(b) SingleHDR \cite{le2023single}}
	\end{minipage}
	\begin{minipage}[t]{0.136\linewidth}
		\centering
		\includegraphics[height=2.1cm,width=1.95cm]{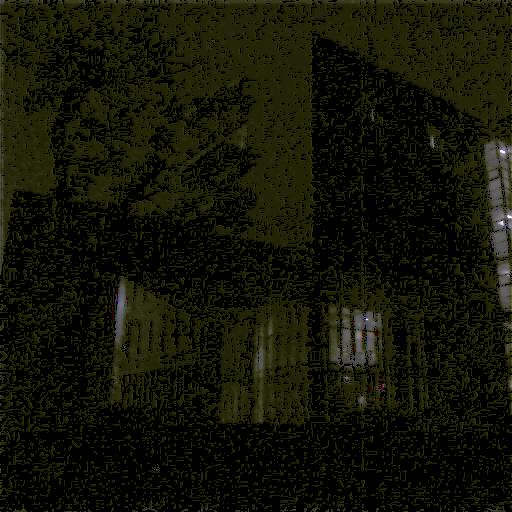}
		\captionsetup{font={tiny}}
         \vspace{-6mm}
 \caption*{(c) LCDP-Net \cite{wang2022local}}
	\end{minipage}
  	\begin{minipage}[t]{0.136\linewidth}
		\centering
		\captionsetup{font={tiny}}
		\includegraphics[height=2.1cm,width=1.95cm]{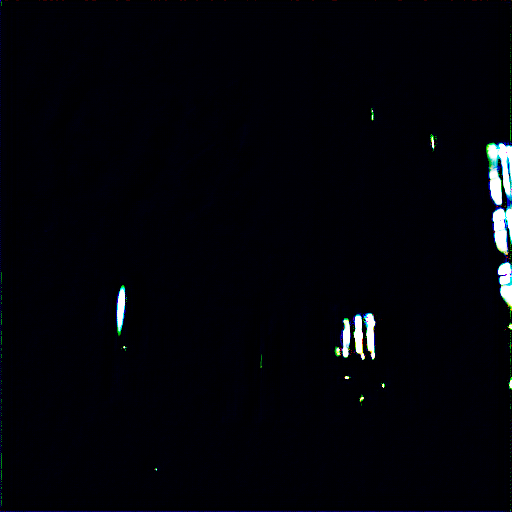}
         \vspace{-6mm}
 \caption*{(d) HDRUNet \cite{chen2021hdrunet}}
	\end{minipage}
 \begin{minipage}[t]{0.136\linewidth}
		\centering
		\captionsetup{font={tiny}}
		\includegraphics[height=2.1cm,width=1.95cm]{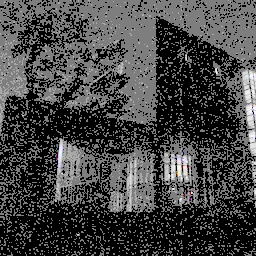}
          \vspace{-6mm}
\caption*{(e) Glow-GAN \cite{wang2023glowgan}}
	\end{minipage}
	\begin{minipage}[t]{0.136\linewidth}
		\centering
		\captionsetup{font={tiny}}
		\includegraphics[height=2.1cm,width=1.95cm]{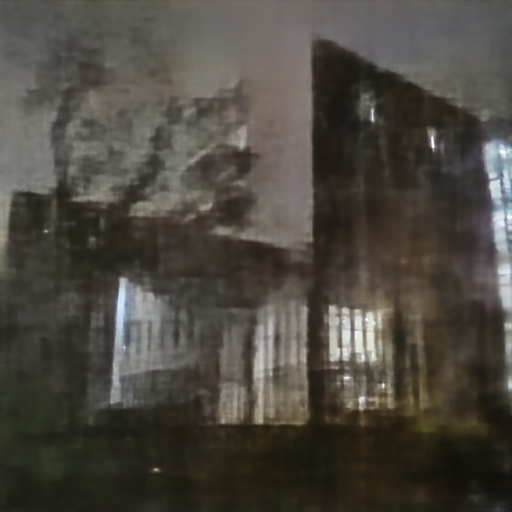}
         \vspace{-6mm}
 \caption*{(f) Latent-SwinIR$_{c}$}
	\end{minipage}
 	\begin{minipage}[t]{0.136\linewidth}
		\centering
		\captionsetup{font={tiny}}
		\includegraphics[height=2.1cm,width=1.95cm]{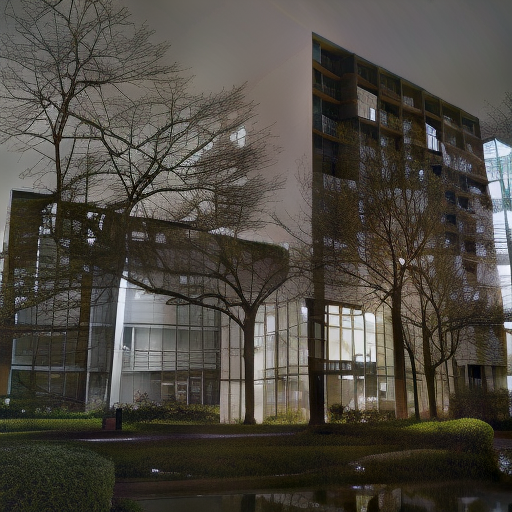}
         \vspace{-6mm}
 \caption*{(g) LS-Sagiri}
	\end{minipage}
	\caption{\textbf{(a-f) Performance of Latent-SwinIR$_c$.} Existing methods often fail to recover content in dynamic range extreme regions. Latent-SwinIR$_{c}$, due to its unique loss function design, captures a more balanced color distribution. \textbf{(g)} Additionally, the Sagiri model excels in generating detailed content in large regions, further improving the overall quality. \textbf{Zoom in the figures for details.}}
 \label{compare_ls}
 \end{figure}

\noindent\textbf{Prompt use in training and inference.} 
For generating prompts, we employ CogVLM \cite{wang2023cogvlm} to summarize the input image. During the fine-tuning of Sagiri, we use prompts generated from the ground truth to adapt the model to the prompt input. For inference on the HDR-Real \jw{testing set} \cite{liu2020single}, we generate prompts from low-quality images. For the HDR-Eye \cite{nemoto2015eye}, Eye-over, Eye-under and NTIRE\cite{gu2022ntire} datasets, we do not input prompts. This design aims to better evaluate Sagiri's adaptability in different scenarios.

\subsection{Results}
\textbf{Datasets.} 
We present quantitative comparison results on the HDR-Real~\cite{liu2020single}, NTIRE~\cite{gu2022ntire}, HDR-Eye~\cite{nemoto2015eye}, Eye-over and Eye-under datasets for evaluation. The latter two datasets are made by uniformly adjusting exposure values
of the HDR-Eye dataset to create over-exposed/under-exposed images. 
\sm{This is because existing datasets doe not contain a large number of images with significant content loss in their test sets, which is not suitable for evaluation in our setting.}

 \begin{figure}[t]
	\centering
 	\begin{minipage}[t]{0.16\linewidth}
		\centering
		\captionsetup{font={tiny}}
		\includegraphics[height=2.3cm,width=2.3cm]{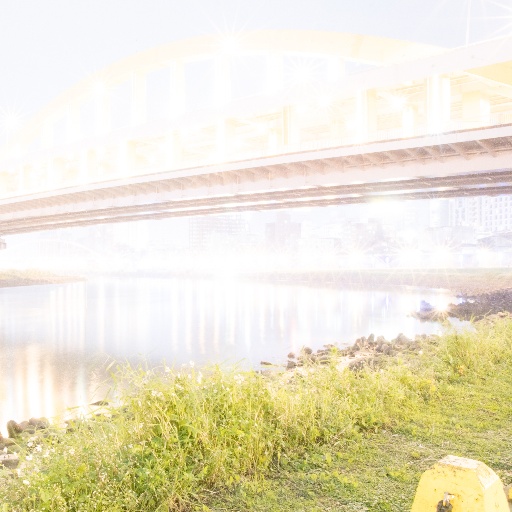}
  \vspace{-6mm}
	\end{minipage}
  	\begin{minipage}[t]{0.16\linewidth}
		\centering
		\captionsetup{font={tiny}}
		\includegraphics[height=2.3cm,width=2.3cm]{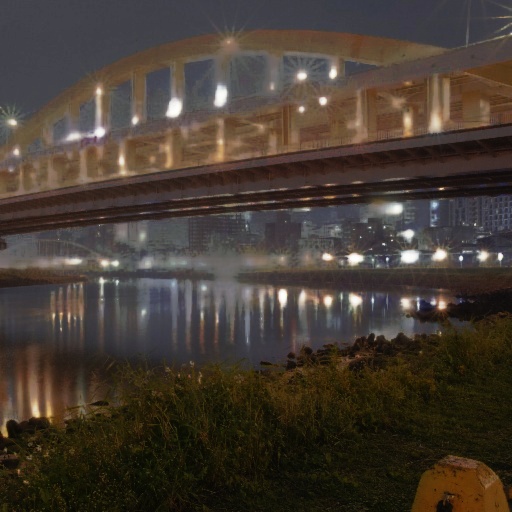}
         \vspace{-6mm}
	\end{minipage}
	\begin{minipage}[t]{0.16\linewidth}
		\centering
		\includegraphics[height=2.3cm,width=2.3cm]{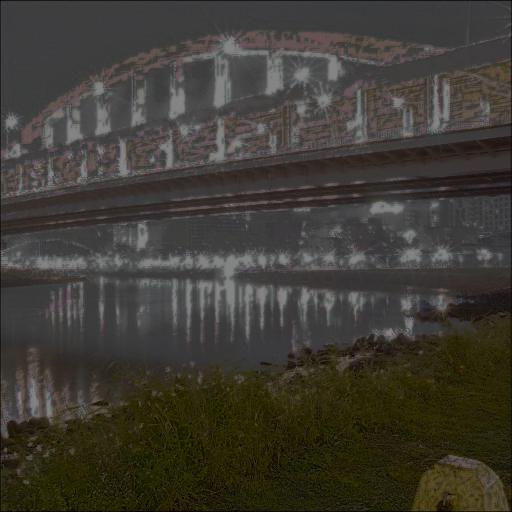}
		\captionsetup{font={tiny}}
         \vspace{-6mm}
	\end{minipage}
  	\begin{minipage}[t]{0.16\linewidth}
		\centering
		\captionsetup{font={tiny}}
		\includegraphics[height=2.3cm,width=2.3cm]{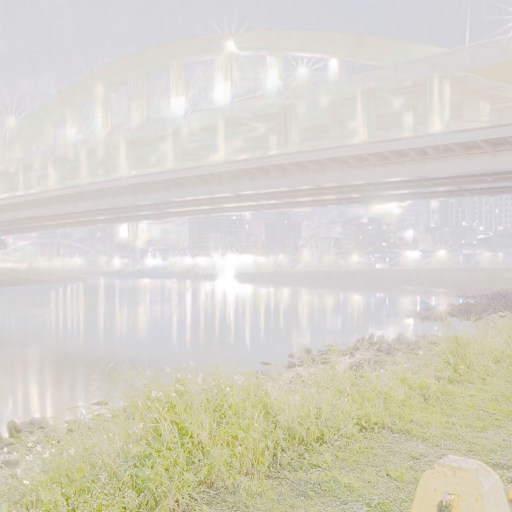}
         \vspace{-6mm}
	\end{minipage}
 	\begin{minipage}[t]{0.16\linewidth}
		\centering
		\captionsetup{font={tiny}}
		\includegraphics[height=2.3cm,width=2.3cm]{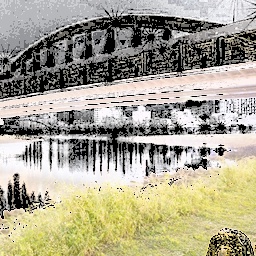}
          \vspace{-6mm}
	\end{minipage}
	\begin{minipage}[t]{0.16\linewidth}
		\centering
		\captionsetup{font={tiny}}
		\includegraphics[height=2.3cm,width=2.3cm]{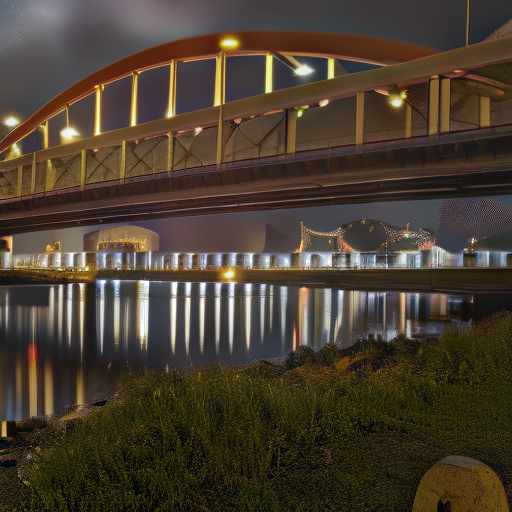}
         \vspace{-6mm}
	\end{minipage}
  	\begin{minipage}[t]{0.16\linewidth}
		\centering
		\captionsetup{font={tiny}}
		\includegraphics[height=2.3cm,width=2.3cm]{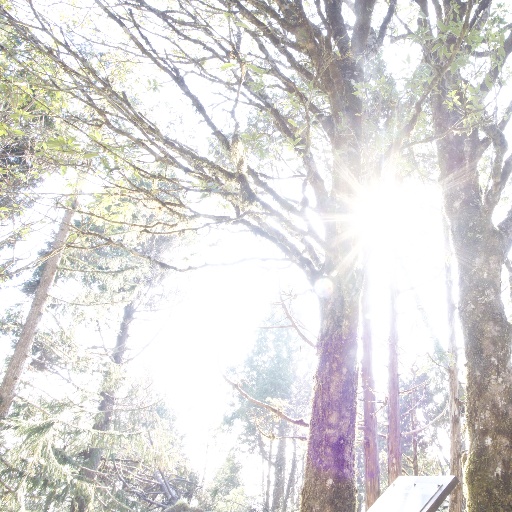}
  \vspace{-6mm}
	\end{minipage}
  	\begin{minipage}[t]{0.16\linewidth}
		\centering
		\captionsetup{font={tiny}}
		\includegraphics[height=2.3cm,width=2.3cm]{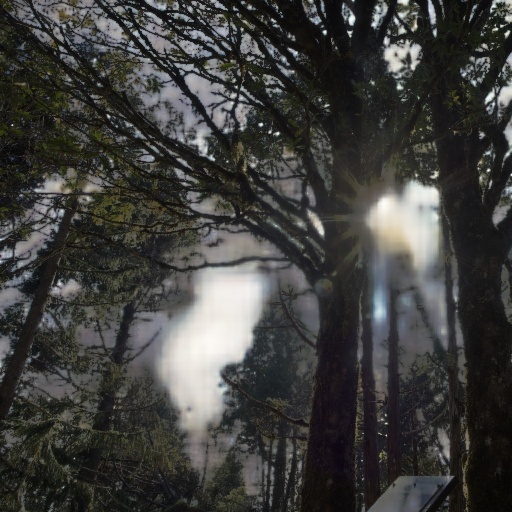}
         \vspace{-6mm}
	\end{minipage}
	\begin{minipage}[t]{0.16\linewidth}
		\centering
		\includegraphics[height=2.3cm,width=2.3cm]{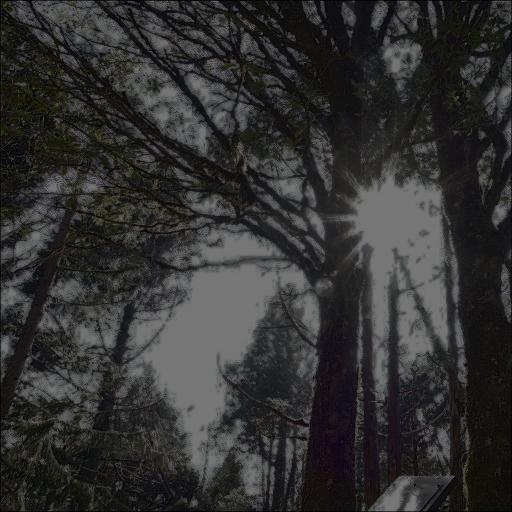}
		\captionsetup{font={tiny}}
         \vspace{-6mm}
	\end{minipage}
  	\begin{minipage}[t]{0.16\linewidth}
		\centering
		\captionsetup{font={tiny}}
		\includegraphics[height=2.3cm,width=2.3cm]{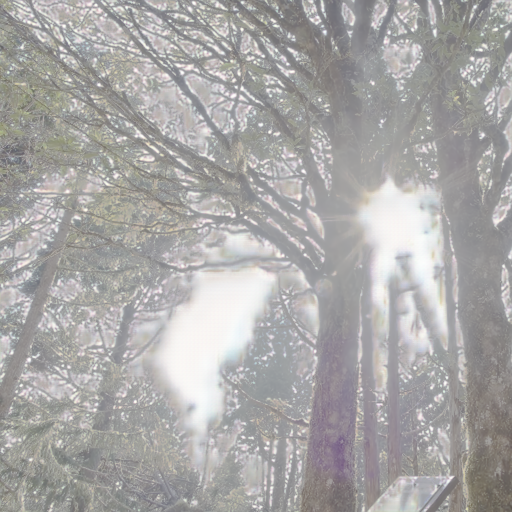}
         \vspace{-6mm}
	\end{minipage}
 	\begin{minipage}[t]{0.16\linewidth}
		\centering
		\captionsetup{font={tiny}}
		\includegraphics[height=2.3cm,width=2.3cm]{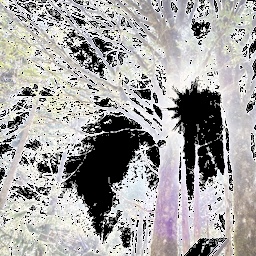}
          \vspace{-6mm}
	\end{minipage}
	\begin{minipage}[t]{0.16\linewidth}
		\centering
		\captionsetup{font={tiny}}
		\includegraphics[height=2.3cm,width=2.3cm]{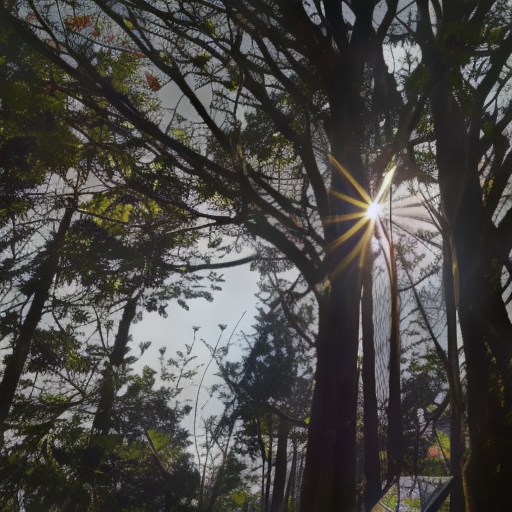}
         \vspace{-6mm}
	\end{minipage}
	\begin{minipage}[t]{0.16\linewidth}
		\centering
		\captionsetup{font={tiny}}
		\includegraphics[height=2.3cm,width=2.3cm]{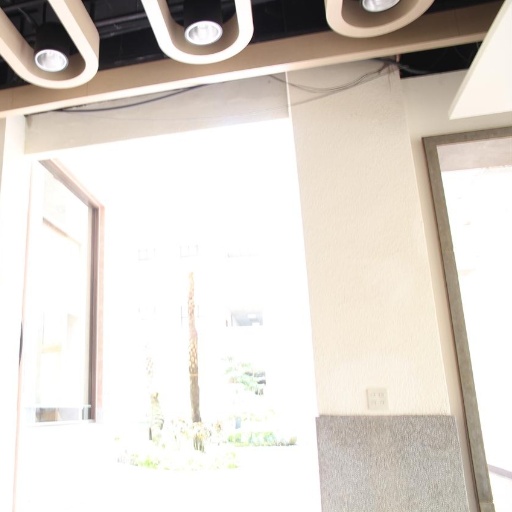}
  \vspace{-6mm}
        \caption*{LQ}
	\end{minipage}
  	\begin{minipage}[t]{0.16\linewidth}
		\centering
		\captionsetup{font={tiny}}
		\includegraphics[height=2.3cm,width=2.3cm]{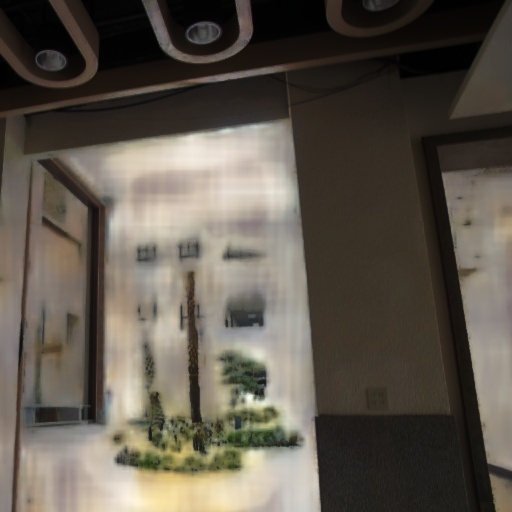}
         \vspace{-6mm}
 \caption*{SingleHDR \cite{le2023single}}
	\end{minipage}
	\begin{minipage}[t]{0.16\linewidth}
		\centering
		\includegraphics[height=2.3cm,width=2.3cm]{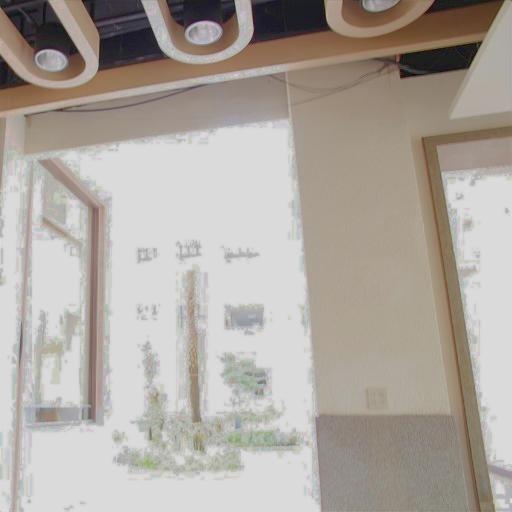}
		\captionsetup{font={tiny}}
         \vspace{-6mm}
 \caption*{LCDP-Net \cite{wang2022local}}
	\end{minipage}
  	\begin{minipage}[t]{0.16\linewidth}
		\centering
		\captionsetup{font={tiny}}
		\includegraphics[height=2.3cm,width=2.3cm]{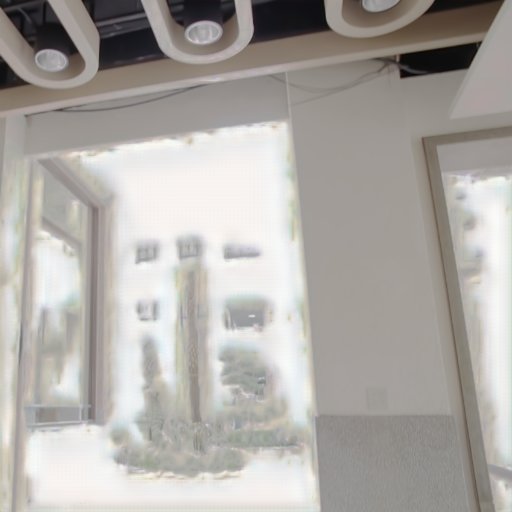}
         \vspace{-6mm}
 \caption*{HDRUNet \cite{chen2021hdrunet}}
	\end{minipage}
 \begin{minipage}[t]{0.16\linewidth}
		\centering
		\captionsetup{font={tiny}}
		\includegraphics[height=2.3cm,width=2.3cm]{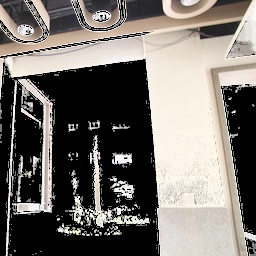}
          \vspace{-6mm}
\caption*{Glow-GAN \cite{wang2023glowgan}}
	\end{minipage}
	\begin{minipage}[t]{0.16\linewidth}
		\centering
		\captionsetup{font={tiny}}
		\includegraphics[height=2.3cm,width=2.3cm]{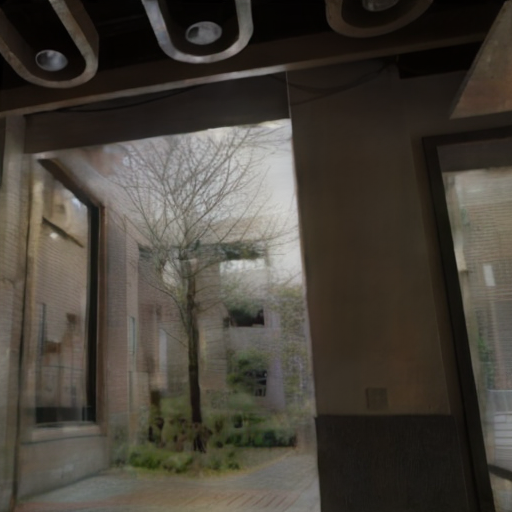}
         \vspace{-6mm}
 \caption*{LS-Sagiri}
	\end{minipage}
	\caption{\textbf{Performance of LS-Sagiri.} Previous restoration-based methods can only restore over-exposed areas to blurry content. Although Glow-GAN \cite{wang2023glowgan} is a generative method, it fails to handle \emph{large} over-exposed regions, often rendering them black. In contrast, our method can generate realistic content based on existing information and specified mask areas. }
 \label{compare_glow}
 \end{figure}
 \begin{figure}[h]
	\centering
	\begin{minipage}[t]{0.16\linewidth}
		\centering
		\captionsetup{font={tiny}}
		\includegraphics[height=2.3cm,width=2.3cm]{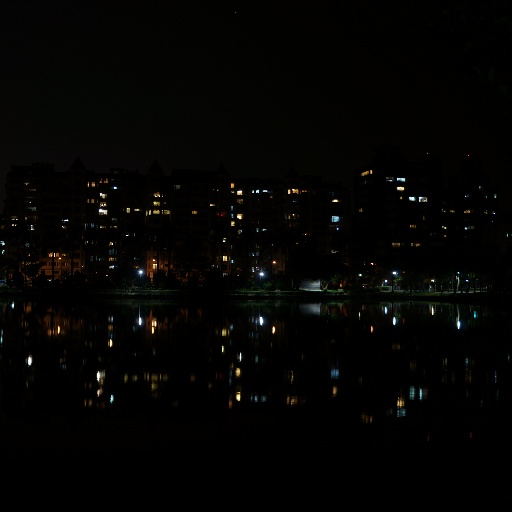}
  \vspace{-6mm}
        \caption*{LQ}
	\end{minipage}
 	\begin{minipage}[t]{0.16\linewidth}
		\centering
		\captionsetup{font={tiny}}
		\includegraphics[height=2.3cm,width=2.3cm]{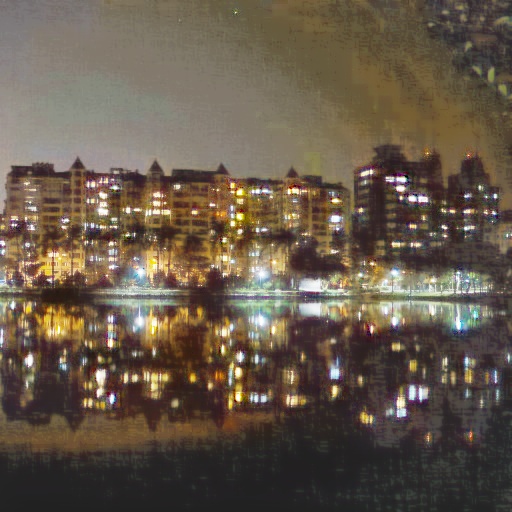}
    \vspace{-6mm}
        \caption*{SingleHDR \cite{liu2020single}}
	\end{minipage}
	\begin{minipage}[t]{0.16\linewidth}
		\centering
		\captionsetup{font={tiny}}
		\includegraphics[height=2.3cm,width=2.3cm]{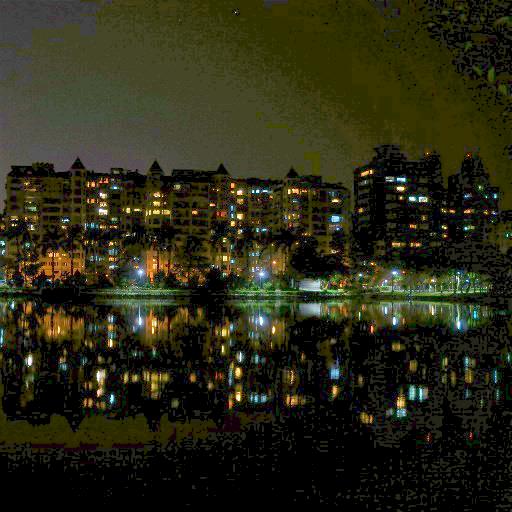}
  \vspace{-6mm}
        \caption*{LCDPNet \cite{wang2022local}}
	\end{minipage}
 	\begin{minipage}[t]{0.16\linewidth}
		\centering
		\captionsetup{font={tiny}}
		\includegraphics[height=2.3cm,width=2.3cm]{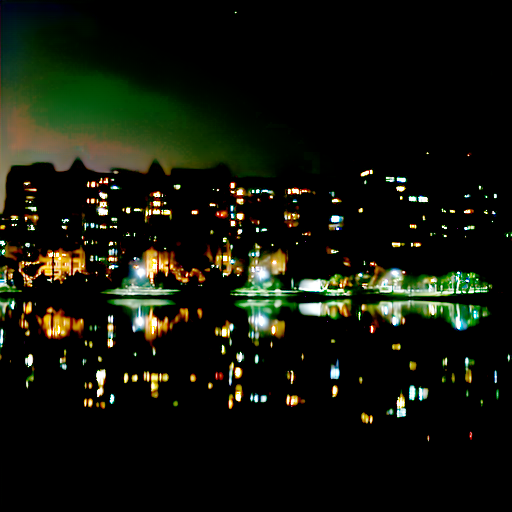}
  \vspace{-6mm}
        \caption*{HDRUNet \cite{chen2021hdrunet}}
	\end{minipage}
  	\begin{minipage}[t]{0.16\linewidth}
		\centering
		\captionsetup{font={tiny}}
		\includegraphics[height=2.3cm,width=2.3cm]{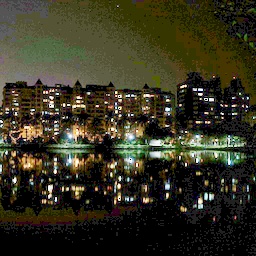}
  \vspace{-6mm}
        \caption*{GlowGAN \cite{wang2023glowgan}}
	\end{minipage}
	\begin{minipage}[t]{0.16\linewidth}
		\centering
		\includegraphics[height=2.3cm,width=2.3cm]{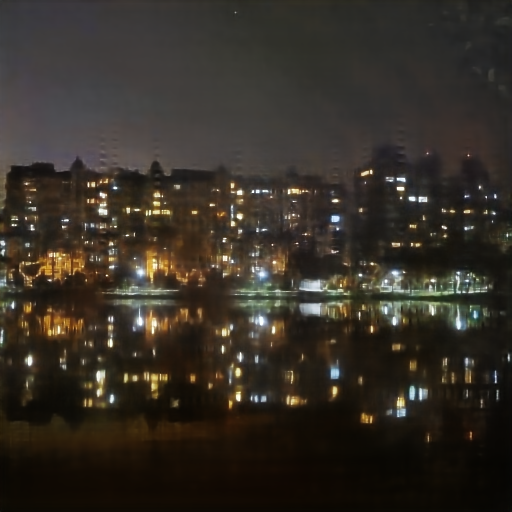}
		\captionsetup{font={tiny}}
  \vspace{-6mm}
        \caption*{Latent-SwinIR$_{c}$(LS, Ours)}
	\end{minipage}\\
 	\begin{minipage}[t]{0.16\linewidth}
		\centering
		\captionsetup{font={tiny}}
		\includegraphics[height=2.3cm,width=2.3cm]{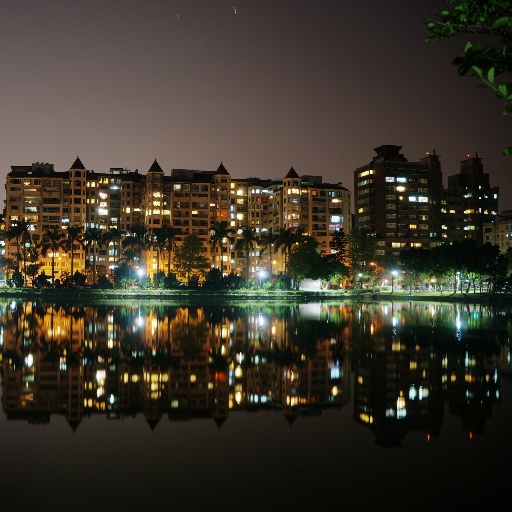}
  \vspace{-6mm}
        \caption*{Reference}
	\end{minipage}
 	\begin{minipage}[t]{0.16\linewidth}
		\centering
		\captionsetup{font={tiny}}
		\includegraphics[height=2.3cm,width=2.3cm]{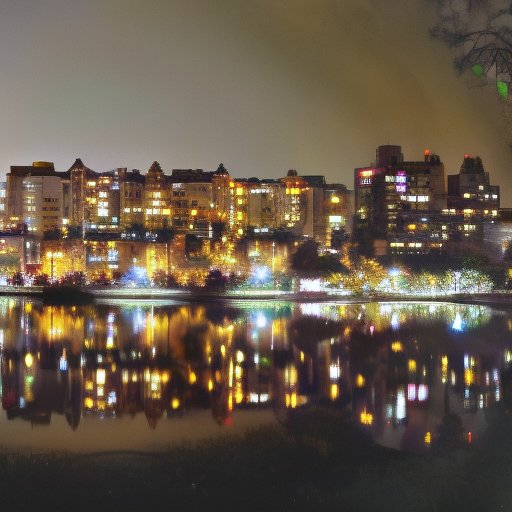}
  \vspace{-6mm}
        \caption*{SingleHDR\cite{liu2020single}+Sagiri}
	\end{minipage}
  	\begin{minipage}[t]{0.16\linewidth}
		\centering
		\captionsetup{font={tiny}}
		\includegraphics[height=2.3cm,width=2.3cm]{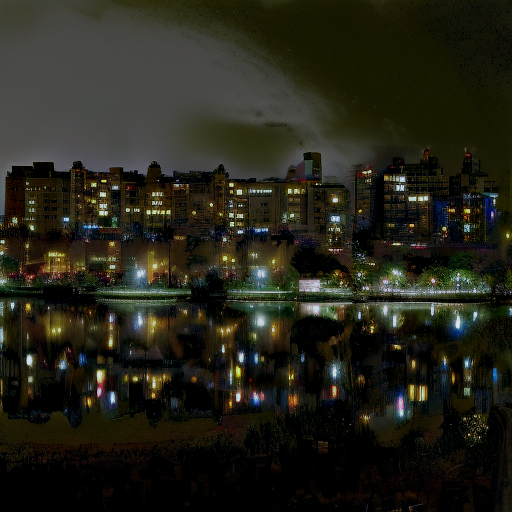}
  \vspace{-6mm}
        \caption*{LCDPNet\cite{wang2022local}+Sagiri}
	\end{minipage}
   	\begin{minipage}[t]{0.16\linewidth}
		\centering
		\captionsetup{font={tiny}}
		\includegraphics[height=2.3cm,width=2.3cm]{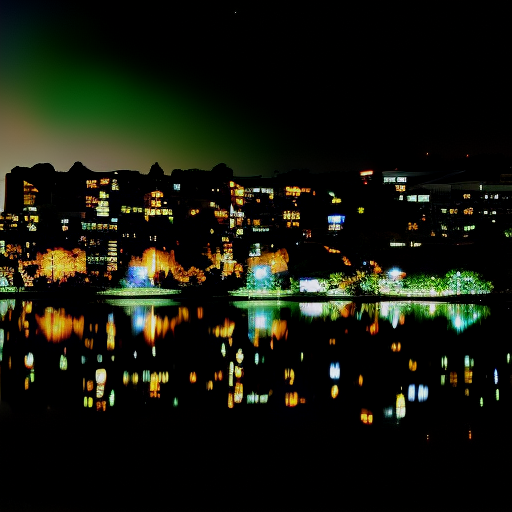}
  \vspace{-6mm}
        \caption*{HDRUNet\cite{chen2021hdrunet}+Sagiri}
	\end{minipage}
	\begin{minipage}[t]{0.16\linewidth}
		\centering
		\captionsetup{font={tiny}}
		\includegraphics[height=2.3cm,width=2.3cm]{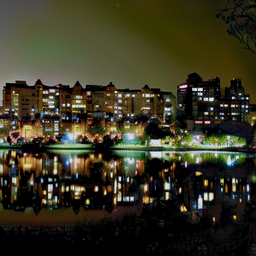}
  \vspace{-6mm}
        \caption*{GlowGAN\cite{wang2023glowgan}+Sagiri}
	\end{minipage}
	\begin{minipage}[t]{0.16\linewidth}
		\centering
		\captionsetup{font={tiny}}
		\includegraphics[height=2.3cm,width=2.3cm]{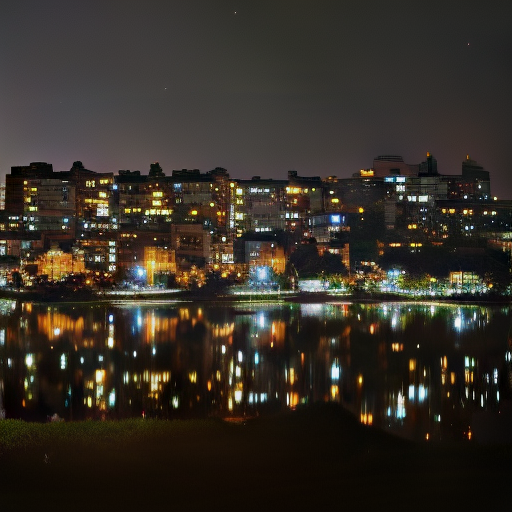}
  \vspace{-6mm}
        \caption*{\jw{LS}+Sagiri(Ours)}
	\end{minipage}
 	\begin{minipage}[t]{0.16\linewidth}
		\centering
		\captionsetup{font={tiny}}
		\includegraphics[height=2.3cm,width=2.3cm]{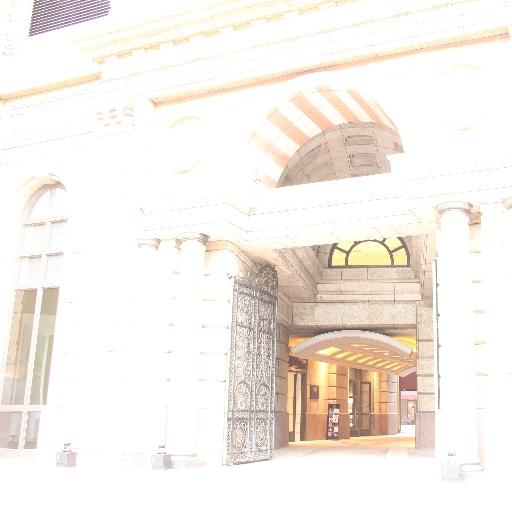}
  \vspace{-6mm}
        \caption*{LQ}
	\end{minipage}
 	\begin{minipage}[t]{0.16\linewidth}
		\centering
		\captionsetup{font={tiny}}
		\includegraphics[height=2.3cm,width=2.3cm]{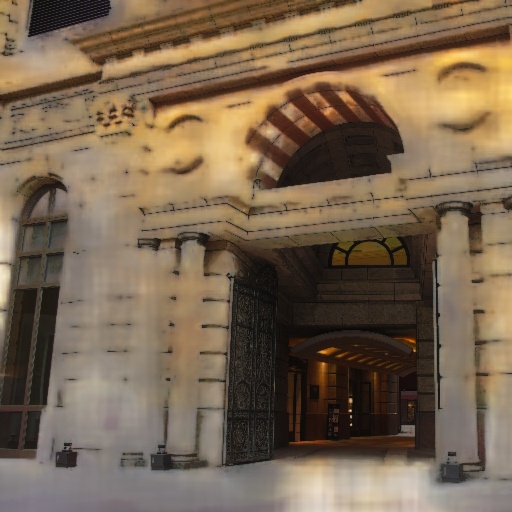}
    \vspace{-6mm}
        \caption*{SingleHDR \cite{liu2020single}}
	\end{minipage}
	\begin{minipage}[t]{0.16\linewidth}
		\centering
		\captionsetup{font={tiny}}
		\includegraphics[height=2.3cm,width=2.3cm]{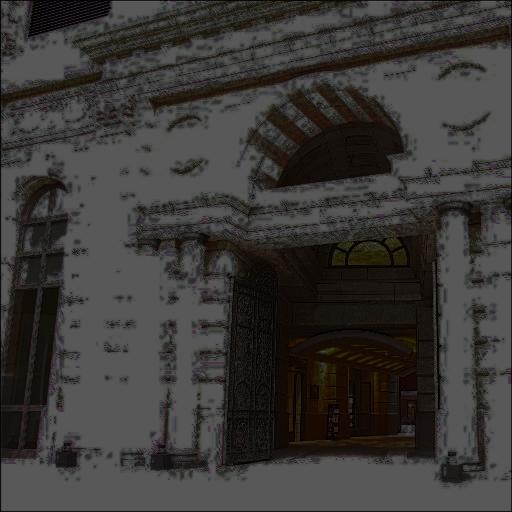}
  \vspace{-6mm}
        \caption*{LCDPNet \cite{wang2022local}}
	\end{minipage}
 	\begin{minipage}[t]{0.16\linewidth}
		\centering
		\captionsetup{font={tiny}}
		\includegraphics[height=2.3cm,width=2.3cm]{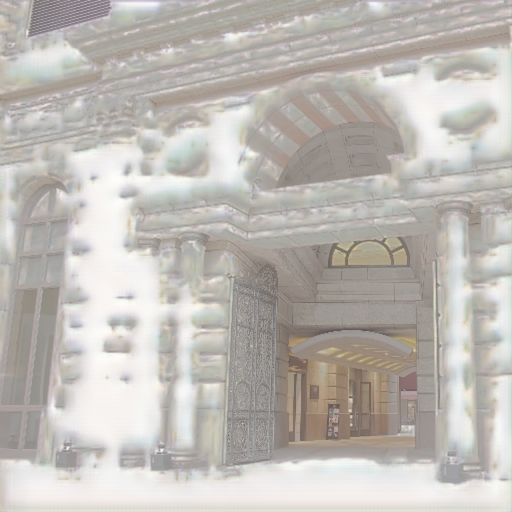}
  \vspace{-6mm}
        \caption*{HDRUNet \cite{chen2021hdrunet}}
	\end{minipage}
  	\begin{minipage}[t]{0.16\linewidth}
		\centering
		\captionsetup{font={tiny}}
		\includegraphics[height=2.3cm,width=2.3cm]{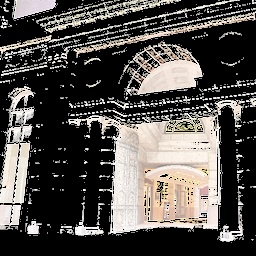}
  \vspace{-6mm}
        \caption*{GlowGAN \cite{wang2023glowgan}}
	\end{minipage}
	\begin{minipage}[t]{0.16\linewidth}
		\centering
		\includegraphics[height=2.3cm,width=2.3cm]{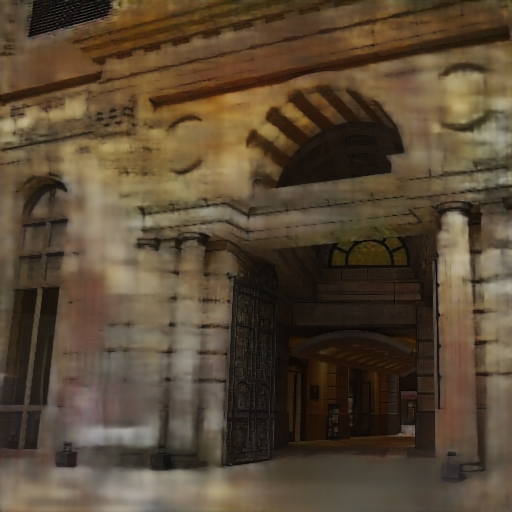}
		\captionsetup{font={tiny}}
  \vspace{-6mm}
        \caption*{Latent-SwinIR$_{c}$(LS, Ours)}
	\end{minipage}\\
 	\begin{minipage}[t]{0.16\linewidth}
		\centering
		\captionsetup{font={tiny}}
		\includegraphics[height=2.3cm,width=2.3cm]{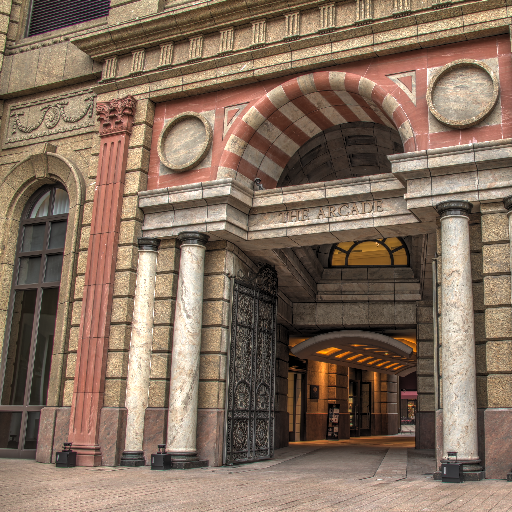}
  \vspace{-6mm}
        \caption*{Reference}
	\end{minipage}
 	\begin{minipage}[t]{0.16\linewidth}
		\centering
		\captionsetup{font={tiny}}
		\includegraphics[height=2.3cm,width=2.3cm]{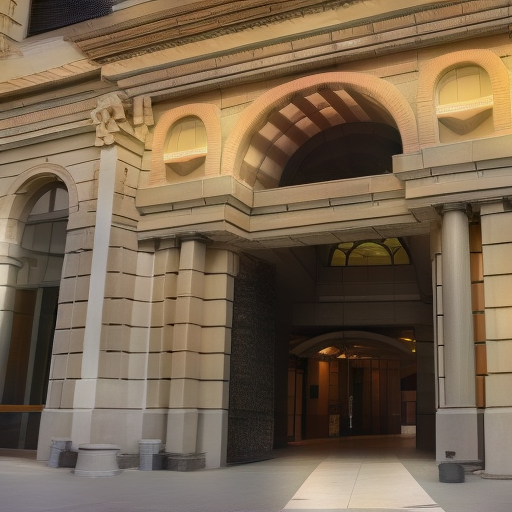}
  \vspace{-6mm}
        \caption*{SingleHDR\cite{liu2020single}+Sagiri}
	\end{minipage}
  	\begin{minipage}[t]{0.16\linewidth}
		\centering
		\captionsetup{font={tiny}}
		\includegraphics[height=2.3cm,width=2.3cm]{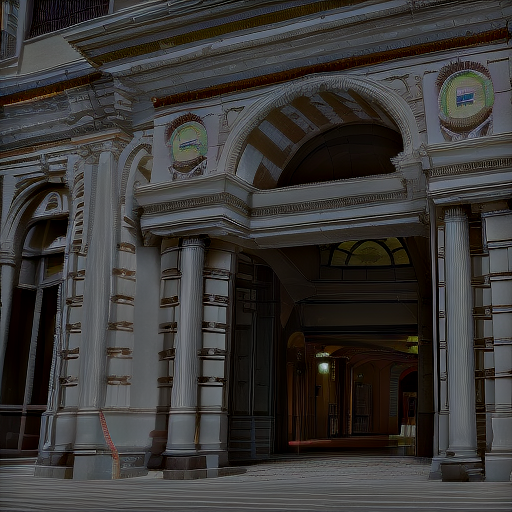}
  \vspace{-6mm}
        \caption*{LCDPNet\cite{wang2022local}+Sagiri}
	\end{minipage}
   	\begin{minipage}[t]{0.16\linewidth}
		\centering
		\captionsetup{font={tiny}}
		\includegraphics[height=2.3cm,width=2.3cm]{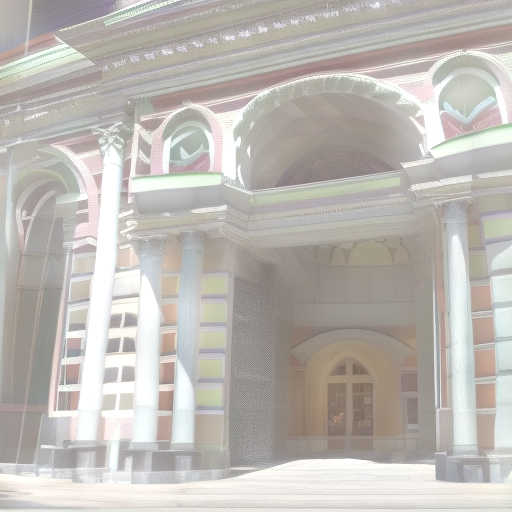}
  \vspace{-6mm}
        \caption*{HDRUNet\cite{chen2021hdrunet}+Sagiri}
	\end{minipage}
	\begin{minipage}[t]{0.16\linewidth}
		\centering
		\captionsetup{font={tiny}}
		\includegraphics[height=2.3cm,width=2.3cm]{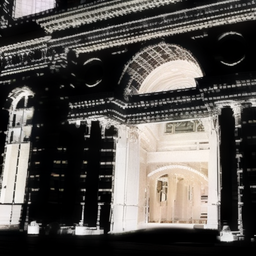}
  \vspace{-6mm}
        \caption*{GlowGAN\cite{wang2023glowgan}+Sagiri}
	\end{minipage}
	\begin{minipage}[t]{0.16\linewidth}
		\centering
		\captionsetup{font={tiny}}
		\includegraphics[height=2.3cm,width=2.3cm]{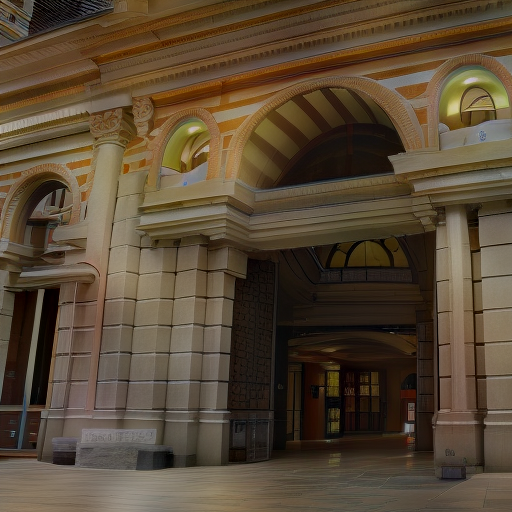}
  \vspace{-6mm}
        \caption*{\jw{LS}+Sagiri(Ours)}
	\end{minipage}
 	\begin{minipage}[t]{0.16\linewidth}
		\centering
		\captionsetup{font={tiny}}
		\includegraphics[height=2.3cm,width=2.3cm]{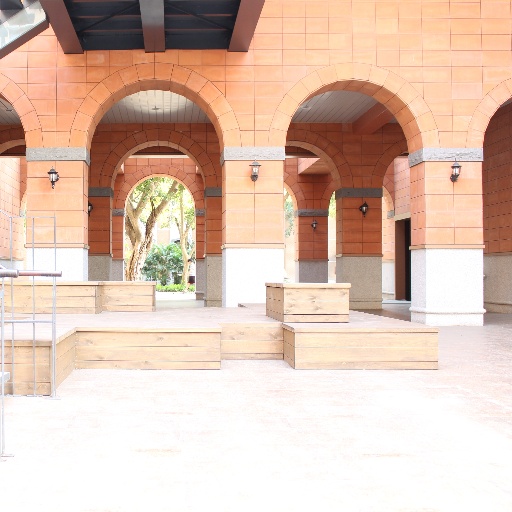}
  \vspace{-6mm}
        \caption*{LQ}
	\end{minipage}
 	\begin{minipage}[t]{0.16\linewidth}
		\centering
		\captionsetup{font={tiny}}
		\includegraphics[height=2.3cm,width=2.3cm]{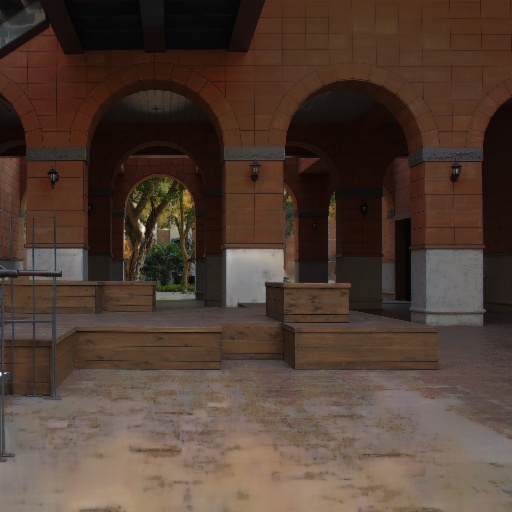}
    \vspace{-6mm}
        \caption*{SingleHDR \cite{liu2020single}}
	\end{minipage}
	\begin{minipage}[t]{0.16\linewidth}
		\centering
		\captionsetup{font={tiny}}
		\includegraphics[height=2.3cm,width=2.3cm]{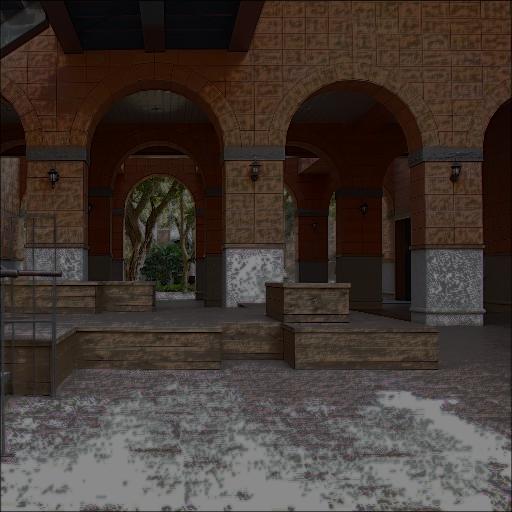}
  \vspace{-6mm}
        \caption*{LCDPNet \cite{wang2022local}}
	\end{minipage}
 	\begin{minipage}[t]{0.16\linewidth}
		\centering
		\captionsetup{font={tiny}}
		\includegraphics[height=2.3cm,width=2.3cm]{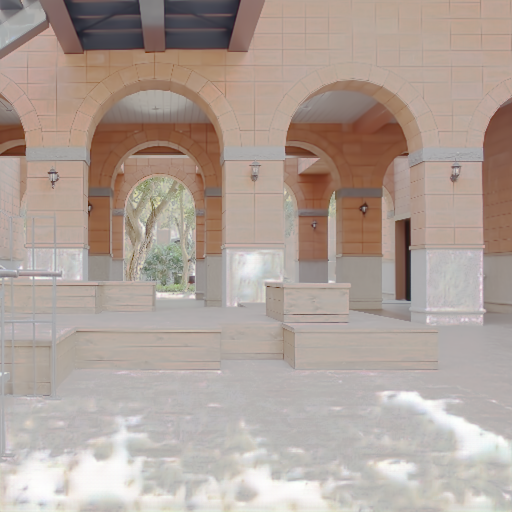}
  \vspace{-6mm}
        \caption*{HDRUNet \cite{chen2021hdrunet}}
	\end{minipage}
  	\begin{minipage}[t]{0.16\linewidth}
		\centering
		\captionsetup{font={tiny}}
		\includegraphics[height=2.3cm,width=2.3cm]{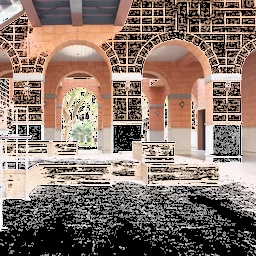}
  \vspace{-6mm}
        \caption*{GlowGAN \cite{wang2023glowgan}}
	\end{minipage}
	\begin{minipage}[t]{0.16\linewidth}
		\centering
		\includegraphics[height=2.3cm,width=2.3cm]{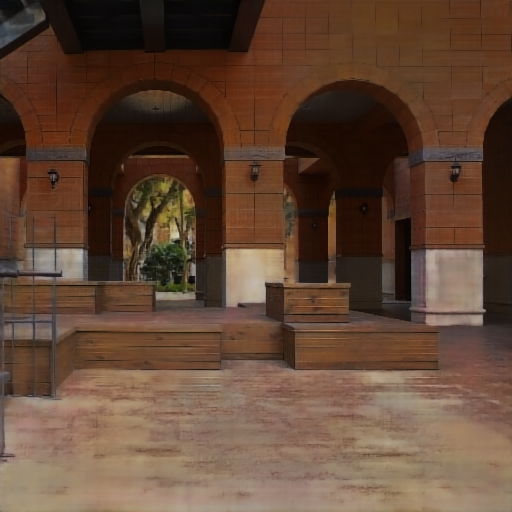}
		\captionsetup{font={tiny}}
  \vspace{-6mm}
        \caption*{Latent-SwinIR$_{c}$(LS, Ours)}
	\end{minipage}\\
 	\begin{minipage}[t]{0.16\linewidth}
		\centering
		\captionsetup{font={tiny}}
		\includegraphics[height=2.3cm,width=2.3cm]{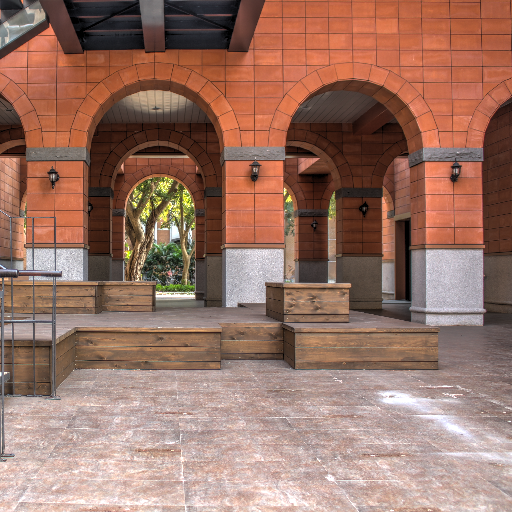}
  \vspace{-6mm}
        \caption*{Reference}
	\end{minipage}
 	\begin{minipage}[t]{0.16\linewidth}
		\centering
		\captionsetup{font={tiny}}
		\includegraphics[height=2.3cm,width=2.3cm]{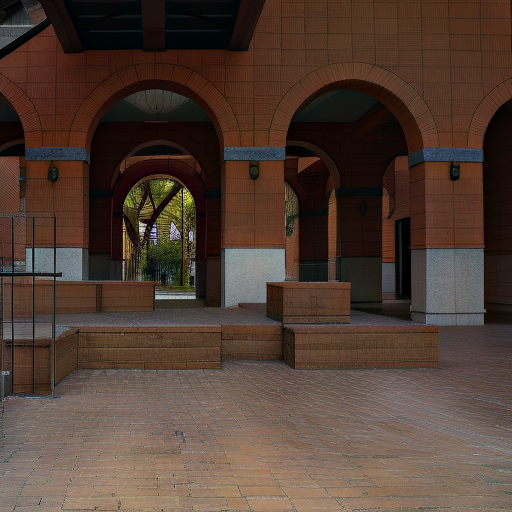}
  \vspace{-6mm}
        \caption*{SingleHDR\cite{liu2020single}+Sagiri}
	\end{minipage}
  	\begin{minipage}[t]{0.16\linewidth}
		\centering
		\captionsetup{font={tiny}}
		\includegraphics[height=2.3cm,width=2.3cm]{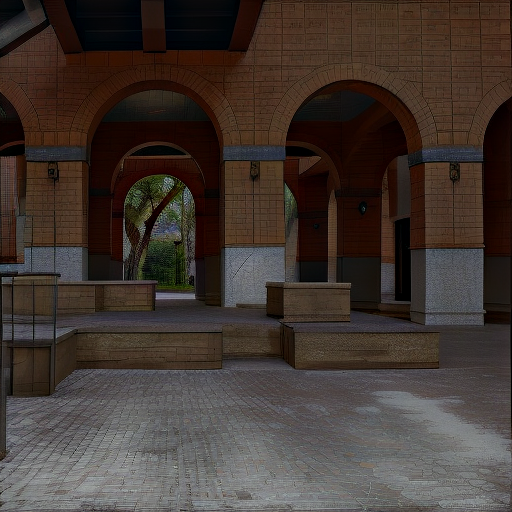}
  \vspace{-6mm}
        \caption*{LCDPNet\cite{wang2022local}+Sagiri}
	\end{minipage}
   	\begin{minipage}[t]{0.16\linewidth}
		\centering
		\captionsetup{font={tiny}}
		\includegraphics[height=2.3cm,width=2.3cm]{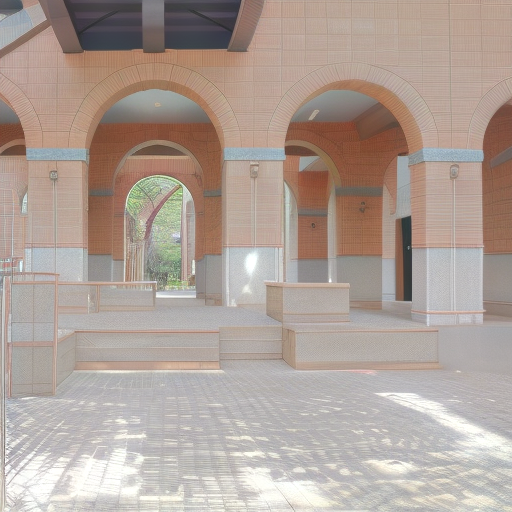}
  \vspace{-6mm}
        \caption*{HDRUNet\cite{chen2021hdrunet}+Sagiri}
	\end{minipage}
	\begin{minipage}[t]{0.16\linewidth}
		\centering
		\captionsetup{font={tiny}}
		\includegraphics[height=2.3cm,width=2.3cm]{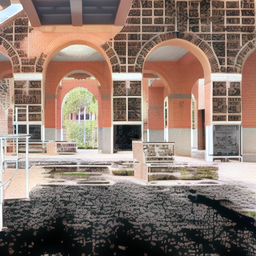}
  \vspace{-6mm}
        \caption*{GlowGAN\cite{wang2023glowgan}+Sagiri}
	\end{minipage}
	\begin{minipage}[t]{0.16\linewidth}
		\centering
		\captionsetup{font={tiny}}
		\includegraphics[height=2.3cm,width=2.3cm]{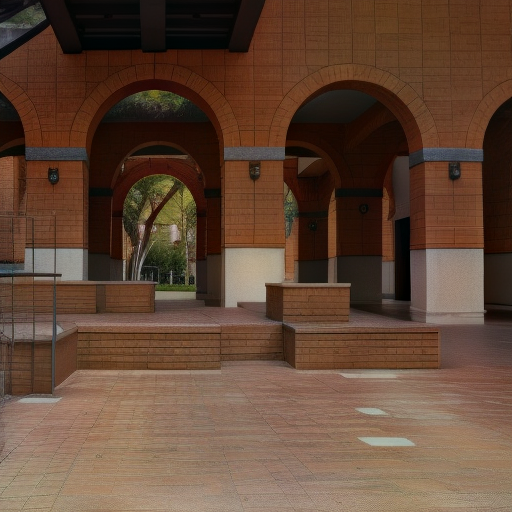}
  \vspace{-6mm}
        \caption*{\jw{LS}+Sagiri(Ours)}
	\end{minipage}
	\caption{
  \textbf{Sagiri as a plug-and-play module.} \sm{Although the images generated by the baselines significantly differ from each other, Sagiri shows strong versatility and improves the visual quality of almost all of them. Additionally, the combination of LS-Sagiri surpasses the performance of other models integrated with Sagiri, confirming the superiority and adaptability of our framework.}
  \vspace{-3mm}
  }
	\label{color_plug}
\end{figure}
\begin{figure}[h]
\vspace{-5mm}
\centering
    \begin{minipage}[t]{0.16\linewidth}
        \centering
        \captionsetup{font={tiny}}
        \includegraphics[height=2.3cm,width=2.3cm]{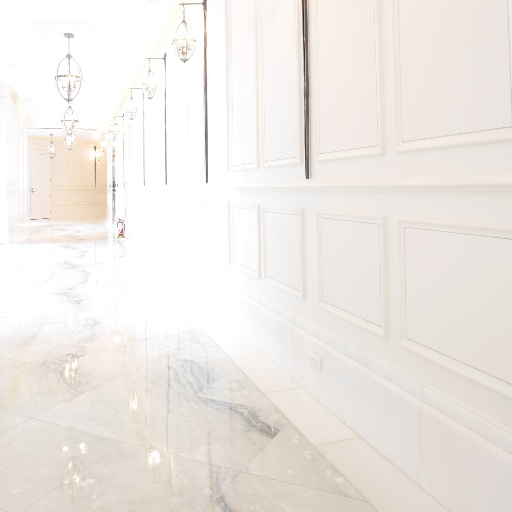}
      \vspace{-6mm}
        \caption*{LQ}
    \end{minipage}
    \begin{minipage}[t]{0.16\linewidth}
        \centering
        \captionsetup{font={tiny}}
        \includegraphics[height=2.3cm,width=2.3cm]{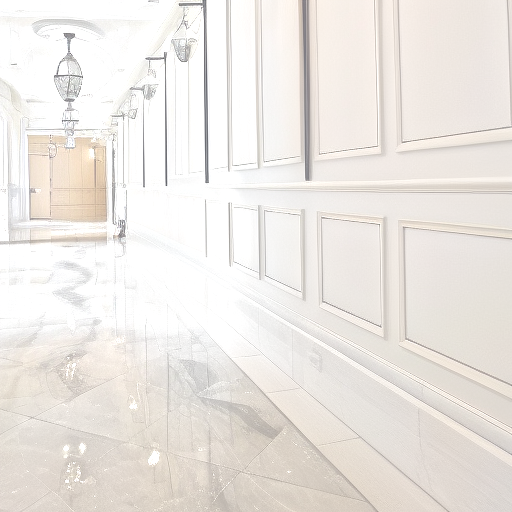}
            \vspace{-6mm}
\caption*{Sagiri}
    \end{minipage}
    \begin{minipage}[t]{0.16\linewidth}
        \centering
        \includegraphics[height=2.3cm,width=2.3cm]{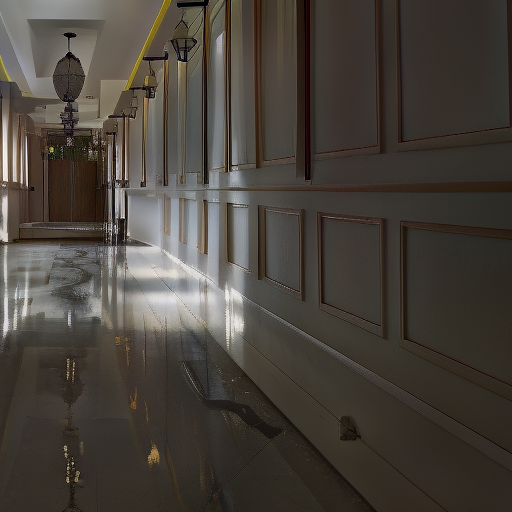}
        \captionsetup{font={tiny}}
           \vspace{-6mm}
 \caption*{LS-Sagiri}
    \end{minipage}
     \begin{minipage}[t]{0.16\linewidth}
        \centering
        \captionsetup{font={tiny}}
        \includegraphics[height=2.3cm,width=2.3cm]{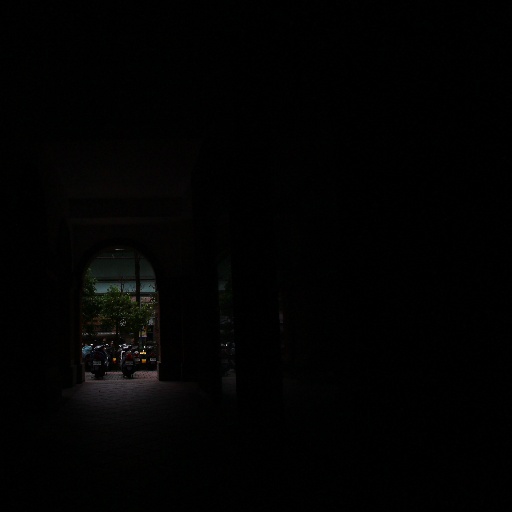}
           \vspace{-6mm}
 \caption*{LQ}
    \end{minipage}
    \begin{minipage}[t]{0.16\linewidth}
        \centering
        \captionsetup{font={tiny}}
        \includegraphics[height=2.3cm,width=2.3cm]{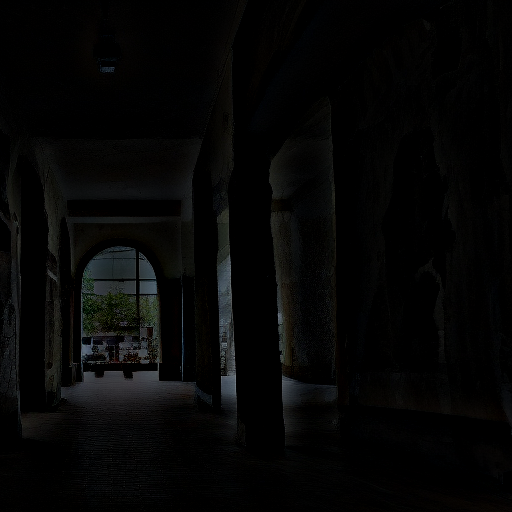}
           \vspace{-6mm}
 \caption*{Sagiri}
    \end{minipage}
    \begin{minipage}[t]{0.16\linewidth}
        \centering
        \includegraphics[height=2.3cm,width=2.3cm]{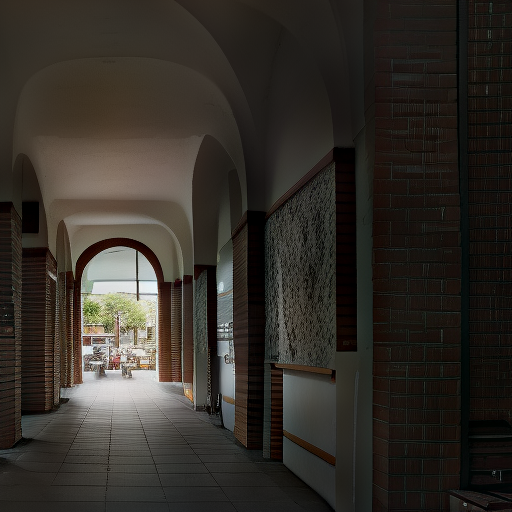}
        \captionsetup{font={tiny}}
           \vspace{-6mm}
 \caption*{LS-Sagiri}
    \end{minipage}
    \vspace{4mm}
    \\
 	\begin{minipage}[t]{0.22\linewidth}
		\centering
		\includegraphics[height=2.8cm,width=3.0cm]{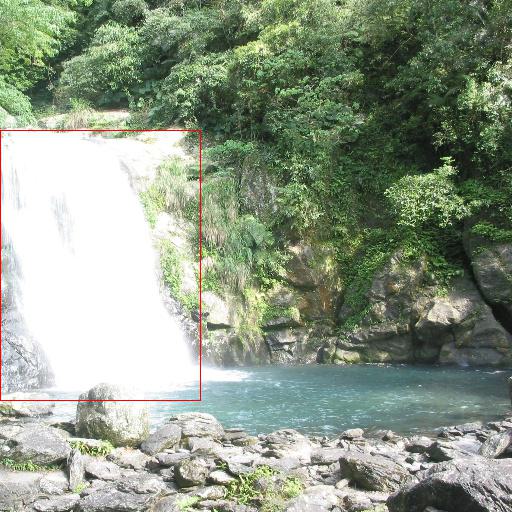}
		\captionsetup{font={tiny}}
           \vspace{-6mm}
 \caption*{LQ}
	\end{minipage}
	\begin{minipage}[t]{0.15\linewidth}
		\centering
		\captionsetup{font={tiny}}
		\includegraphics[height=2.8cm,width=2.1cm]{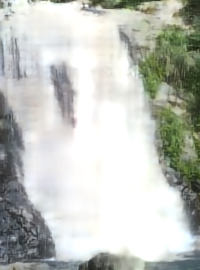}
      \vspace{-6mm}
        \caption*{Latent-SwinIR$_{c}$}
	\end{minipage}
	\begin{minipage}[t]{0.15\linewidth}
		\centering
		\captionsetup{font={tiny}}
		\includegraphics[height=2.8cm,width=2.1cm]{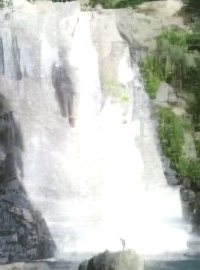}
            \vspace{-6mm}
\caption*{w/o pretrain}

	\end{minipage}
	\begin{minipage}[t]{0.15\linewidth}
		\centering
		\includegraphics[height=2.8cm,width=2.1cm]{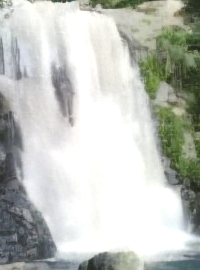}
		\captionsetup{font={tiny}}
           \vspace{-6mm}
 \caption*{w/o prompt}
	\end{minipage}
 	\begin{minipage}[t]{0.15\linewidth}
		\centering
		\captionsetup{font={tiny}}
		\includegraphics[height=2.8cm,width=2.1cm]{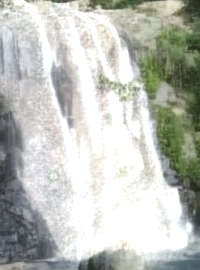}
           \vspace{-6mm}
  \caption*{w/o ConRLoss}
	\end{minipage}
	\begin{minipage}[t]{0.15\linewidth}
		\centering
		\captionsetup{font={tiny}}
		\includegraphics[height=2.8cm,width=2.1cm]{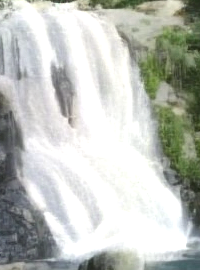}
           \vspace{-6mm}
 \caption*{LS-Sagiri}
	\end{minipage}\\
 \vspace{4mm}
 \begin{minipage}[t]{0.16\linewidth}
        \centering
        \captionsetup{font={tiny}}
        \includegraphics[height=2.3cm,width=2.3cm]{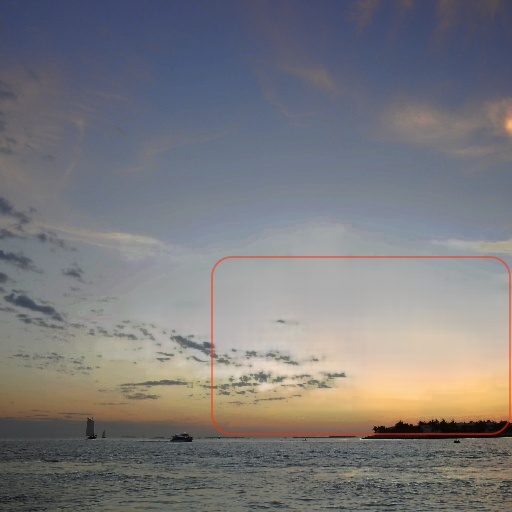}
      \vspace{-6mm}
        \caption*{SingleHDR \cite{liu2020single}}    \end{minipage}
    \begin{minipage}[t]{0.16\linewidth}
        \centering
        \captionsetup{font={tiny}}
        \includegraphics[height=2.3cm,width=2.3cm]{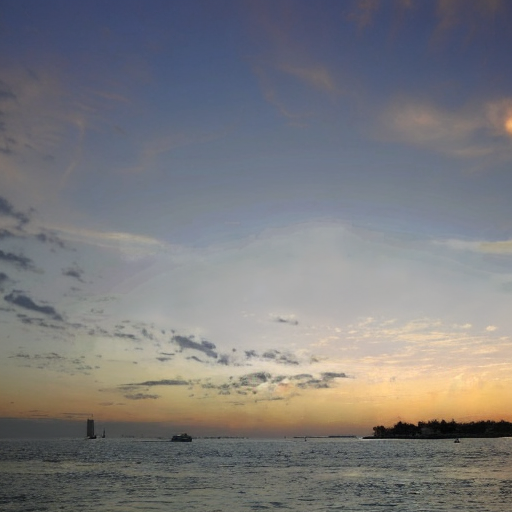}
            \vspace{-6mm}
\caption*{+Sagiri (Prompt a)}
    \end{minipage}
    \begin{minipage}[t]{0.16\linewidth}
        \centering
        \includegraphics[height=2.3cm,width=2.3cm]{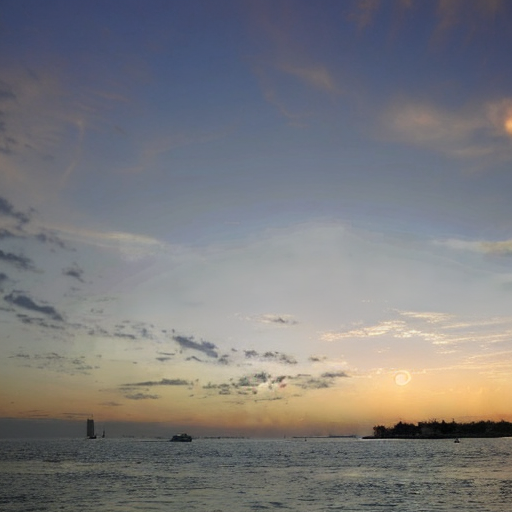}
        \captionsetup{font={tiny}}
           \vspace{-6mm}
\caption*{+Sagiri (Prompt b)}
    \end{minipage}
     \begin{minipage}[t]{0.16\linewidth}
        \centering
        \captionsetup{font={tiny}}
        \includegraphics[height=2.3cm,width=2.3cm]{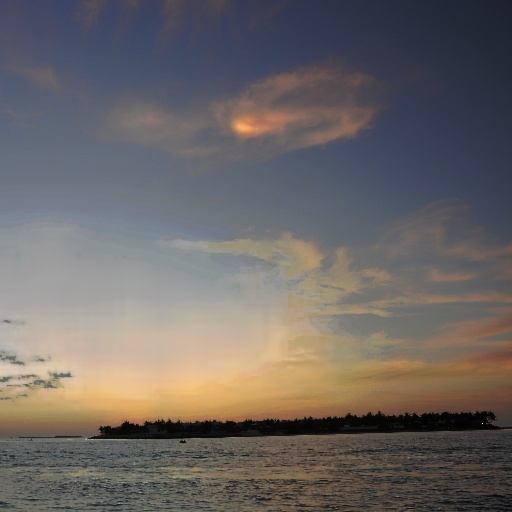}
      \vspace{-6mm}
        \caption*{SingleHDR \cite{liu2020single}}
    \end{minipage}
    \begin{minipage}[t]{0.16\linewidth}
        \centering
        \captionsetup{font={tiny}}
        \includegraphics[height=2.3cm,width=2.3cm]{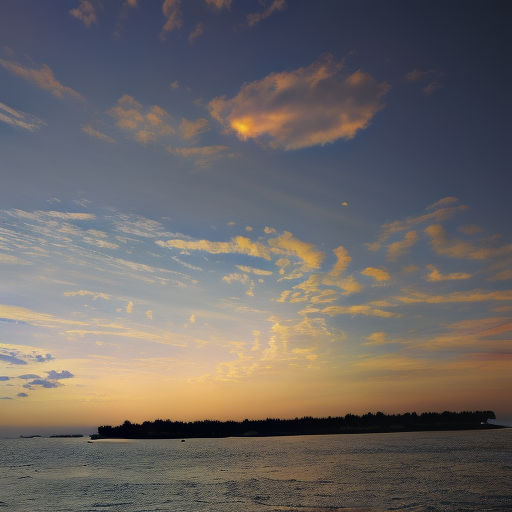}
            \vspace{-6mm}
\caption*{+Sagiri (Prompt a)}
    \end{minipage}
    \begin{minipage}[t]{0.16\linewidth}
        \centering
        \includegraphics[height=2.3cm,width=2.3cm]{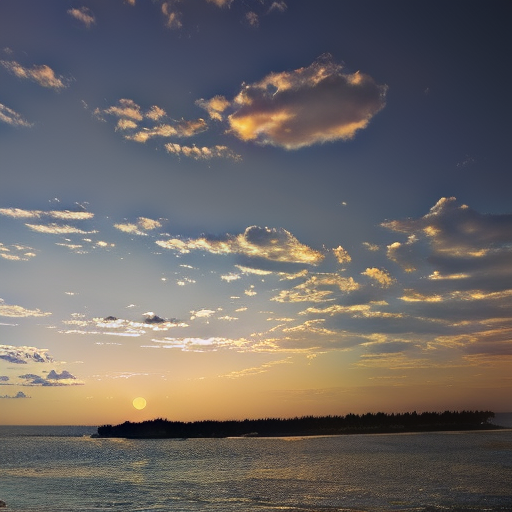}
        \captionsetup{font={tiny}}
           \vspace{-6mm}
 \caption*{+Sagiri (Prompt b)}
    \end{minipage}

    \caption{\textbf{Ablation studies. (Top)} We attempted to force Sagiri to learn both color distribution correction and details generation, which led to weak color mapping capabilities. \textbf{(Middle)} Ablation of pretraining, text prompt and content reconstruction loss. Prompts generated by CogVLM \cite{wang2023cogvlm}: ``A white waterfall is flowing down from the cliff, surrounded by rocks and trees.'' \textbf{(Bottom)} We use different user-defined unknown region mask and different prompts on Sagiri to refine SingleHDR's~\cite{liu2020single} results. Left: We manually select the red box. Right: We select the entire image. Prompt a: ``The sky is filled with clouds.'' Prompt b: ``The sun is setting, and the sky is filled with clouds.''}
    \label{ablation}
 \vspace{-2mm}
\end{figure}
\noindent\textbf{Baseline methods.} 
 We compare our method with SingleHDR~\cite{liu2020single}, LCDPNet~\cite{wang2022local},
 HDRUNet~\cite{chen2021hdrunet}, GlowGAN~\cite{wang2023glowgan},
 (and GDP~\cite{fei2023generative} in the supplementary material). 

 \textbf{Metrics.} To evaluate the overall performance of our LS-Sagiri model, as well as Sagiri's generalizability and adaptability in refining outputs from different models, we utilize non-reference metrics such as BRISQUE~\cite{brisque}, NIQE~\cite{niqe}, MANIQA~\cite{maniqa} and CLIP-IQA~\cite{clipiqa}, which primarily assess visual effects. We do not use PSNR, SSIM~\cite{ssim} and LPIPS~\cite{lpips} for comparing Sagiri's performance, as these metrics have limitations in evaluating generative models, as demonstrated by previous studies~\cite{jinjin2020pipal,gu2022ntire,blau2018perception,yu2024scaling}.
 


\textbf{Performance of Latent-SwinIR$_c$.} \sm{Figure~\ref{compare_ls}(a-f) shows Latent-SwinIR$_c$'s capability of correcting the color distribution in an image. Existing methods suffer from low contrast, low brightness, quantization-like artifacts or wrong colors. Due to the proposed loss functions, Latent-SwinIR$_c$ can map the brightness to the right range with correct colors, achieving the best visual quality.}

\textbf{Performance of LS-Sagiri. }
\sm{Figure~\ref{compare_ls}(g) and Figure~\ref{compare_glow} shows the performance of the whole pipeline LS-Sagiri. While existing methods struggle in generating the content in saturated regions, LS-Sagiri can fill in the details even in large saturated regions.
Quantitative results for the comparisons are presented in Table~\ref{tab1}. 
The proposed LS-Sagiri achieves the best score on almost every metric, showing its superior performance in enhancing LDR images across various datasets. Notice that using Latent-SwinIR$_c$ alone does not always lead to good scores. Our hypothesis is that current non-reference image quality metrics do not take the overall brightness distribution into consideration, which can be a potential area for future work. }

\textbf{Sagiri as a plug-and-play module.} \sm{In addition to working as 
a refine step for Latent-SwinIR$_c$, Figure~\ref{color_plug} shows that the proposed
Sagiri model can also work as a plug-and-play module for existing LDR enhancement models. Although 
the output images of different method vary widely in quality, Sagiri is able to fix and generate details in the dynamic range extremes, enhancing their perceptual quality. This versatility of Sagiri is also demonstrated in Table~\ref{tab1}, where Sagiri significantly improves the outputs of almost every baseline method. The only exception is GlowGAN, for which we give a detailed explanation in Appendix~\ref{sec:glowgan_performance}.}

\subsection{Ablation Studies}

\textbf{The use of \jw{two-stage} model.}
To assess the necessity of having a two-stage pipeline, we attempt to use only Sagiri for both color reconstruction and content fine adjustment. The results in Figure~\ref{ablation} (Top) indicate that Sagiri alone lacks sufficient capabilities for color restoration and brightness adjustment.

\textbf{The use of pre-training and prompt.} 
To illustrate the impact of our pre-training strategy and prompts on guiding details generation, we provide visual results in Figure~\ref{ablation} (Middle). 
\sm{The pre-training strategy and the auto-generated text prompt leads to clear improvements in image quality. The proposed content reconstruction loss (ConRLoss) also plays an important role, which is quantitatively evaluated in Appendix~\ref{sec:ablation_losses}. 
Additionally, Figure~\ref{ablation} (Bottom) shows that our approach allows users to determine (1) where  to generate the contents by replacing the unknown region mask with a user-defined region of interest, and (2) what content to generate by providing a user-defined text prompt. }

\section{Conclusion}
We introduce a pioneering pipeline for low dynamic range (LDR) image enhancement, centered around our robust and adaptable model, Sagiri, which seamlessly integrates with a variety of restoration methods to deliver visually compelling results. More specifically, the pipeline includes two stages, where stage 1 Latent-SwinIR$_c$ (LS) corrects the brightness and color distribution and stage 2 Sagiri generates content for missing areas and enhances details. Sagiri is trained in a way that it can also be directly plugged into other methods to enhance their results. Our comprehensive experiments validate the superior performance of the LS-Sagiri framework and demonstrate Sagiri's exceptional detail generation capabilities. Currently Sagiri outputs an LDR image due to limitation that Stable Diffusion is trained on LDR images only. Potential future direction includes adapting Sagiri such that it outputs HDR images (files), allowing more freedom in user-defined tone-mapping to generate the final output.  Another direction is to adopt the recent progress on  Stable Diffusion speedup and model size reduction to make the task more accessible to mobile devices.
\\
\small
\bibliographystyle{plain}
\bibliography{hdr}
\appendix
\clearpage

\section{Comparing Latent-SwinIR$_c$ with Other LDR Enhancement Methods}
\begin{figure}[t]
	\centering
 	\begin{minipage}[t]{0.13\linewidth}
		\centering
		\captionsetup{font={tiny}}
		\includegraphics[height=2.1cm,width=1.9cm]{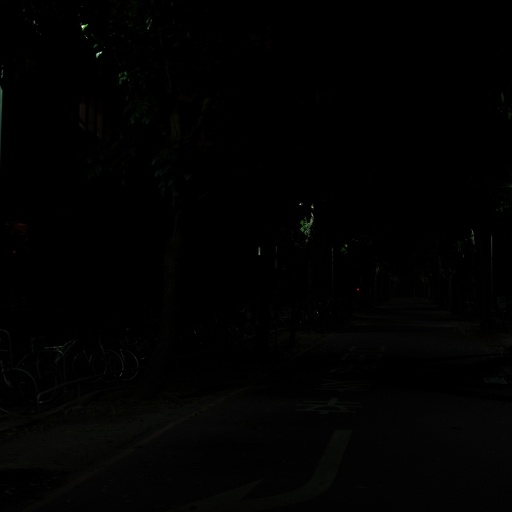}
  \vspace{-6mm}
	\end{minipage}
  	\begin{minipage}[t]{0.13\linewidth}
		\centering
		\captionsetup{font={tiny}}
		\includegraphics[height=2.1cm,width=1.9cm]{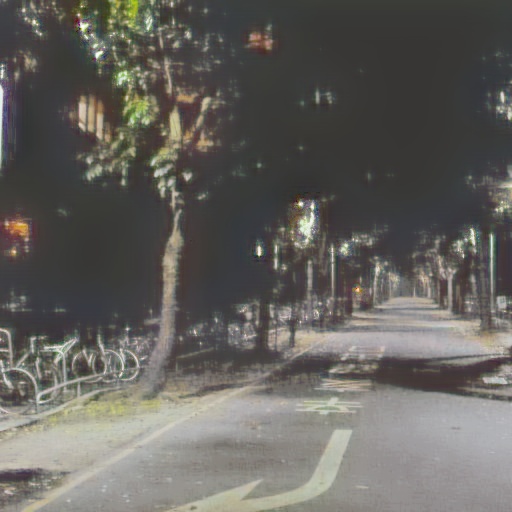}
         \vspace{-6mm}
	\end{minipage}
	\begin{minipage}[t]{0.13\linewidth}
		\centering
		\includegraphics[height=2.1cm,width=1.9cm]{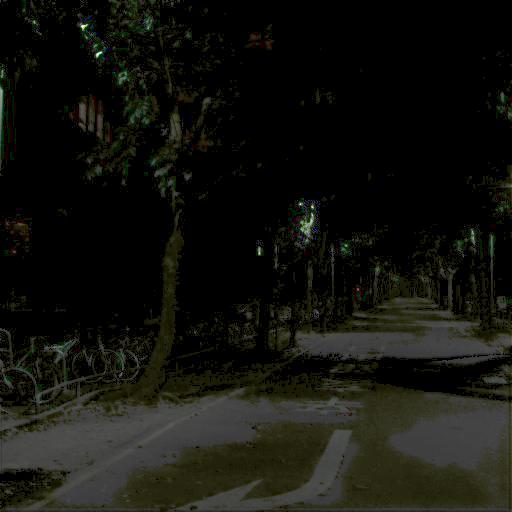}
		\captionsetup{font={tiny}}
         \vspace{-6mm}
	\end{minipage}
  	\begin{minipage}[t]{0.13\linewidth}
		\centering
		\captionsetup{font={tiny}}
		\includegraphics[height=2.1cm,width=1.9cm]{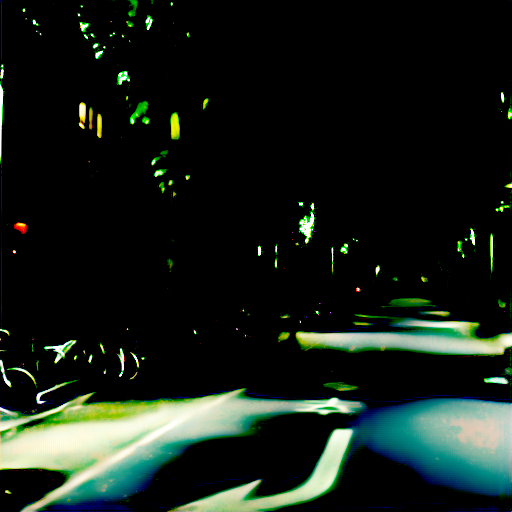}
         \vspace{-6mm}
	\end{minipage}
 	\begin{minipage}[t]{0.13\linewidth}
		\centering
		\captionsetup{font={tiny}}
		\includegraphics[height=2.1cm,width=1.9cm]{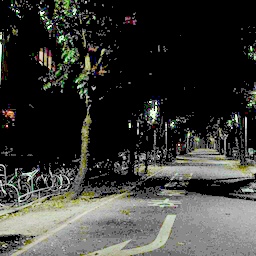}
          \vspace{-6mm}
	\end{minipage}
	\begin{minipage}[t]{0.13\linewidth}
		\centering
		\captionsetup{font={tiny}}
		\includegraphics[height=2.1cm,width=1.9cm]{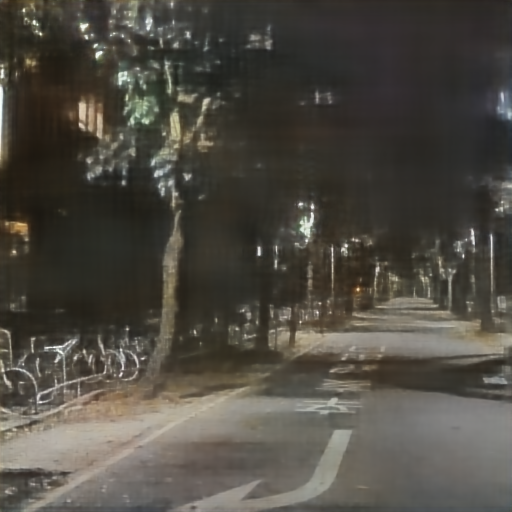}
         \vspace{-6mm}
	\end{minipage}
 	\begin{minipage}[t]{0.13\linewidth}
		\centering
		\captionsetup{font={tiny}}
		\includegraphics[height=2.1cm,width=1.9cm]{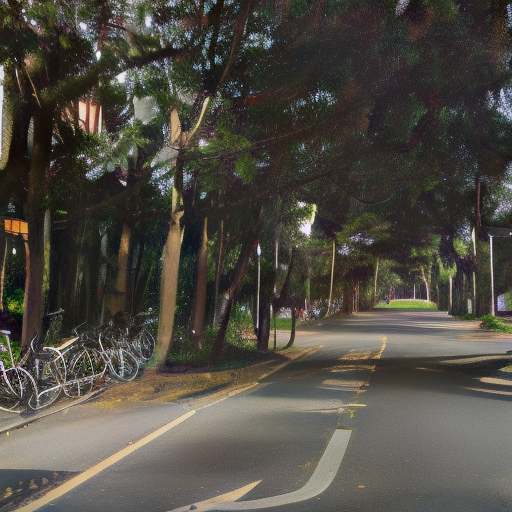}
         \vspace{-6mm}
	\end{minipage}
 	\begin{minipage}[t]{0.13\linewidth}
		\centering
		\captionsetup{font={tiny}}
		\includegraphics[height=2.1cm,width=1.9cm]{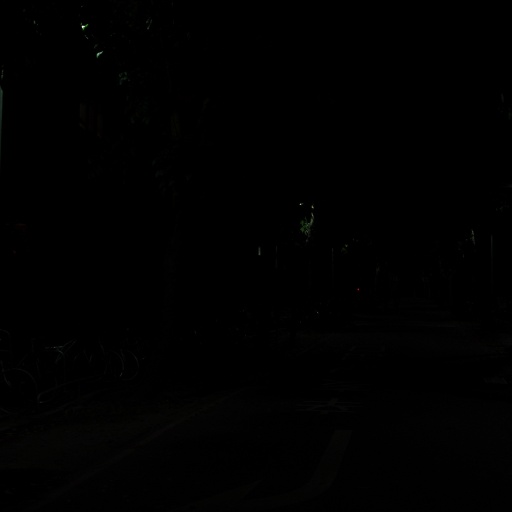}
  \vspace{-6mm}
	\end{minipage}
  	\begin{minipage}[t]{0.13\linewidth}
		\centering
		\captionsetup{font={tiny}}
		\includegraphics[height=2.1cm,width=1.9cm]{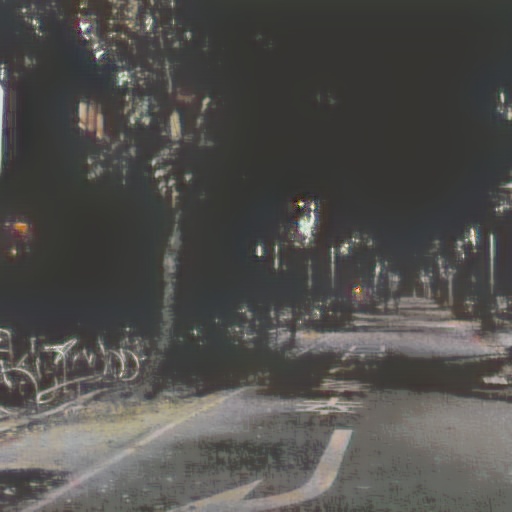}
         \vspace{-6mm} 
	\end{minipage}
	\begin{minipage}[t]{0.13\linewidth}
		\centering
		\includegraphics[height=2.1cm,width=1.9cm]{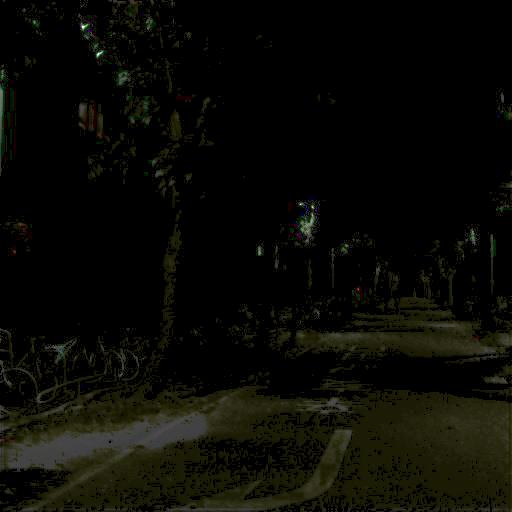}
		\captionsetup{font={tiny}}
         \vspace{-6mm}
	\end{minipage}
  	\begin{minipage}[t]{0.13\linewidth}
		\centering
		\captionsetup{font={tiny}}
		\includegraphics[height=2.1cm,width=1.9cm]{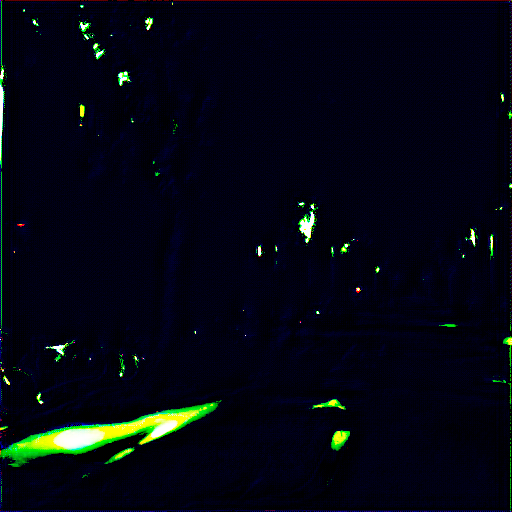}
         \vspace{-6mm}
	\end{minipage}
 	\begin{minipage}[t]{0.13\linewidth}
		\centering
		\captionsetup{font={tiny}}
		\includegraphics[height=2.1cm,width=1.9cm]{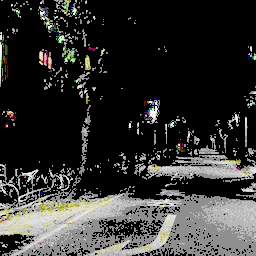}
          \vspace{-6mm}
	\end{minipage}
	\begin{minipage}[t]{0.13\linewidth}
		\centering
		\captionsetup{font={tiny}}
		\includegraphics[height=2.1cm,width=1.9cm]{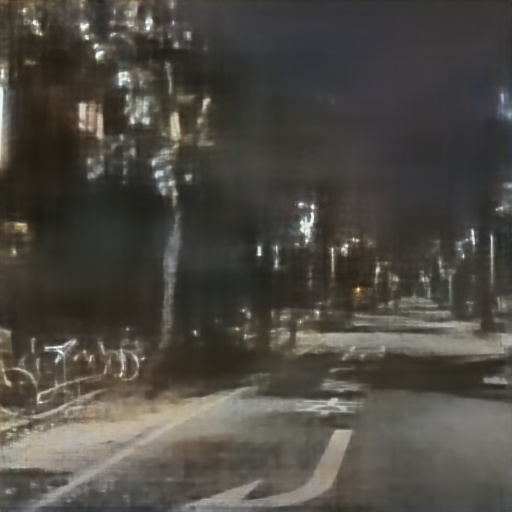}
         \vspace{-6mm}
	\end{minipage}
 	\begin{minipage}[t]{0.13\linewidth}
		\centering
		\captionsetup{font={tiny}}
		\includegraphics[height=2.1cm,width=1.9cm]{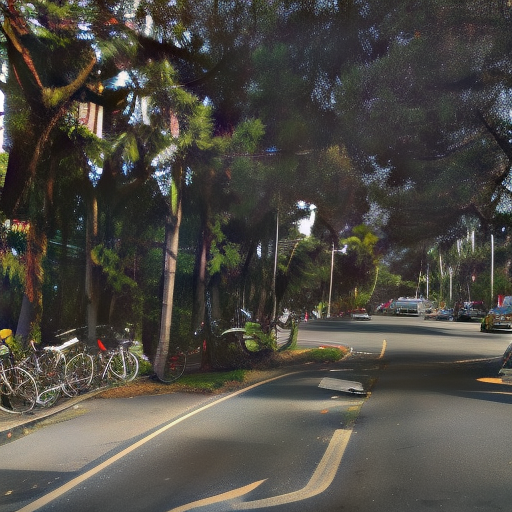}
         \vspace{-6mm}
	\end{minipage}
 	\begin{minipage}[t]{0.13\linewidth}
		\centering
		\captionsetup{font={tiny}}
		\includegraphics[height=2.1cm,width=1.9cm]{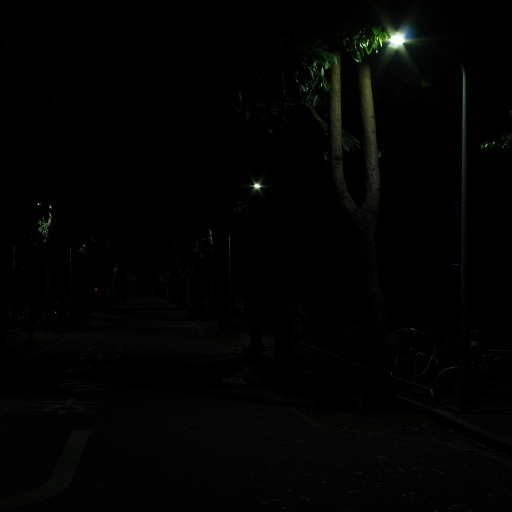}
  \vspace{-6mm}
	\end{minipage}
  	\begin{minipage}[t]{0.13\linewidth}
		\centering
		\captionsetup{font={tiny}}
		\includegraphics[height=2.1cm,width=1.9cm]{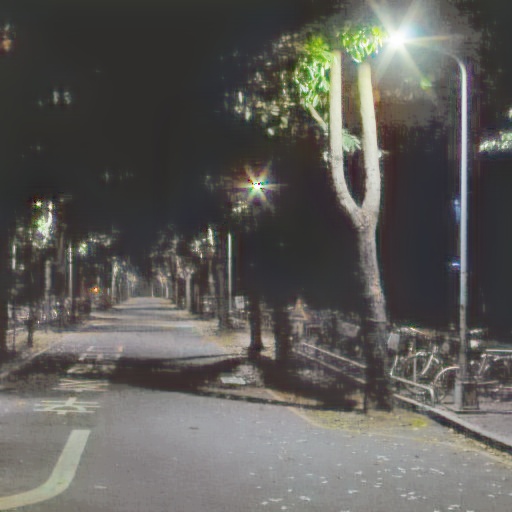}
         \vspace{-6mm}
	\end{minipage}
	\begin{minipage}[t]{0.13\linewidth}
		\centering
		\includegraphics[height=2.1cm,width=1.9cm]{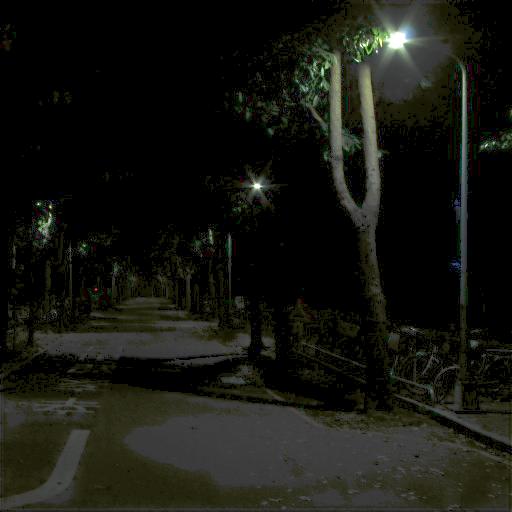}
		\captionsetup{font={tiny}}
         \vspace{-6mm}
	\end{minipage}
  	\begin{minipage}[t]{0.13\linewidth}
		\centering
		\captionsetup{font={tiny}}
		\includegraphics[height=2.1cm,width=1.9cm]{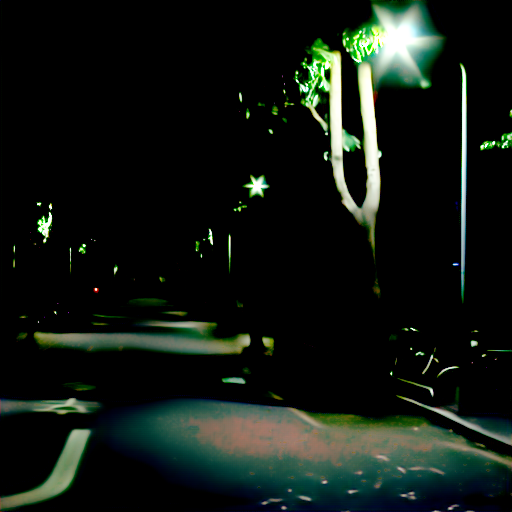}
         \vspace{-6mm}
	\end{minipage}
 	\begin{minipage}[t]{0.13\linewidth}
		\centering
		\captionsetup{font={tiny}}
		\includegraphics[height=2.1cm,width=1.9cm]{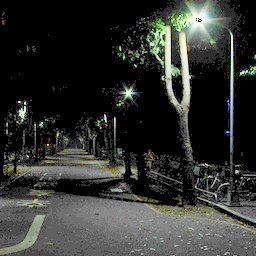}
          \vspace{-6mm}
	\end{minipage}
	\begin{minipage}[t]{0.13\linewidth}
		\centering
		\captionsetup{font={tiny}}
		\includegraphics[height=2.1cm,width=1.9cm]{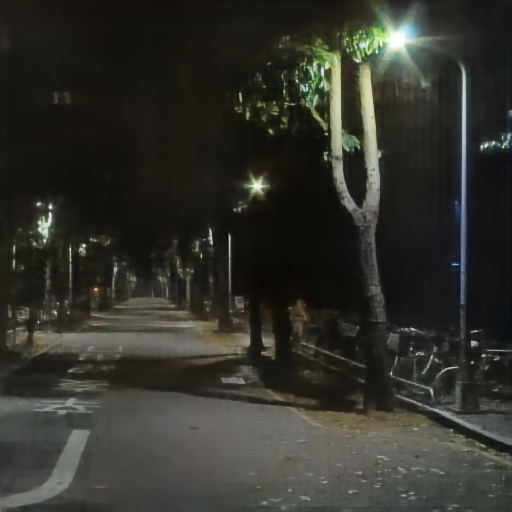}
         \vspace{-6mm}
	\end{minipage}
 	\begin{minipage}[t]{0.13\linewidth}
		\centering
		\captionsetup{font={tiny}}
		\includegraphics[height=2.1cm,width=1.9cm]{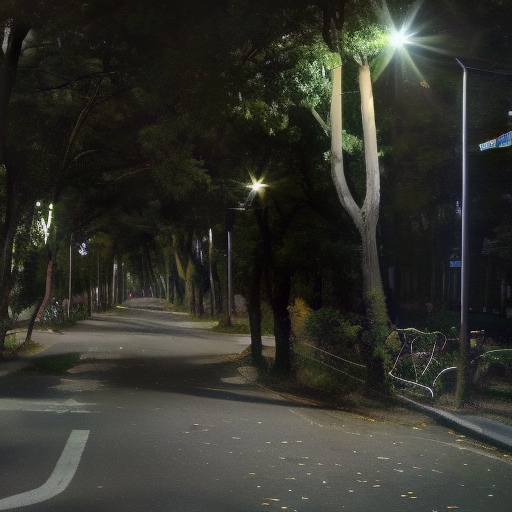}
         \vspace{-6mm}
	\end{minipage}
 	\begin{minipage}[t]{0.13\linewidth}
		\centering
		\captionsetup{font={tiny}}
		\includegraphics[height=2.1cm,width=1.9cm]{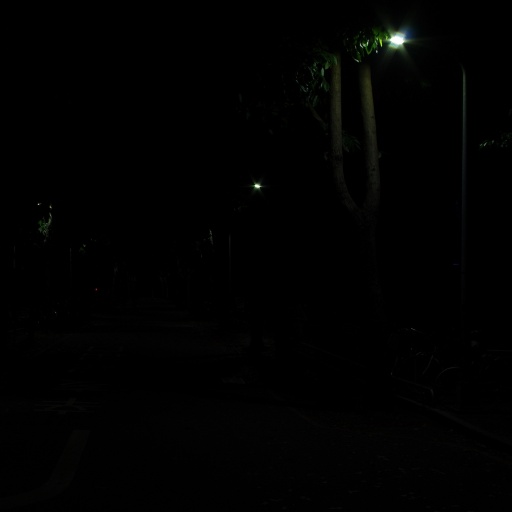}
  \vspace{-6mm}
	\end{minipage}
  	\begin{minipage}[t]{0.13\linewidth}
		\centering
		\captionsetup{font={tiny}}
		\includegraphics[height=2.1cm,width=1.9cm]{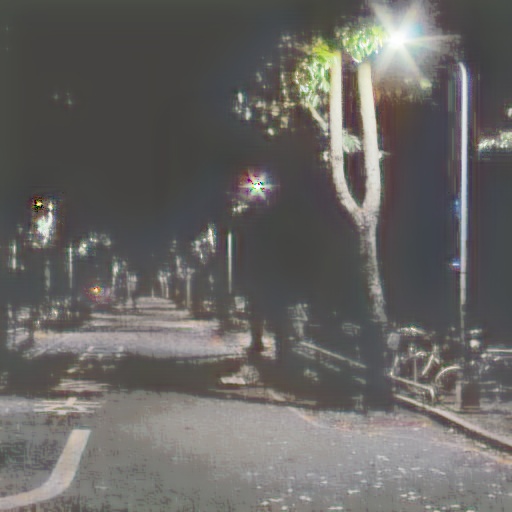}
         \vspace{-6mm}
	\end{minipage}
	\begin{minipage}[t]{0.13\linewidth}
		\centering
		\includegraphics[height=2.1cm,width=1.9cm]{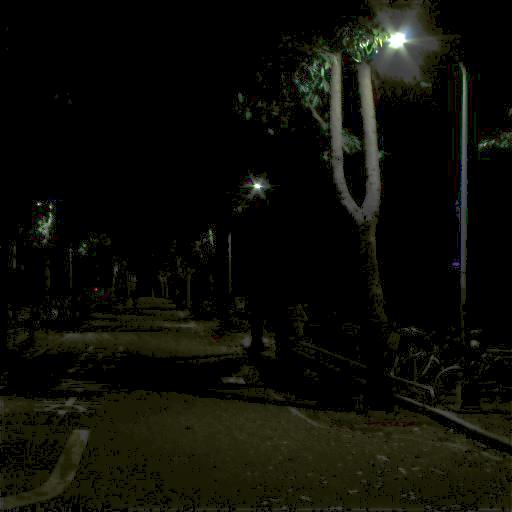}
		\captionsetup{font={tiny}}
         \vspace{-6mm}
	\end{minipage}
  	\begin{minipage}[t]{0.13\linewidth}
		\centering
		\captionsetup{font={tiny}}
		\includegraphics[height=2.1cm,width=1.9cm]{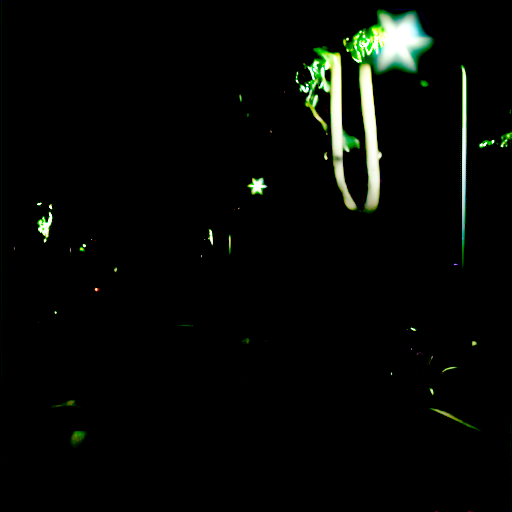}
         \vspace{-6mm}
	\end{minipage}
 	\begin{minipage}[t]{0.13\linewidth}
		\centering
		\captionsetup{font={tiny}}
		\includegraphics[height=2.1cm,width=1.9cm]{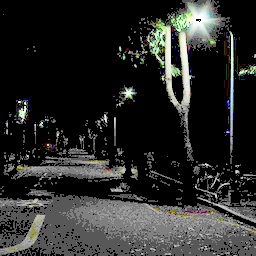}
          \vspace{-6mm}
	\end{minipage}
	\begin{minipage}[t]{0.13\linewidth}
		\centering
		\captionsetup{font={tiny}}
		\includegraphics[height=2.1cm,width=1.9cm]{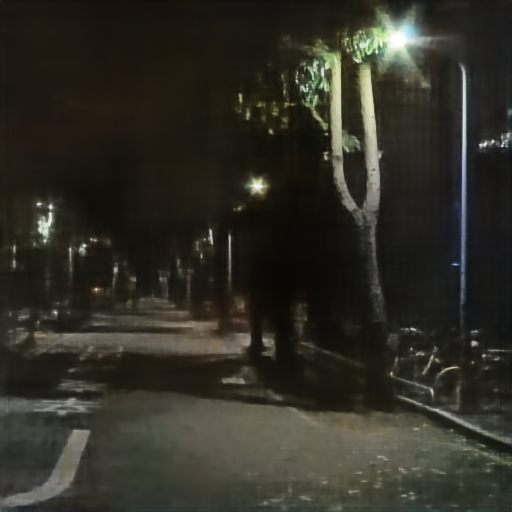}
         \vspace{-6mm}
	\end{minipage}
 	\begin{minipage}[t]{0.13\linewidth}
		\centering
		\captionsetup{font={tiny}}
		\includegraphics[height=2.1cm,width=1.9cm]{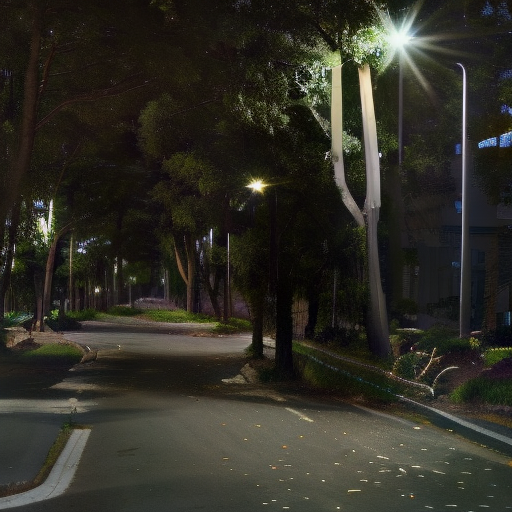}
         \vspace{-6mm}
	\end{minipage}
	\begin{minipage}[t]{0.13\linewidth}
		\centering
		\captionsetup{font={tiny}}
		\includegraphics[height=2.1cm,width=1.9cm]{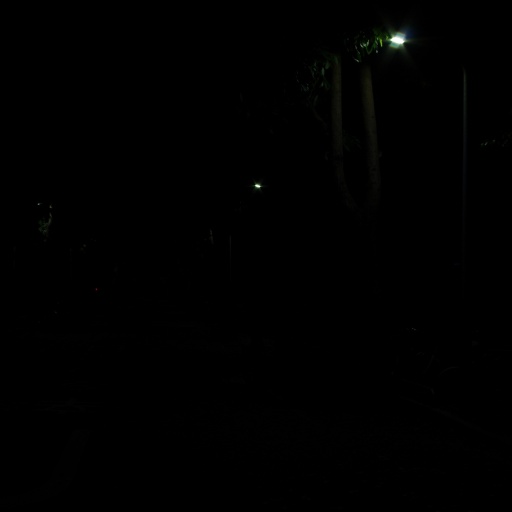}
  \vspace{-6mm}
        \caption*{(a) LQ}
	\end{minipage}
  	\begin{minipage}[t]{0.13\linewidth}
		\centering
		\captionsetup{font={tiny}}
		\includegraphics[height=2.1cm,width=1.9cm]{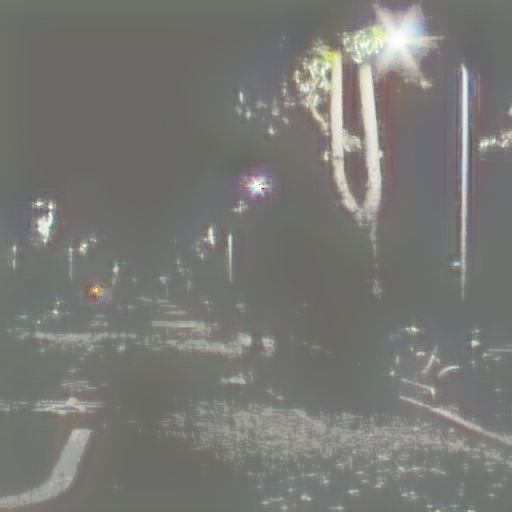}
         \vspace{-6mm}
 \caption*{(b) SingleHDR \cite{le2023single}}
	\end{minipage}
	\begin{minipage}[t]{0.13\linewidth}
		\centering
		\includegraphics[height=2.1cm,width=1.9cm]{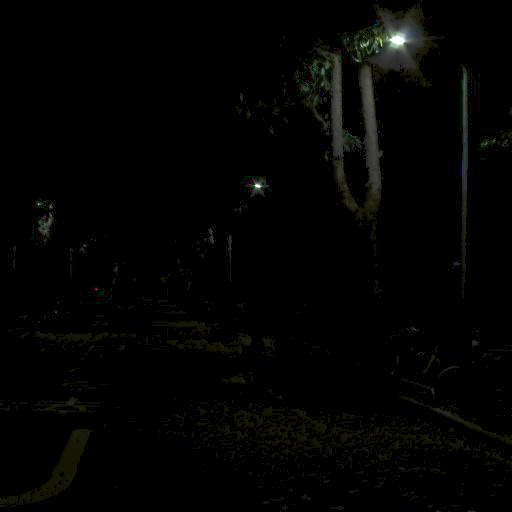}
		\captionsetup{font={tiny}}
         \vspace{-6mm}
 \caption*{(c) LCDP-Net \cite{wang2022local}}
	\end{minipage}
  	\begin{minipage}[t]{0.13\linewidth}
		\centering
		\captionsetup{font={tiny}}
		\includegraphics[height=2.1cm,width=1.9cm]{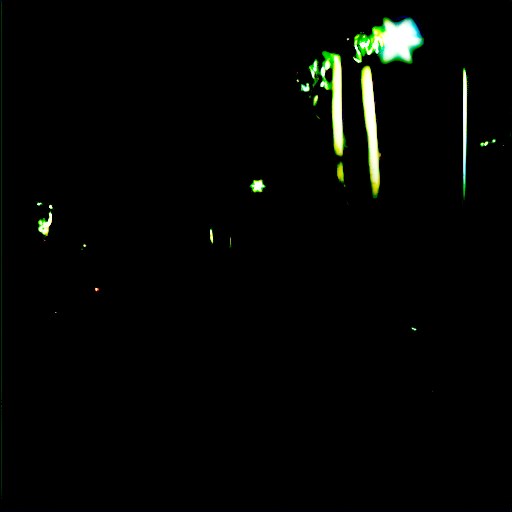}
         \vspace{-6mm}
 \caption*{(d) HDRUNet \cite{chen2021hdrunet}}
	\end{minipage}
 \begin{minipage}[t]{0.13\linewidth}
		\centering
		\captionsetup{font={tiny}}
		\includegraphics[height=2.1cm,width=1.9cm]{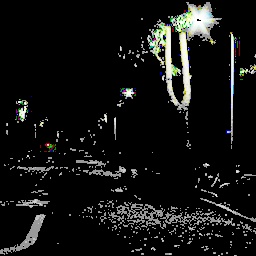}
          \vspace{-6mm}
\caption*{(e) Glow-GAN \cite{wang2023glowgan}}
	\end{minipage}
	\begin{minipage}[t]{0.13\linewidth}
		\centering
		\captionsetup{font={tiny}}
		\includegraphics[height=2.1cm,width=1.9cm]{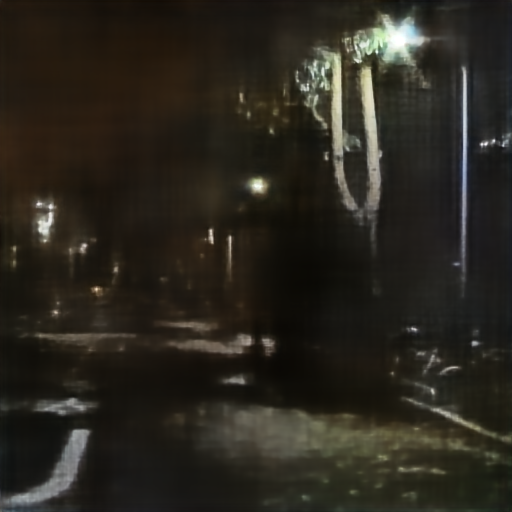}
         \vspace{-6mm}
 \caption*{(f) Latent-SwinIR$_{c}$}
	\end{minipage}
 	\begin{minipage}[t]{0.13\linewidth}
		\centering
		\captionsetup{font={tiny}}
		\includegraphics[height=2.1cm,width=1.9cm]{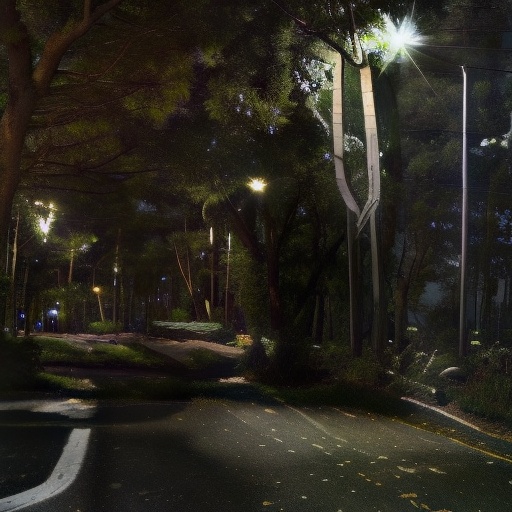}
         \vspace{-6mm}
 \caption*{(g) LS-Sagiri}
	\end{minipage}
 \caption{\sm{Comparison of our Stage 1 model Latent-SwinIR$_{c}$ (LS) to other methods. \textbf{(a)} Input LQ images, with a gradual decrease in exposure. \textbf{(b-f)} The performance of existing methods is affected by the exposure. While SingleHDR achieves the closest performance to our method, it still produces hazy results for low exposures. \textbf{(f)} LS achieves robust color and brightness preservation as the exposure decreases. \textbf{(g)} Sagiri enhances LS's results with generated details. }}
 \label{compare_robust}
 \end{figure}

Figure \ref{compare_robust} shows the comparison between Latent-SwinIR$_c$ and other LDR enhancement methods. 
Our Latent-SwinIR$_{c}$ (LS) can maintain relatively robust color and brightness preservation, and further be enhanced through Sagiri. However, other methods are affected by the exposure of the original image and cannot maintain a good balance of the brightness. SingleHDR \cite{le2023single} is the strongest competitor, but it still generates hazy results.

\section{Comparing LS-Sagiri with Other Generative Methods}
Several generative approaches have been developed for tasks akin to ours, notable among them being Generative Diffusion Prior (GDP) \cite{fei2023generative} and GlowGAN \cite{wang2023glowgan}. Yet, these methods exhibit limitations when dealing with dynamic range extremes and require long inference times. Specifically, GlowGAN is constrained by the generative capabilities of GANs, impairing its effectiveness in content recovery within dynamic range extremes. Additionally, the unsupervised training model of GDP does not ensure high fidelity, which compromises its performance in correcting overexposed images to achieve normal illumination levels. Moreover, GDP lacks the capability to reconstruct HDR images from single LDR inputs since it requires multi-exposure inputs, which further limiting its applicability in achieving our desired outcomes.
Unlike previous approaches, our method offers several advantages:

 \begin{table}[h]
\caption{Comparison of inference time with other generative methods.}
    \centering
    \footnotesize
    \begin{tabular}{cccc}
\toprule
\multicolumn{1}{c}{Methods} & \multicolumn{1}{c}{GDP~\cite{fei2023generative}} & \multicolumn{1}{c}{GlowGAN~\cite{wang2023glowgan}}& \multicolumn{1}{c}{LS-Sagiri}  \\ 
\midrule
\multicolumn{1}{c}{Infer time(per image)} & \multicolumn{1}{c}{900s} & \multicolumn{1}{c}{90s} & \multicolumn{1}{c}{4s} \\
\bottomrule
    \end{tabular}
    \label{table_infer}
\end{table}

 \begin{figure}[t]
	\centering
	\begin{minipage}[t]{0.19\linewidth}
		\centering
		\captionsetup{font={tiny}}
		\includegraphics[height=2.7cm,width=2.7cm]{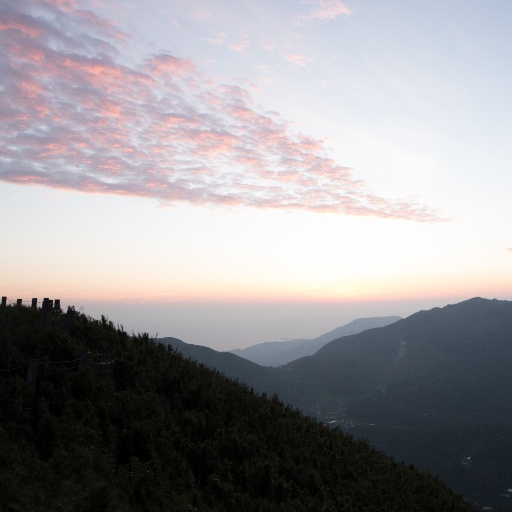}
  \vspace{-7mm}
	\end{minipage}
	\begin{minipage}[t]{0.19\linewidth}
		\centering
		\captionsetup{font={tiny}}
		\includegraphics[height=2.7cm,width=2.7cm]{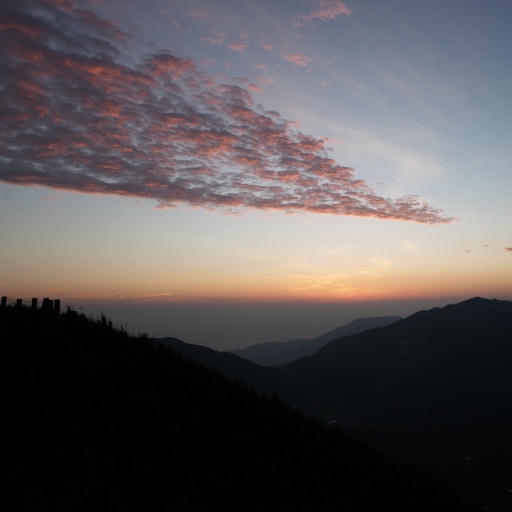}
          \vspace{-7mm}
	\end{minipage}
	\begin{minipage}[t]{0.19\linewidth}
		\centering
		\includegraphics[height=2.7cm,width=2.7cm]{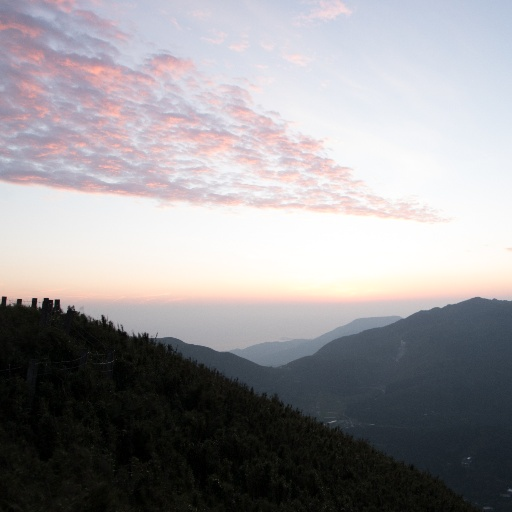}
		\captionsetup{font={tiny}}
         \vspace{-7mm}
	\end{minipage}
 	\begin{minipage}[t]{0.19\linewidth}
		\centering
		\captionsetup{font={tiny}}
		\includegraphics[height=2.7cm,width=2.7cm]{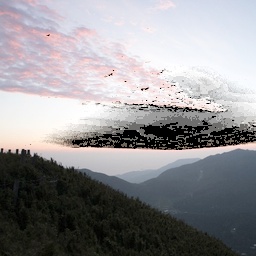}
         \vspace{-7mm}
	\end{minipage}
	\begin{minipage}[t]{0.19\linewidth}
		\centering
		\captionsetup{font={tiny}}
		\includegraphics[height=2.7cm,width=2.7cm]{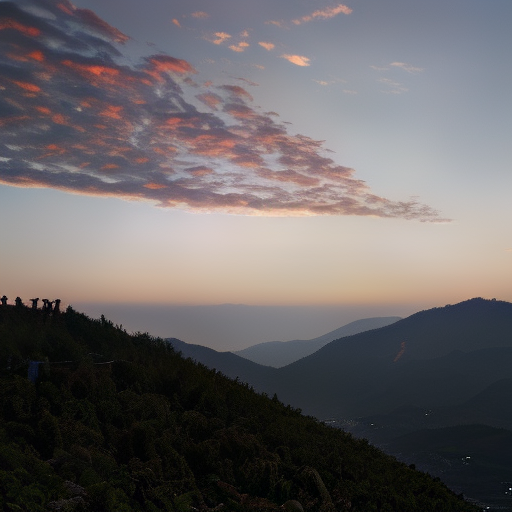}
  	\end{minipage}
  \\
  	\begin{minipage}[t]{0.19\linewidth}
		\centering
		\captionsetup{font={tiny}}
		\includegraphics[height=2.7cm,width=2.7cm]{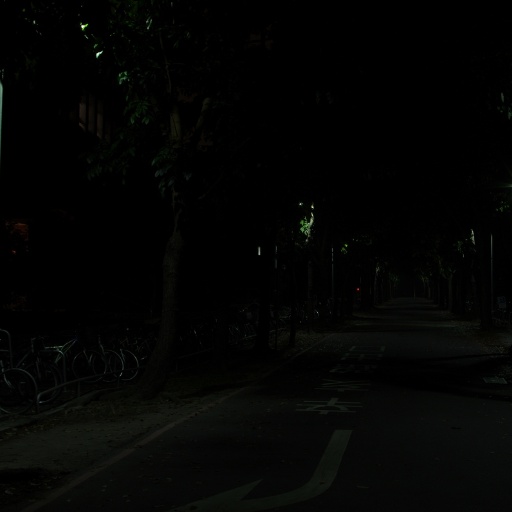}
  \vspace{-7mm}
	\end{minipage}
	\begin{minipage}[t]{0.19\linewidth}
		\centering
		\captionsetup{font={tiny}}
		\includegraphics[height=2.7cm,width=2.7cm]{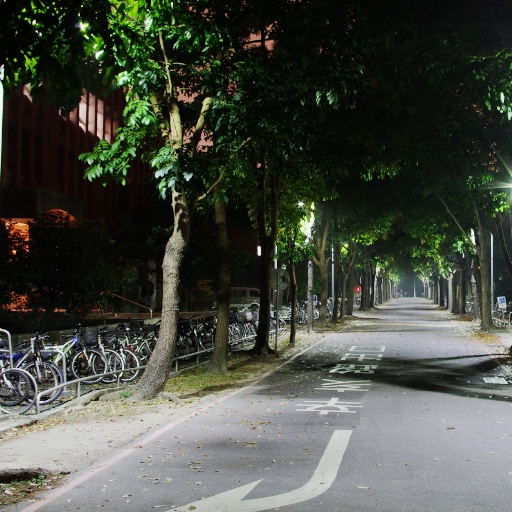}
          \vspace{-7mm}
	\end{minipage}
	\begin{minipage}[t]{0.19\linewidth}
		\centering
		\includegraphics[height=2.7cm,width=2.7cm]{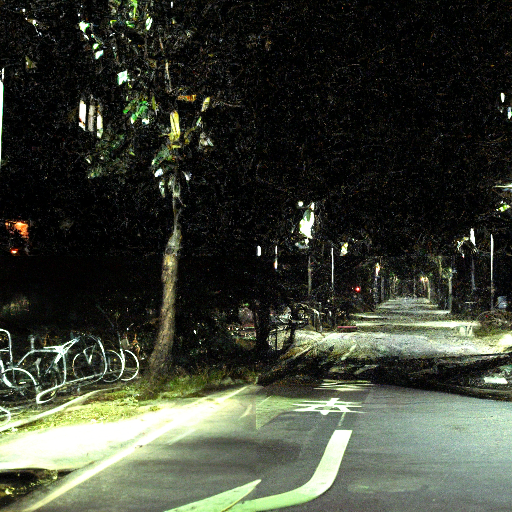}
		\captionsetup{font={tiny}}
         \vspace{-7mm}
	\end{minipage}
 	\begin{minipage}[t]{0.19\linewidth}
		\centering
		\captionsetup{font={tiny}}
		\includegraphics[height=2.7cm,width=2.7cm]{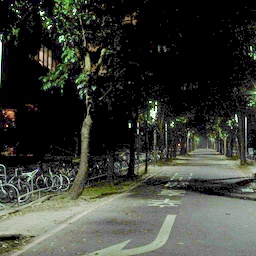}
         \vspace{-7mm}
	\end{minipage}
	\begin{minipage}[t]{0.19\linewidth}
		\centering
		\captionsetup{font={tiny}}
		\includegraphics[height=2.7cm,width=2.7cm]{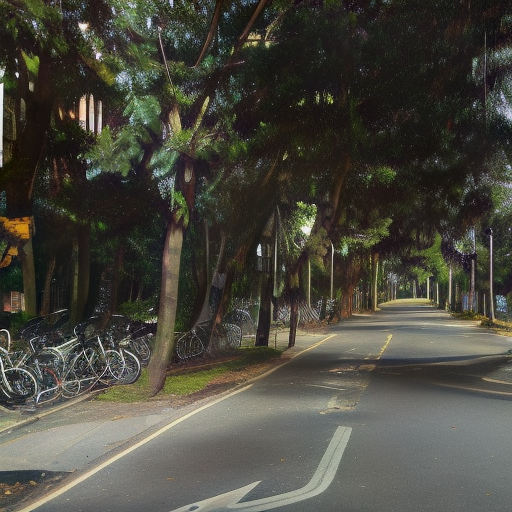}
  	\end{minipage}\\
  	\begin{minipage}[t]{0.19\linewidth}
		\centering
		\captionsetup{font={tiny}}
		\includegraphics[height=2.7cm,width=2.7cm]{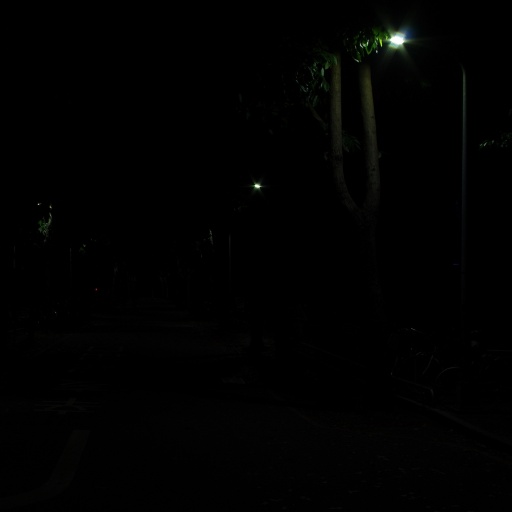}
  \vspace{-6mm}
        \caption*{LQ}
	\end{minipage}
	\begin{minipage}[t]{0.19\linewidth}
		\centering
		\captionsetup{font={tiny}}
		\includegraphics[height=2.7cm,width=2.7cm]{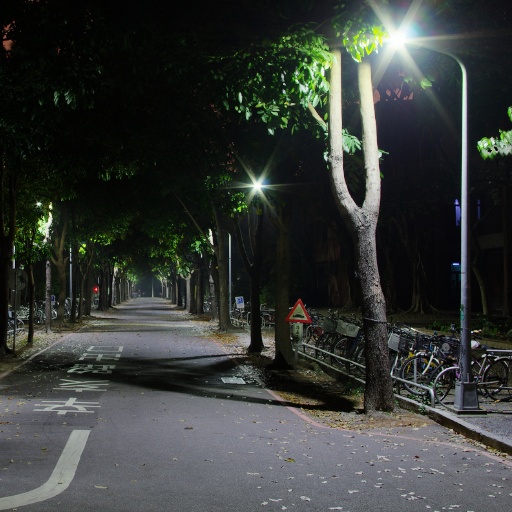}
          \vspace{-6mm}
\caption*{Reference}
	\end{minipage}
	\begin{minipage}[t]{0.19\linewidth}
		\centering
		\includegraphics[height=2.7cm,width=2.7cm]{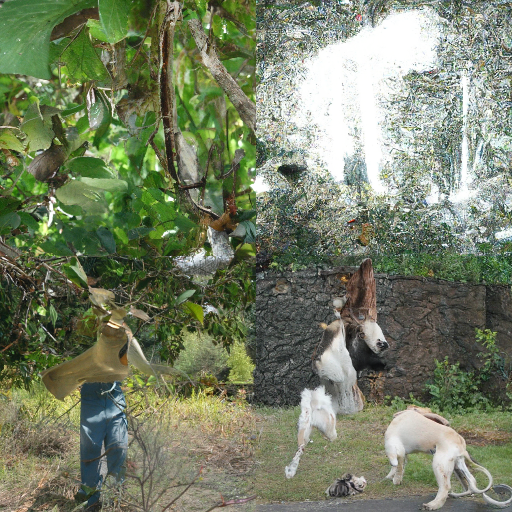}
		\captionsetup{font={tiny}}
         \vspace{-6mm}
 \caption*{GDP \cite{fei2023generative}}
	\end{minipage}
 	\begin{minipage}[t]{0.19\linewidth}
		\centering
		\captionsetup{font={tiny}}
		\includegraphics[height=2.7cm,width=2.7cm]{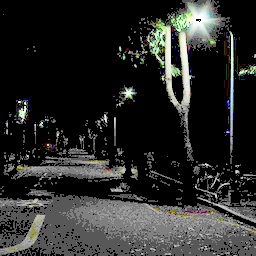}
         \vspace{-6mm}
 \caption*{GlowGAN \cite{wang2023glowgan}}
	\end{minipage}
	\begin{minipage}[t]{0.19\linewidth}
		\centering
		\captionsetup{font={tiny}}
		\includegraphics[height=2.7cm,width=2.7cm]{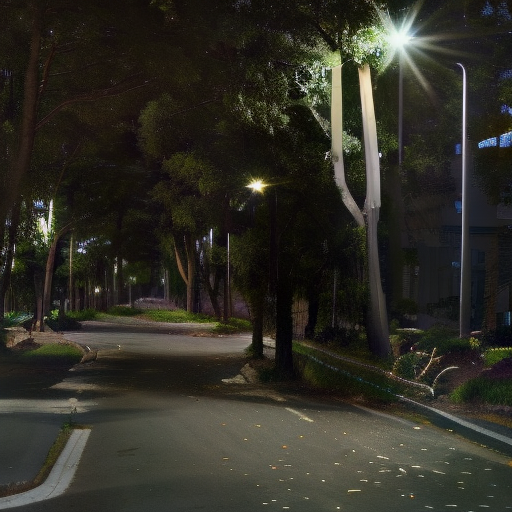}
         \vspace{-6mm}
 \caption*{LS-sagiri}
	\end{minipage}
	\caption{Comparison with other generative methods. GDP \cite{fei2023generative} lacks the ability to handle single LDR input effectively. (Top) When dealing with an overexposed image, GDP fails to adjust it to a normal brightness distribution. (Middle) It fails to generate foliage in dark areas. (Bottom) It fails to output a restored image and instead produce a completely new image. GlowGAN \cite{wang2023glowgan} often turns unrecoverable areas into black, failing to restore meaningful details. Our method can restore the image to its normal brightness and generate reasonable details in both over-/under-exposed regions.}
 \label{figure_generative}
 \end{figure}
(1) Our model excels in correcting dynamic range extremes by utilizing available information to restore lost details in both overexposed and underexposed areas. Furthermore, it 
sharpens
well-exposed regions with additional details, showcasing its comprehensive ability to improve image quality across various exposure levels.

(2) 
Our approach is designed to function as a versatile plug-in model, offering the capability to refine existing methods and bolster their efficacy in tasks related to HDR reconstruction and LDR enhancement.

(3) 
Our method requires a relatively short inference time, making it significantly more feasible for real-world applications.

We have evaluated the performance of these methods using one NVIDIA A100 GPU, with inference time comparison presented in Table \ref{table_infer} and visual comparisons shown in Figure \ref{figure_generative}.
\begin{figure}[t]
	\centering
	\begin{minipage}[t]{0.32\linewidth}
		\centering
		\captionsetup{font={tiny}}
\includegraphics[height=4.4cm,width=4.4cm]{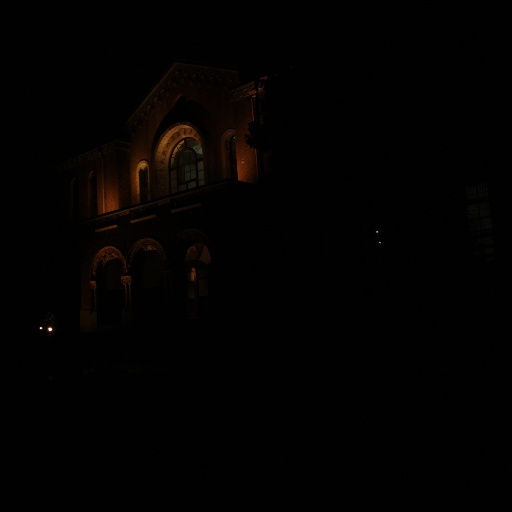}
      \vspace{-6mm}
        \caption*{LQ}
	\end{minipage}
	\begin{minipage}[t]{0.32\linewidth}
		\centering
		\captionsetup{font={tiny}}\includegraphics[height=4.4cm,width=4.4cm]{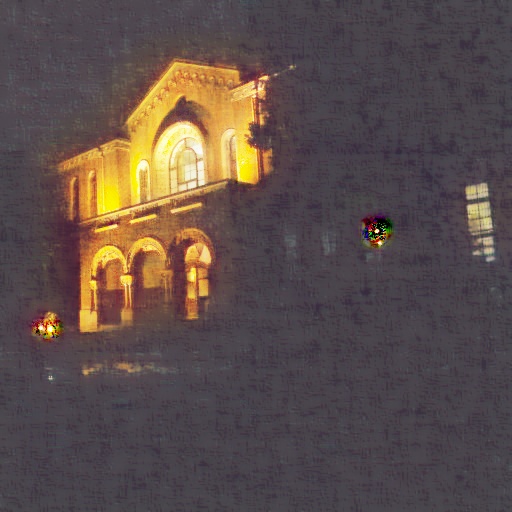}
            \vspace{-6mm}
\caption*{SingleHDR \cite{liu2020single}}
	\end{minipage}
	\begin{minipage}[t]{0.32\linewidth}
		\centering
		\includegraphics[height=4.4cm,width=4.4cm]{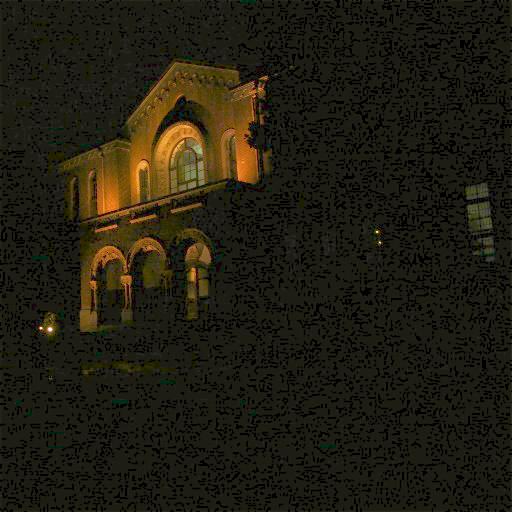}
		\captionsetup{font={tiny}}
           \vspace{-6mm}
 \caption*{LCDPNet \cite{wang2022local}}
	\end{minipage}
 	\begin{minipage}[t]{0.32\linewidth}
		\centering
		\captionsetup{font={tiny}}
		\includegraphics[height=4.4cm,width=4.4cm]{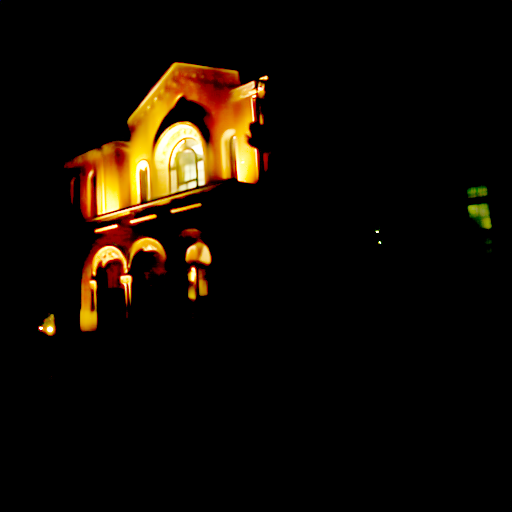}
      \vspace{-6mm}
        \caption*{HDRUNet \cite{chen2021hdrunet}}
	\end{minipage}
	\begin{minipage}[t]{0.32\linewidth}
		\centering
		\captionsetup{font={tiny}}
		\includegraphics[height=4.4cm,width=4.4cm]{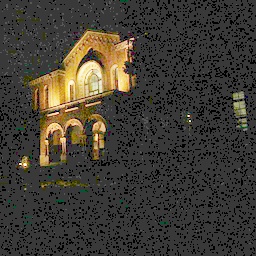}
            \vspace{-6mm}
\caption*{GlowGAN \cite{wang2023glowgan}}
	\end{minipage}
	\begin{minipage}[t]{0.32\linewidth}
		\centering
		\includegraphics[height=4.4cm,width=4.4cm]{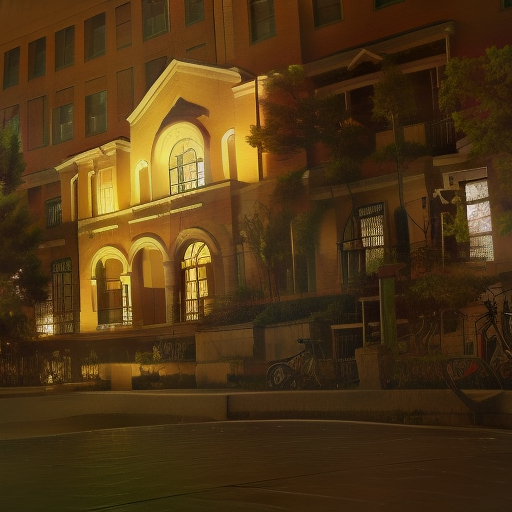}
		\captionsetup{font={tiny}}
           \vspace{-6mm}
 \caption*{LS-Sagiri}
	\end{minipage}\\
	\caption{Visual comparison with SingleHDR \cite{liu2020single}, LCDPNet \cite{wang2022local}, HDRUNet \cite{chen2021hdrunet} and GlowGAN \cite{wang2023glowgan}. It is also one comparison example used in user study.}
	\label{compare_others}
 \vspace{-5mm}
\end{figure}
\section{User Study}
To conduct a broader and more thorough evaluation of our results, we designed a user study aimed at gauging user preferences across various LDR enhancement methods. This includes LS-Sagiri, SingleHDR~\cite{liu2020single}, LCDPNet~\cite{wang2022local}, HDRUNet~\cite{chen2021hdrunet}, and GlowGAN~\cite{wang2023glowgan}. 30 participants were asked to assess (1) which output offered a superior visual experience, and (2) which output aligns more closely with their expectations for ideally restoring the Low-Quality (LQ) image. The results are compiled in Table~\ref{table_user}. Besides, we also show one comparison example used in our user study in Figure~\ref{compare_others}.
The percentage values in the table indicate the share of users who favored each method. Our analysis demonstrates that our strategy outperforms the alternatives on visual performance.
\begin{table}[h]
\caption{User study on preference for our method over existing methods. The data indicates the percentage of users who chose each method, with our method being the dominant choice.}
\label{table_user}
    \centering
    \footnotesize
    \begin{tabular}{ccccccc}
\toprule
- & SingleHDR & LCDPNet & HDRUNet & GlowGAN & LS-Sagiri \\ 
\midrule
Image 1  & 0.0\%       & 10.0\%                  & 0.0\%     & 0.0\%     & \textbf{90.0\%}                   \\
Image 2  & 0.0\%       & 30.0\%               & 0.0\%     & 0.0\%       & \textbf{70.0\%}         \\
Image 3  & 0.0\%       & 0.0\%                  & 0.0\%     & 0.0\%       & \textbf{100.0\%}                      \\
Image 4  & 0.0\%       & 0.0\%                   & 0.0\%     & 0.0\%     & \textbf{100.0\%}                     \\
Image 5  & 3.3\%     & 0.0\%                  & 0.0\%     & 0.0\%     & \textbf{96.7\%}                     \\
Image 6  & 0.0\%       & 6.7\%                  & 3.3\%   & 0.0\%     & \textbf{90.0\%}\\
\bottomrule
    \end{tabular}

\end{table}
\section{Loss Function}
In this section, we discuss the loss function we specifically designed for our training pipeline.
\subsection{Color Reconstruction Loss}
Direct application of Mean Squared Error (MSE) loss for color mapping in the enhanced image is insufficient, as it primarily focuses on pixel-wise intensity differences. While MSE ensures overall similarity, it neglects important aspects such as color distribution and frequency-based details, which are crucial for maintaining color fidelity and texture details. Furthermore, our model aims to retain the existing color and content features of the LDR image as much as possible for subsequent operations, necessitating a more comprehensive loss function that addresses these concerns. Thus we introduce a Color Distribution Loss ($L_{cd}$) to enhance color fidelity by ensuring that the color distribution of the enhanced image matches that of the target. This loss is defined as the sum of the absolute differences between the histogram bins of the predicted and target images:
\begin{equation}
L_{cd} = \sum_{i=1}^{N} | H_{pred}(i) - H_{target}(i)|,
\end{equation}
where $N$ is the number of histogram bins, and $H(i)$ represents the value of the $i$-th bin in the histogram.

Additionally, we employ a Frequency Domain Preservation Loss ($L_{fdp}$) to capture and preserve the frequency components of the normalized images, which is crucial for maintaining texture details and adjusting lighting:
\begin{equation}
L_{fdp} = Avg(|FFT(pred)-FFT(target)|),
\end{equation}
where $FFT(\cdot)$ denotes the operation of computing the 2D Fast Fourier Transform of an image, transforming it from the spatial domain to the frequency domain.

The overall intensity differences across the image are minimized using MSE loss ($L_{mse}$), which helps to produce an enhanced image closely matching the target in terms of brightness, contrast, and overall appearance:

\begin{equation}
L_{mse} = \frac{1}{N} \sum_{i=1}^{N} (pred(i) - target(i))^2,
\end{equation}
where $pred(i)$ and $target(i)$ represent the pixel values of the predicted result and target image, respectively.
The total loss for color restoration is thus expressed as a weighted sum of these components:

\begin{equation}
L_{color} = \lambda_{1}L_{mse} + \lambda_{2}L_{cd} + \lambda_{3}L_{fdp},
\end{equation}
where $\lambda_{1}$, $\lambda_{2}$, and $\lambda_{3}$ are the weights assigned to each loss component.
\subsection{Content Reconstruction Loss}
To ensure that Sagiri effectively generates realistic content details, it is essential to constrain the model from multiple perspectives. Firstly, we employ the Structural Similarity Index Measure (SSIM) Loss to prioritize the structural fidelity and perceptual similarity of the generated images. The SSIM Loss is defined as follows:
\vspace{-1mm}
\begin{align}
\text{SSIM}(x, y) &= \frac{(2\mu_x \mu_y + C_1)(2\sigma_{xy} + C_2)}{(\mu_x^2 + \mu_y^2 + C_1)(\sigma_x^2 + \sigma_y^2 + C_2)},\\
\
L_{ssim} &= 1 - \text{SSIM}(pred, target),
\vspace{-1mm}
\end{align}
where $\mu_{x}$ and $\mu_{y}$ represent the average pixel values of $x$ and $y$, respectively; $\sigma_x^2$ and $\sigma_y^2$ denote the variances of $x$ and $y$, respectively; and $\sigma_{xy}$ is the covariance of $x$ and $y$. Constants $C_{1}$ and $C_{2}$ are included to stabilize the division with a small denominator. This loss focuses on the changes in contrast and structure between the predicted and target images.

In addition to SSIM Loss, we utilize the Frequency Domain Preservation Loss ($L_{fdp}$) to preserve textures and fine details that are often lost when focusing solely on pixel intensity differences. The MSE Loss ($L_{mse}$) is also employed to maintain global consistency across the image.

The overall loss function during the training of Sagiri can be summarized as:
\begin{equation}
L_{content} = \lambda_{4}L_{mse} + \lambda_{5}L_{ssim} + \lambda_{6}L_{fdp}.
\end{equation}

This combination of loss functions ensures that Sagiri generates content with high structural fidelity, realistic
textures, and overall consistency, leading to more authentic and visually pleasing results. \sm{During training, the weights $\lambda_{1}$, $\lambda_{2}$, $\lambda_{3}$, $\lambda_{4}$, $\lambda_{5}$, and $\lambda_{6}$ are set to 10, 1, 0.1, 1, 1, and 0.01, respectively. }

\section{Ablation Study of Losses}
\label{sec:ablation_losses}
\begin{table}
    \vspace{-4mm}
\caption{Ablation study on HDR-Real and HDR-Eye datasets. LS-MSEloss represents using MSE loss for restoration stage, while LS-ColRloss represents using our color reconstruction loss for better color and brightness adjustment.}
    \centering
    \footnotesize
    \begin{tabular}{ccccccc}
\toprule
\multicolumn{1}{c}{Datasets} & \multicolumn{3}{c}{HDR-Real} & \multicolumn{3}{c}{HDR-Eye}  \\ 
\midrule
\multicolumn{1}{c}{Metrics} & \multicolumn{1}{c}{PSNR} & \multicolumn{1}{c}{SSIM} & \multicolumn{1}{c}{LPIPS} & \multicolumn{1}{c}{PSNR} & \multicolumn{1}{c}{SSIM} & \multicolumn{1}{c}{LPIPS}  \\ \midrule
LS-MSEloss & 20.543 & 0.674 & 0.351 & 19.733 & 0.688 & 0.260  \\ 
LS-ColRloss & \textbf{20.954} & \textbf{0.694} & \textbf{0.293} & \textbf{19.975}& \textbf{0.707} & \textbf{0.200}\\ 
\bottomrule
    \end{tabular}
    \vspace{-3mm}
    \label{tab_ablation}
\end{table}
We demonstrate the effectiveness of our color reconstruction loss by evaluating reference metrics in the restoration stage, with 
results shown in Table~\ref{tab_ablation}. We select PSNR, SSIM~\cite{ssim} and LPIPS~\cite{lpips} to test its 
recovery performance. For Sagiri's ablation study, visualizations are presented in Figure~\ref{ablation}. It is evident that, compared to the pipeline without content reconstruction loss (w/o ConRloss), our approach generates more high-quality texture details.\\
\section{Performance of GlowGAN+Sagiri}
\label{sec:glowgan_performance}
While Sagiri demonstrates powerful generation capabilities \sm{to apply fine adjustment to results from existing reconstruction models},
it faces challenges when applied to generative models like GlowGAN \cite{wang2023glowgan}. In scenarios 
where unrecoverable parts are processed as completely black, without any content or texture hints, our model 
struggles to refine them into satisfactory results. This limitation is evident in Figure \ref{compare_glow} and Table \ref{tab1}, where we cannot achieve better results in such cases. 

\section{More Visual Results}
\begin{figure}[h]
	\centering
	\begin{minipage}[t]{0.32\linewidth}
		\centering
		\captionsetup{font={tiny}}
\includegraphics[height=4.4cm,width=4.4cm]{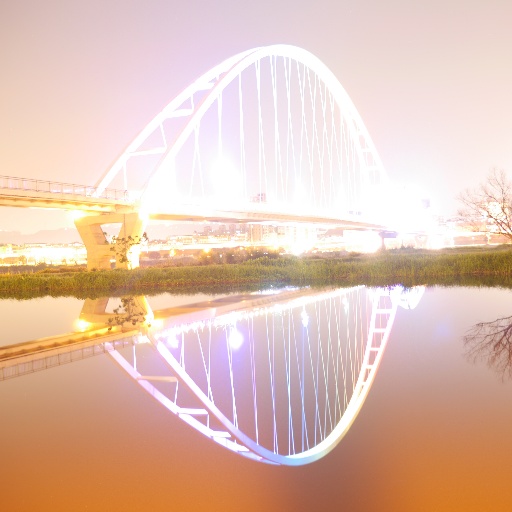}
      \vspace{-5mm}
	\end{minipage}
	\begin{minipage}[t]{0.32\linewidth}
		\centering
		\captionsetup{font={tiny}}\includegraphics[height=4.4cm,width=4.4cm]{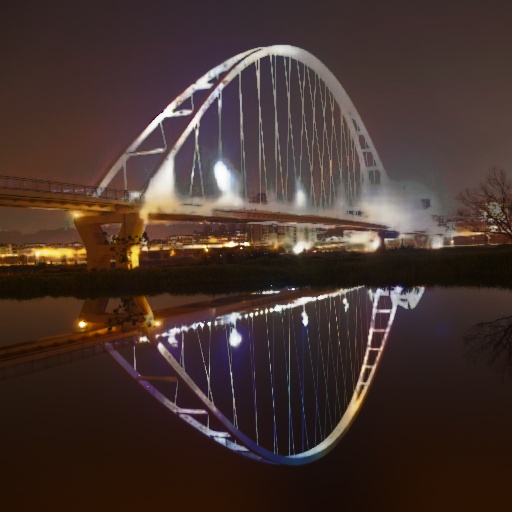}
            \vspace{-5mm}
	\end{minipage}
	\begin{minipage}[t]{0.32\linewidth}
		\centering
		\includegraphics[height=4.4cm,width=4.4cm]{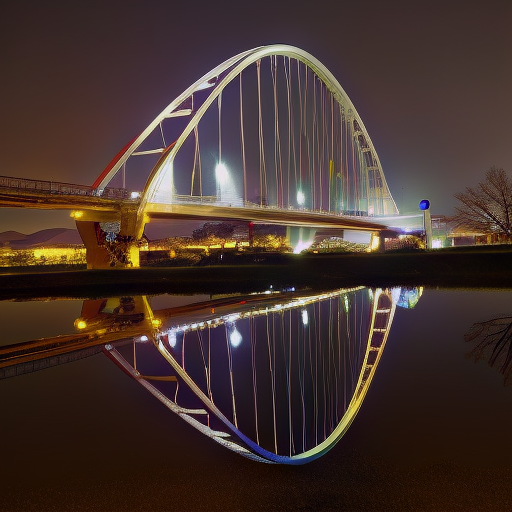}
		\captionsetup{font={tiny}}
           \vspace{-5mm}
	\end{minipage}
 	\begin{minipage}[t]{0.32\linewidth}
		\centering
		\captionsetup{font={tiny}}
\includegraphics[height=4.4cm,width=4.4cm]{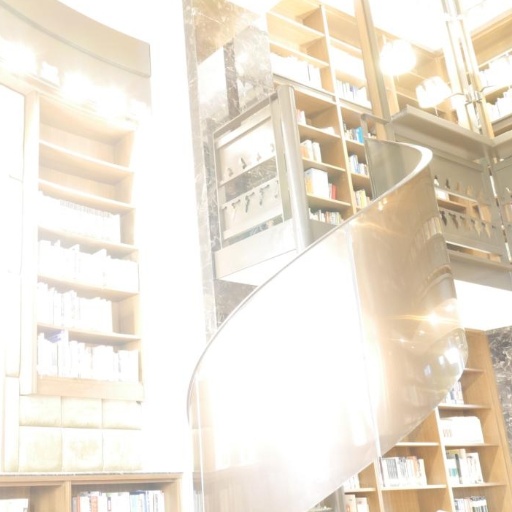}
      \vspace{-5mm}
	\end{minipage}
	\begin{minipage}[t]{0.32\linewidth}
		\centering
		\captionsetup{font={tiny}}\includegraphics[height=4.4cm,width=4.4cm]{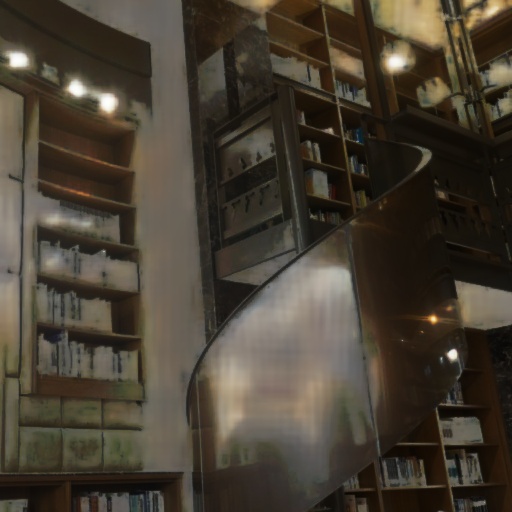}
            \vspace{-5mm}
	\end{minipage}
	\begin{minipage}[t]{0.32\linewidth}
		\centering
		\includegraphics[height=4.4cm,width=4.4cm]{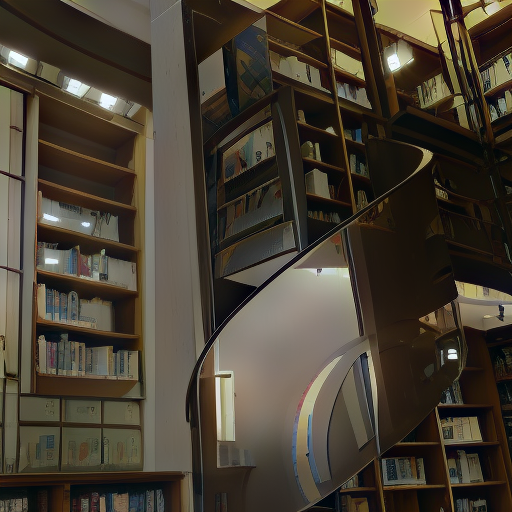}
		\captionsetup{font={tiny}}
           \vspace{-5mm}
	\end{minipage}
 	\begin{minipage}[t]{0.32\linewidth}
		\centering
		\captionsetup{font={tiny}}
		\includegraphics[height=4.4cm,width=4.4cm]{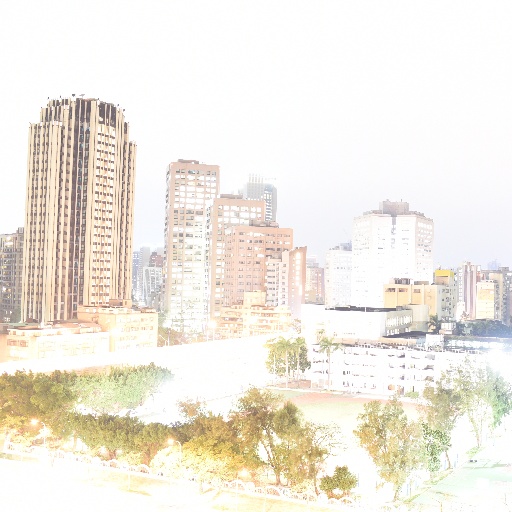}
      \vspace{-5mm}
	\end{minipage}
	\begin{minipage}[t]{0.32\linewidth}
		\centering
		\captionsetup{font={tiny}}
		\includegraphics[height=4.4cm,width=4.4cm]{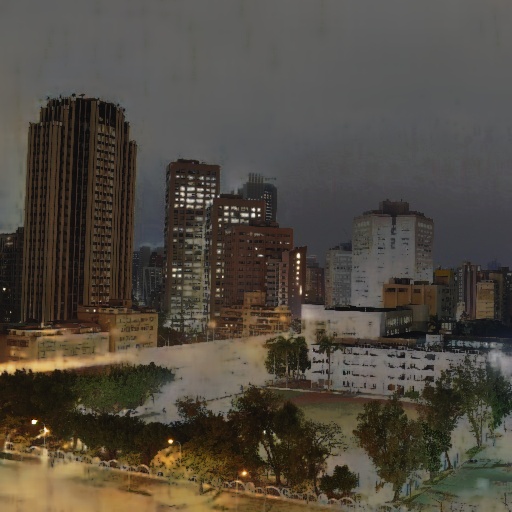}
            \vspace{-5mm}
	\end{minipage}
	\begin{minipage}[t]{0.32\linewidth}
		\centering
		\includegraphics[height=4.4cm,width=4.4cm]{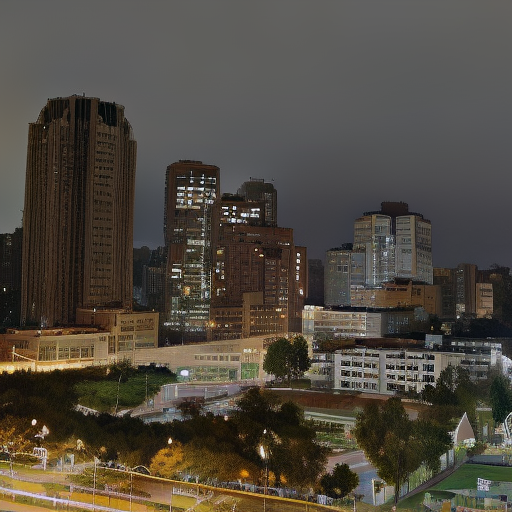}
		\captionsetup{font={tiny}}
           \vspace{-5mm}
	\end{minipage}\\
 \begin{minipage}[t]{0.32\linewidth}
		\centering
		\captionsetup{font={tiny}}
\includegraphics[height=4.4cm,width=4.4cm]{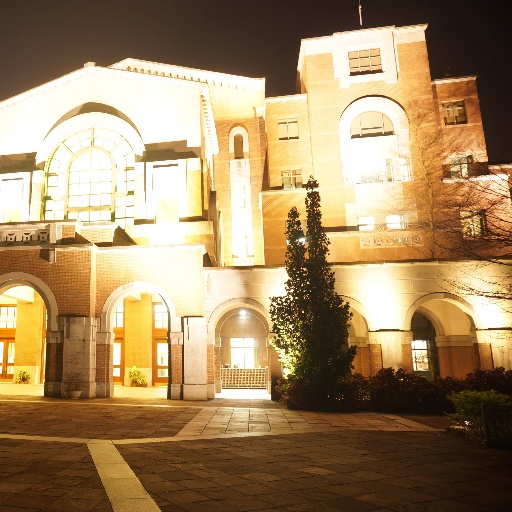}
      \vspace{-6mm}
        \caption*{LQ}
	\end{minipage}
	\begin{minipage}[t]{0.32\linewidth}
		\centering
		\captionsetup{font={tiny}}\includegraphics[height=4.4cm,width=4.4cm]{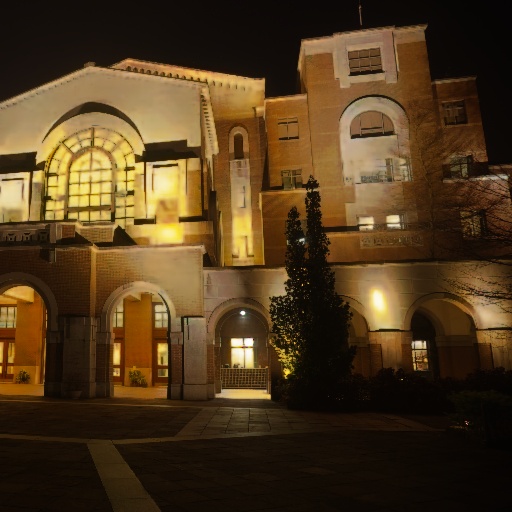}
            \vspace{-6mm}
\caption*{SingleHDR}
	\end{minipage}
	\begin{minipage}[t]{0.32\linewidth}
		\centering
		\includegraphics[height=4.4cm,width=4.4cm]{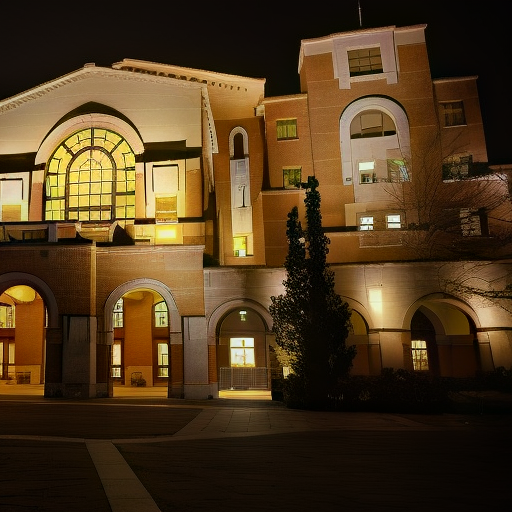}
		\captionsetup{font={tiny}}
           \vspace{-6mm}
 \caption*{SingleHDR+Sagiri}
	\end{minipage}
	\caption{Sagiri is a plug-and-play module and can enhance the results of SingleHDR \cite{liu2020single}  which is a relatively strong baseline.}
	\label{single_sarigi1}
 \vspace{-5mm}
\end{figure}

\begin{figure}[h]
	\centering
	\begin{minipage}[t]{0.32\linewidth}
		\centering
		\captionsetup{font={tiny}}
\includegraphics[height=4.4cm,width=4.4cm]{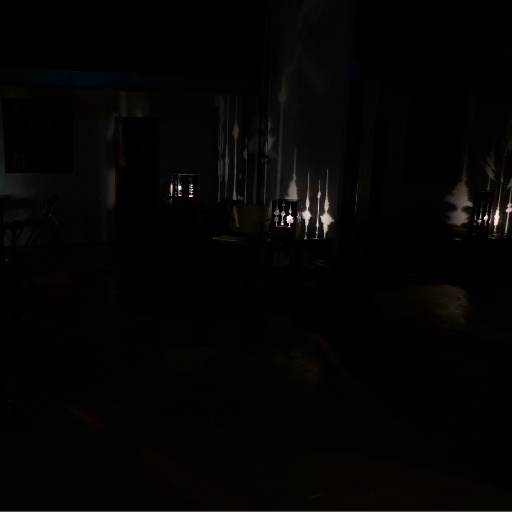}
      \vspace{-5mm}
	\end{minipage}
	\begin{minipage}[t]{0.32\linewidth}
		\centering
		\captionsetup{font={tiny}}\includegraphics[height=4.4cm,width=4.4cm]{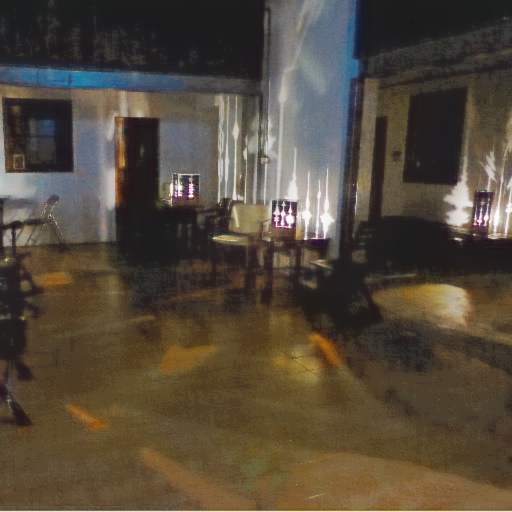}
            \vspace{-5mm}
	\end{minipage}
	\begin{minipage}[t]{0.32\linewidth}
		\centering
		\includegraphics[height=4.4cm,width=4.4cm]{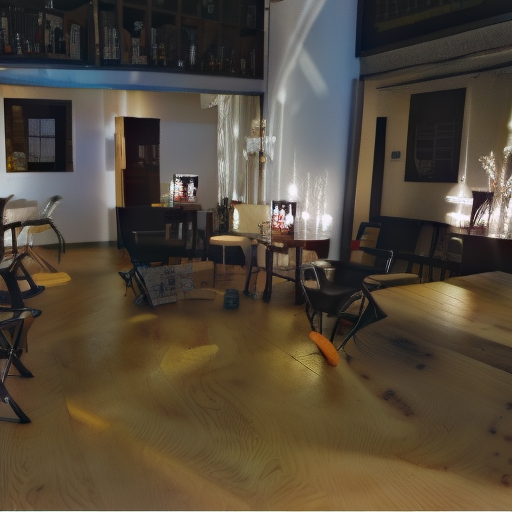}
		\captionsetup{font={tiny}}
           \vspace{-5mm}
	\end{minipage}\\
  	\begin{minipage}[t]{0.32\linewidth}
		\centering
		\captionsetup{font={tiny}}
		\includegraphics[height=4.4cm,width=4.4cm]{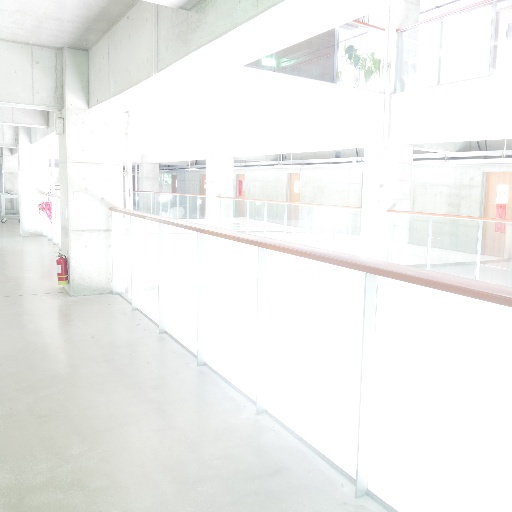}
      \vspace{-5mm}
	\end{minipage}
	\begin{minipage}[t]{0.32\linewidth}
		\centering
		\captionsetup{font={tiny}}
		\includegraphics[height=4.4cm,width=4.4cm]{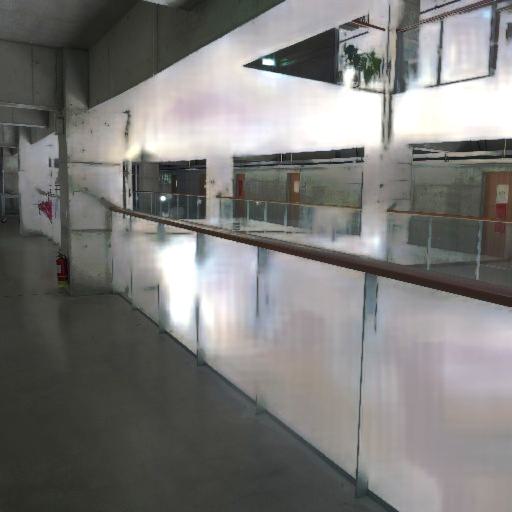}
            \vspace{-5mm}
	\end{minipage}
	\begin{minipage}[t]{0.32\linewidth}
		\centering
		\includegraphics[height=4.4cm,width=4.4cm]{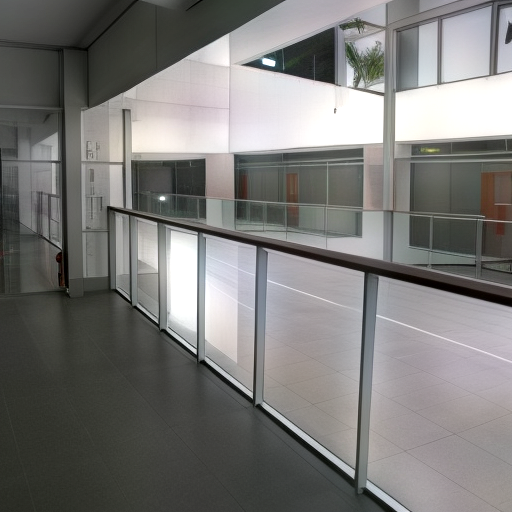}
		\captionsetup{font={tiny}}
           \vspace{-5mm}
	\end{minipage}\\
   	\begin{minipage}[t]{0.32\linewidth}
		\centering
		\captionsetup{font={tiny}}
		\includegraphics[height=4.4cm,width=4.4cm]{shows/00221_lq.jpg}
      \vspace{-5mm}
	\end{minipage}
	\begin{minipage}[t]{0.32\linewidth}
		\centering
		\captionsetup{font={tiny}}
		\includegraphics[height=4.4cm,width=4.4cm]{shows/00221_single.jpg}
            \vspace{-5mm}
	\end{minipage}
	\begin{minipage}[t]{0.32\linewidth}
		\centering
		\includegraphics[height=4.4cm,width=4.4cm]{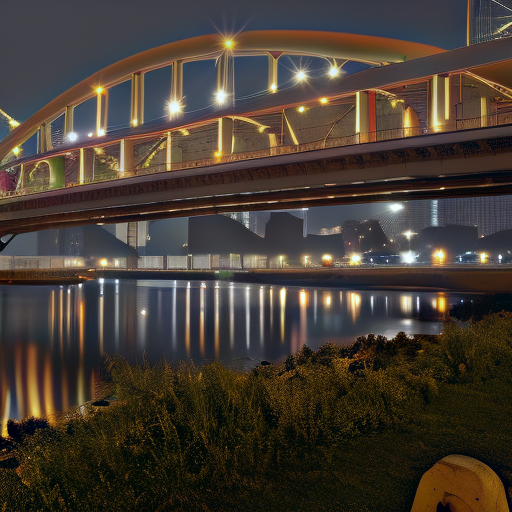}
		\captionsetup{font={tiny}}
           \vspace{-5mm}
	\end{minipage}\\
 	\begin{minipage}[t]{0.32\linewidth}
		\centering
		\captionsetup{font={tiny}}
		\includegraphics[height=4.4cm,width=4.4cm]{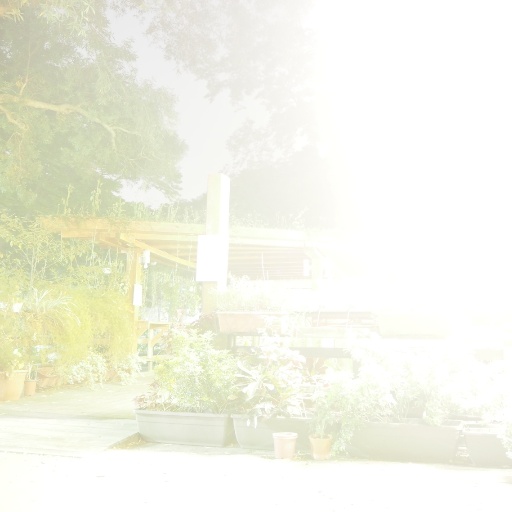}
      \vspace{-6mm}
        \caption*{LQ}
	\end{minipage}
	\begin{minipage}[t]{0.32\linewidth}
		\centering
		\captionsetup{font={tiny}}
		\includegraphics[height=4.4cm,width=4.4cm]{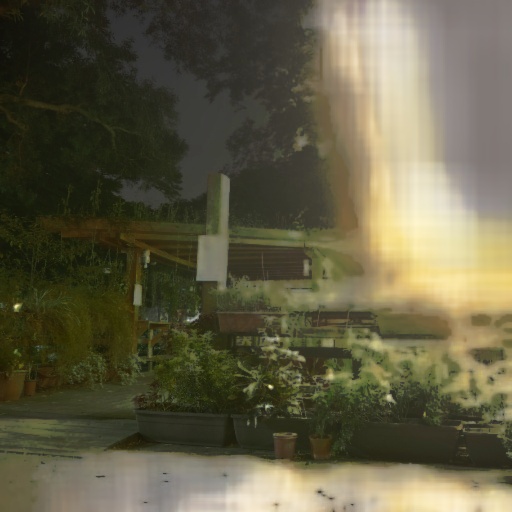}
            \vspace{-6mm}
\caption*{SingleHDR}
	\end{minipage}
	\begin{minipage}[t]{0.32\linewidth}
		\centering
		\includegraphics[height=4.4cm,width=4.4cm]{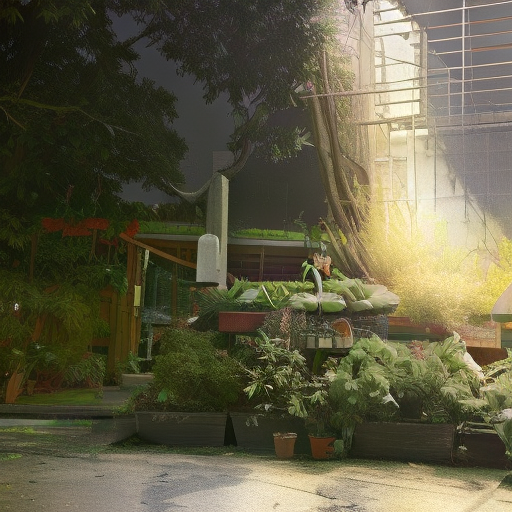}
		\captionsetup{font={tiny}}
           \vspace{-6mm}
 \caption*{SingleHDR+Sagiri}
	\end{minipage}\\
	\caption{Sagiri is a plug-and-play module and can enhance the results of SingleHDR \cite{liu2020single} which is a relatively strong baseline.}
	\label{single_sagiri2}
 \vspace{-5mm}
\end{figure}

\begin{figure}[t]
	\centering
	\begin{minipage}[t]{0.32\linewidth}
		\centering
		\captionsetup{font={tiny}}
\includegraphics[height=4.4cm,width=4.4cm]{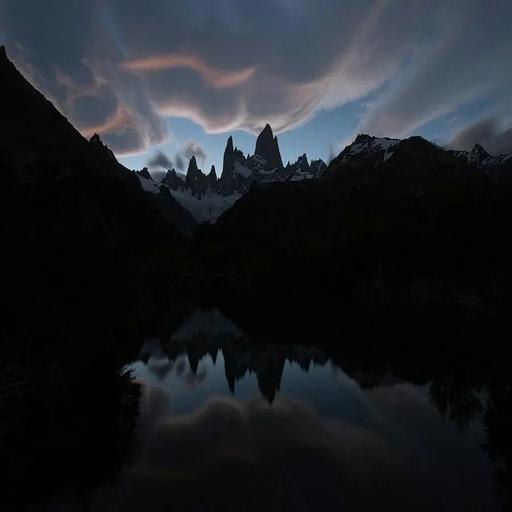}
      \vspace{-5mm}
	\end{minipage}
	\begin{minipage}[t]{0.32\linewidth}
		\centering
		\captionsetup{font={tiny}}\includegraphics[height=4.4cm,width=4.4cm]{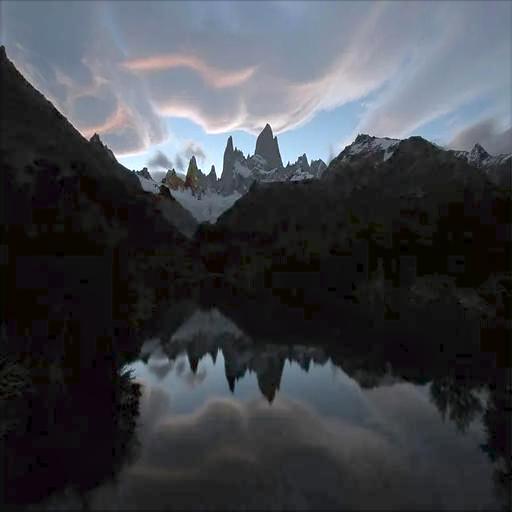}
            \vspace{-5mm}
	\end{minipage}
	\begin{minipage}[t]{0.32\linewidth}
		\centering
		\includegraphics[height=4.4cm,width=4.4cm]{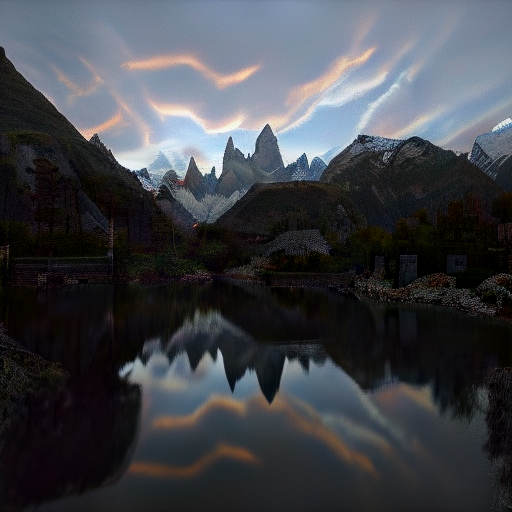}
		\captionsetup{font={tiny}}
           \vspace{-5mm}
	\end{minipage}\\
 	\begin{minipage}[t]{0.32\linewidth}
		\centering
		\captionsetup{font={tiny}}
		\includegraphics[height=4.4cm,width=4.4cm]{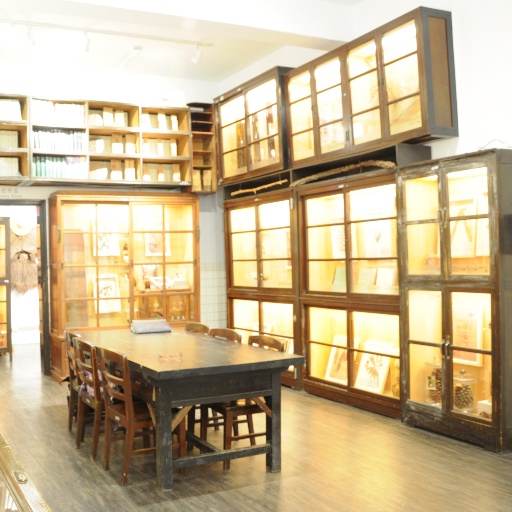}
      \vspace{-6mm}
        \caption*{LQ}
	\end{minipage}
	\begin{minipage}[t]{0.32\linewidth}
		\centering
		\captionsetup{font={tiny}}
		\includegraphics[height=4.4cm,width=4.4cm]{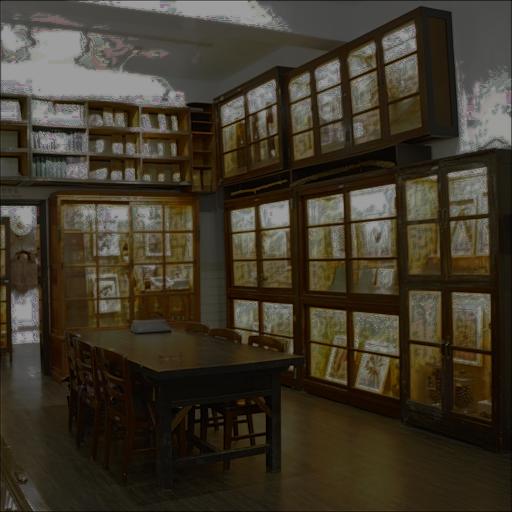}
            \vspace{-6mm}
\caption*{LCDPNet~\cite{wang2022local}}
	\end{minipage}
	\begin{minipage}[t]{0.32\linewidth}
		\centering
		\includegraphics[height=4.4cm,width=4.4cm]{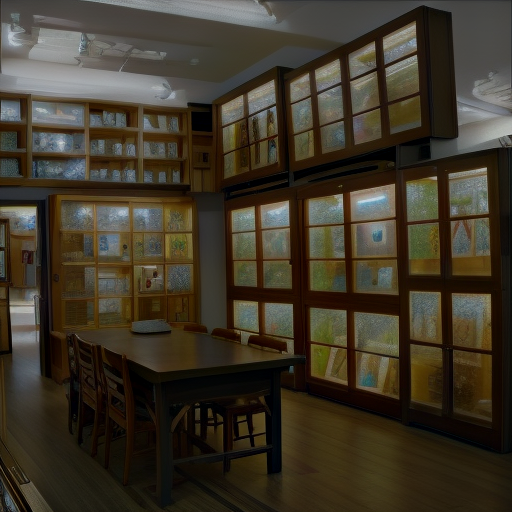}
		\captionsetup{font={tiny}}
           \vspace{-6mm}
 \caption*{LCDPNet+Sagiri}
	\end{minipage}\\
	\caption{Sagiri is a plug-and-play module and can enhance the results of LCDPNet \cite{wang2022local}.}
	\label{lcdpsagiri}
 \vspace{-5mm}
\end{figure}

\begin{figure}[t]
	\centering
	\begin{minipage}[t]{0.32\linewidth}
		\centering
		\captionsetup{font={tiny}}
\includegraphics[height=4.4cm,width=4.4cm]{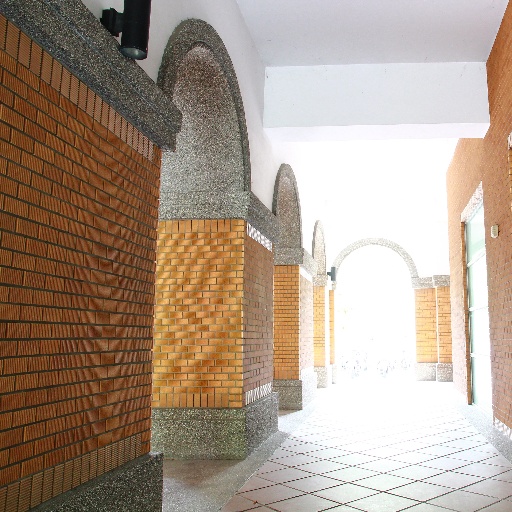}
      \vspace{-5mm}
	\end{minipage}
	\begin{minipage}[t]{0.32\linewidth}
		\centering
		\captionsetup{font={tiny}}\includegraphics[height=4.4cm,width=4.4cm]{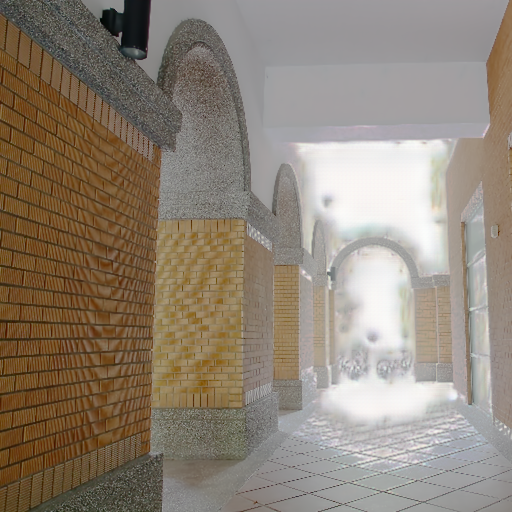}
            \vspace{-5mm}
	\end{minipage}
	\begin{minipage}[t]{0.32\linewidth}
		\centering
		\includegraphics[height=4.4cm,width=4.4cm]{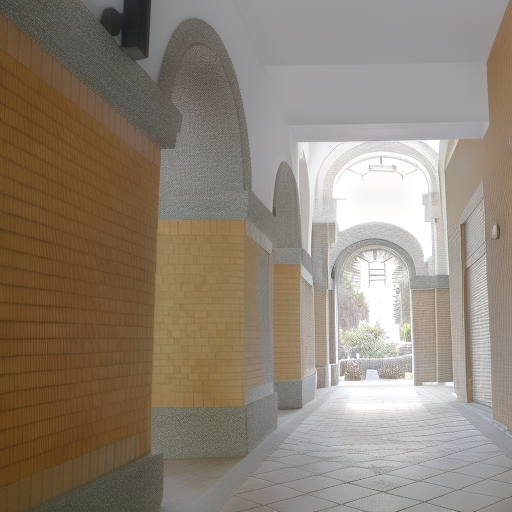}
		\captionsetup{font={tiny}}
           \vspace{-5mm}
	\end{minipage}
 	\begin{minipage}[t]{0.32\linewidth}
		\centering
		\captionsetup{font={tiny}}
		\includegraphics[height=4.4cm,width=4.4cm]{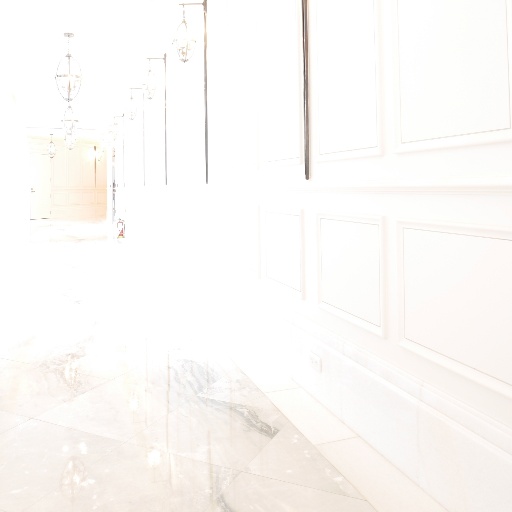}
      \vspace{-5mm}
        \caption*{LQ}
	\end{minipage}
	\begin{minipage}[t]{0.32\linewidth}
		\centering
		\captionsetup{font={tiny}}
		\includegraphics[height=4.4cm,width=4.4cm]{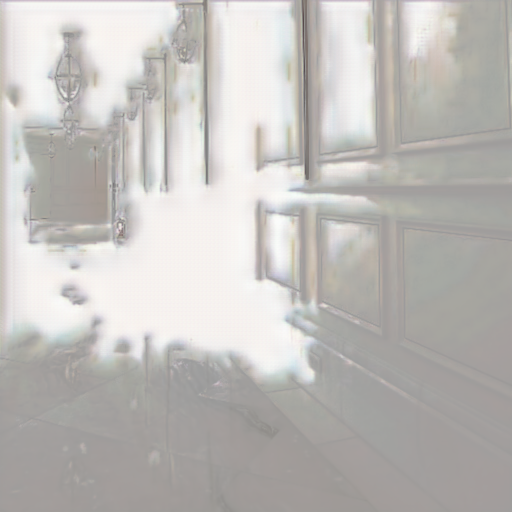}
            \vspace{-5mm}
\caption*{HDRUNet}
	\end{minipage}
	\begin{minipage}[t]{0.32\linewidth}
		\centering
		\includegraphics[height=4.4cm,width=4.4cm]{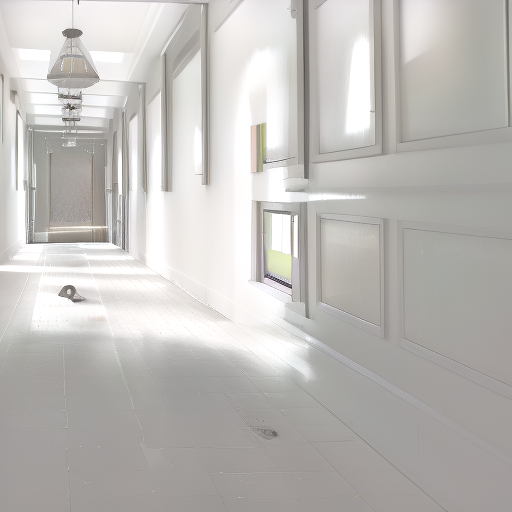}
		\captionsetup{font={tiny}}
           \vspace{-5mm}
 \caption*{HDRUNet+Sagiri}
	\end{minipage}\\
	\caption{Sagiri is a plug-and-play module and can enhance the results of HDRUNet \cite{chen2021hdrunet}.}
	\label{hdru_sagiri}
 \vspace{-5mm}
\end{figure}

We integrated Sagiri into various models to further validate its robust plug-and-play capabilities.
These visual results can be seen in Figures \ref{single_sarigi1} and \ref{single_sagiri2} on SingleHDR \cite{liu2020single}, Figure \ref{lcdpsagiri} on LCDPNet \cite{wang2022local}, and Figure \ref{hdru_sagiri} on HDRUNet \cite{chen2021hdrunet}. 

\begin{figure*}[t]
	\centering
	\begin{minipage}[t]{0.24\linewidth}
		\centering
		\captionsetup{font={tiny}}
\includegraphics[height=3.4cm,width=3.4cm]{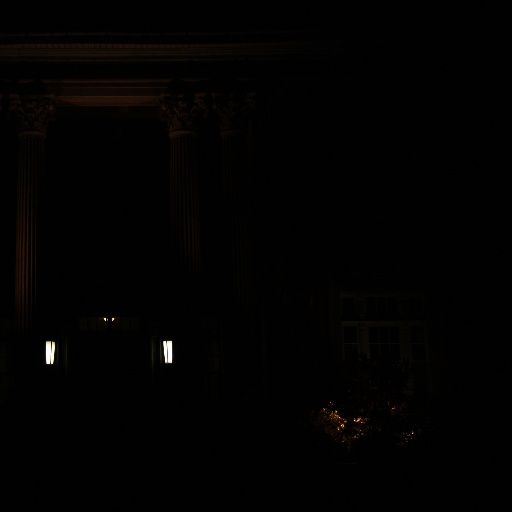}
      \vspace{-5mm}
        \caption*{LQ}
	\end{minipage}
	\begin{minipage}[t]{0.24\linewidth}
		\centering
		\captionsetup{font={tiny}}\includegraphics[height=3.4cm,width=3.4cm]{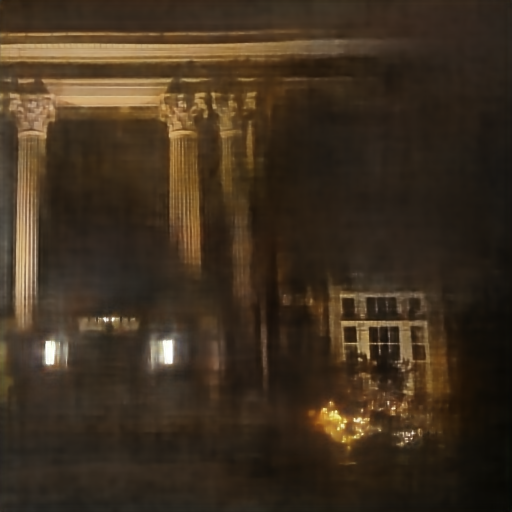}
            \vspace{-5mm}
\caption*{LatentSwinIR$_{c}$}
	\end{minipage}
	\begin{minipage}[t]{0.24\linewidth}
		\centering
		\includegraphics[height=3.4cm,width=3.4cm]{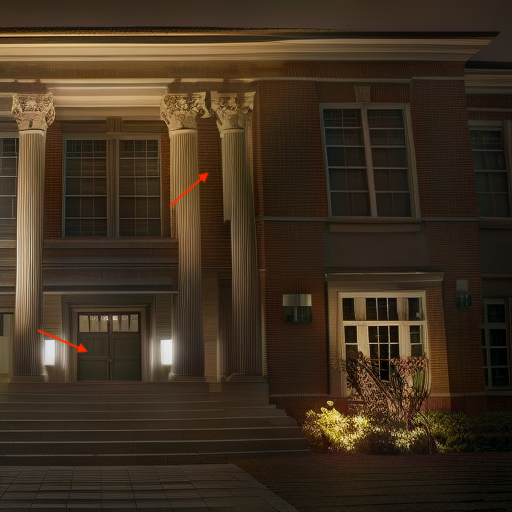}
		\captionsetup{font={tiny}}
           \vspace{-5mm}
 \caption*{LS-Sagiri (Prompt a)}
	\end{minipage}
 	\begin{minipage}[t]{0.24\linewidth}
		\centering
		\captionsetup{font={tiny}}
		\includegraphics[height=3.4cm,width=3.4cm]{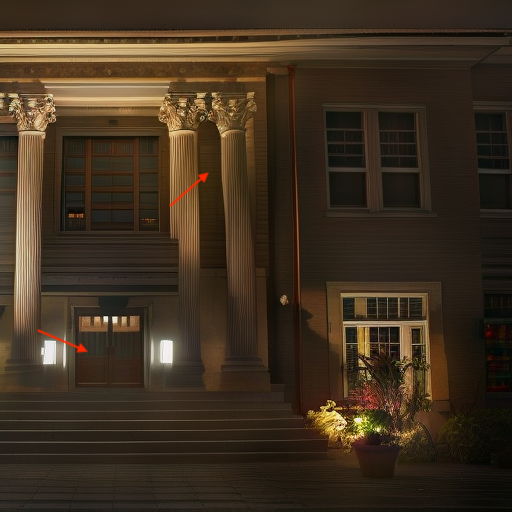}
      \vspace{-5mm}
        \caption*{LS-Sagiri (Prompt b)}
	\end{minipage}\\
	\caption{Use different prompts to control the generated results. Prompt a: `A building with a \textbf{red} brick exterior, white columns, and a \textbf{black} door...' Prompt b: `A building with a \textbf{black} brick exterior, white columns, and a \textbf{red} door...'. Please zoom in to see more details.}
	\label{prompt_supp}
 \vspace{-5mm}
\end{figure*}

\begin{figure*}[t]
	\centering
	\begin{minipage}[t]{0.24\linewidth}
		\centering
		\captionsetup{font={tiny}}
\includegraphics[height=3.4cm,width=3.4cm]{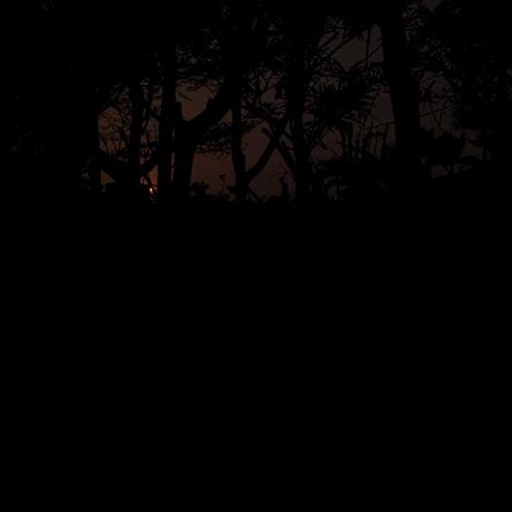}
      \vspace{-5mm}
        \caption*{LQ}
	\end{minipage}
	\begin{minipage}[t]{0.24\linewidth}
		\centering
		\captionsetup{font={tiny}}\includegraphics[height=3.4cm,width=3.4cm]{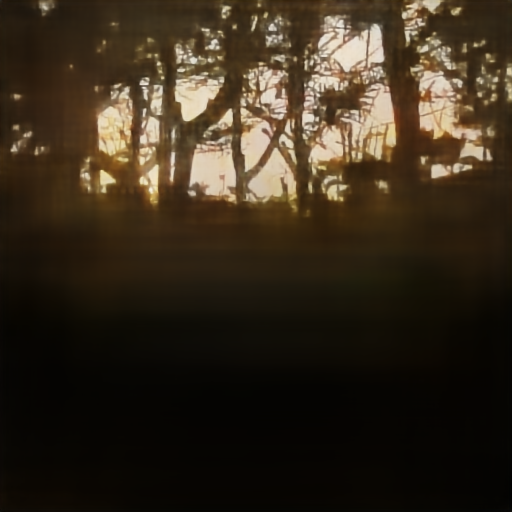}
            \vspace{-5mm}
\caption*{LatentSwinIR$_{c}$}
	\end{minipage}
	\begin{minipage}[t]{0.24\linewidth}
		\centering
		\includegraphics[height=3.4cm,width=3.4cm]{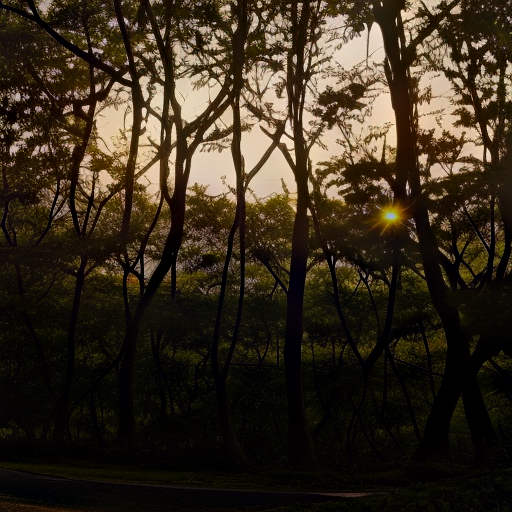}
		\captionsetup{font={tiny}}
           \vspace{-5mm}
 \caption*{LS-Sagiri(prompt a)}
	\end{minipage}
 	\begin{minipage}[t]{0.24\linewidth}
		\centering
		\captionsetup{font={tiny}}
		\includegraphics[height=3.4cm,width=3.4cm]{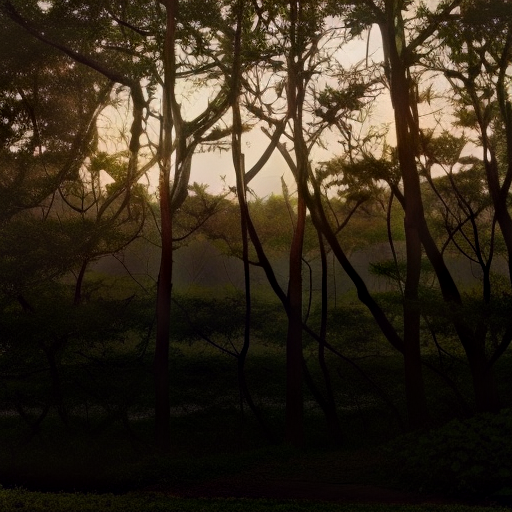}
      \vspace{-5mm}
        \caption*{LS-Sagiri(prompt b)}
	\end{minipage}\\
	\caption{The model has poor responsiveness to prompts that do not fit the current context, as we found. Prompt a: `The \textbf{sun} is setting in the forest, and the trees are \textbf{black}.' Prompt b: `The \textbf{moon} is setting in the forest, and the trees are \textbf{green}'. Please zoom in to see more details.}
	\label{prompt_suppv3}
 \vspace{-5mm}
\end{figure*}

We also show more visual results of using prompts to control the results of dynamic range extreme region generation, as shown in Figures \ref{prompt_supp} and \ref{prompt_suppv3}.


\clearpage
\clearpage

\clearpage
\end{document}